\documentclass[reprint, amsmath,amssymb,aps,onecolumn]{revtex4-2}

\usepackage{graphicx}
\usepackage{bm, slashed}
\usepackage{tikz} 
\usepackage{hyperref}
\usepackage{comment}
\usepackage{mathdots,bbold}
\usepackage{soul}
\usepackage[normalem]{ulem}
\setcitestyle{square,numbers}

\begin{document}

\title{Quantum simulator of link models using spinor dipolar ultracold atoms}

\author{Pierpaolo Fontana}
 \email{pfontana@sissa.it}
\affiliation{
 SISSA and INFN, Sezione di Trieste, Via Bonomea 265, I-34136 Trieste, Italy
}

\author{Joao C. Pinto Barros}
\affiliation{
	Institut für Theoretische Physik, ETH Zürich, Wolfgang-Pauli-Str. 27, 8093 Zürich, Switzerland
}

\author{Andrea Trombettoni}
\affiliation{%
 Department of Physics, University of Trieste, Strada Costiera 11, I-34151 Trieste, Italy
}
\affiliation{SISSA and INFN, Sezione di Trieste, Via Bonomea 265,
I-34136 Trieste, Italy}

\date{\today}

\begin{abstract}
    We propose a scheme for the quantum simulation of quantum link models in two-dimensional lattices. Our approach considers spinor dipolar gases on a suitably shaped lattice, where the dynamics of particles in the different hyperfine levels of the gas takes place in one-dimensional chains coupled by the dipolar interactions. We show that at least four levels are needed. The present scheme does not require any particular fine-tuning of the parameters. We perform the  derivation of the parameters of the quantum link models by means of two different approaches, a non-perturbative one tied to angular momentum conservation, and a perturbative one.
    A comparison with other schemes for $(2+1)$-dimensional quantum link models present in literature is discussed. Finally, the extension to three-dimensional lattices is presented, and its subtleties are pointed out.
\end{abstract}

\maketitle

\section{\label{intro}Introduction}
Quantum simulators are of fundamental importance in the realm of quantum and science technologies: they are quantum systems having properties that can be controlled and used to simulate some target system,
whose study is currently hindered by lack of proper classical computational, experimental or analytical tools \cite{feynmanQS1982,Trabesinger2012}. In the last decades, there has been a formidable development in the fields of quantum optics and atomic physics, allowing for the realization of highly precise and controllable platforms by means of trapped ions \cite{Blatt_trappedions2012}, superconducting circuits \cite{Houck_SCcircuits2012}, Rydberg atoms \cite{Wu_2021} and ultracold atoms in optical lattices \cite{Dalibard_UCatoms2012}. For these reasons, quantum simulators play a key role in various areas, including quantum chemistry, condensed matter and high energy physics \cite{Wiese,Dalmonte-Montangero,Zohar-Cirac-Reznik,Preskill2018,banuls_QTreview2020,aidelsburger_RSP2022,Zohar2022,Klco2022,davoudi_QSforHEP2022}.
Various many-body quantum systems have been analyzed as quantum simulators \cite{Martinez2016,Kokail2019,Schweizer2019,Mil2020,Yang2019,Semeghini2021,Zhou2022,RiechertPRB2022,Kalonowski2022}, a typical example being provided by atomic systems loaded in optical lattices, which are described at low energies
by (extended) Hubbard models \cite{Lewenstein_book_2012}.

Over the past years, the idea and use of quantum simulators to study gauge theories has gained increased relevance. These theories are at the basis of the Standard Model in the field of particle physics, and describe the electroweak and strong interactions through a non-Abelian gauge theory \cite{Peskin,Scwhartz,Maggiore}. In condensed matter and statistical physics, often gauge theories arise as low energies effective descriptions of strongly correlated phenomena, such as quantum spin liquids, quantum Hall effect and frustrated magnets \cite{Wen}. The standard approaches to study gauge theories may present various drawbacks, depending on the regimes and properties of interest. A major example is the analysis of quantum chromodynamics through Monte Carlo simulations: due to the well-known sign problem, this numerical method can not reliably approach the analysis of the theory with finite chemical potential, for example \cite{troyer_wiese2005,fukushima2010}. 

Quantum simulators based on ultracold atomic platforms emerge as a promising alternative to investigate such phenomena for lattice gauge theories (LGTs), by circumventing some limitations of classical simulators
\cite{Wiese,Zohar-Cirac-Reznik,Dalmonte-Montangero}. A first point to be addressed is related to the implementation of the Hilbert space of dynamical gauge fields in a quantum simulator, since it is infinite-dimensional for a single link in the Wilson formulation of LGTs \cite{Wilson}. To overcome this difficulty, one could replace the continuum gauge groups with discrete ones that approximate the latter in the proper limit \cite{Cirac_Banuls_PhysRevA,Ercolessi_PhysRevD}, or replace the link variables with discrete degrees of freedom, discretizing the Hilbert space considering the so-called {\em quantum link} formulation of gauge theories. Even if they possess a finite number of states, quantum link models (QLMs) preserve the gauge symmetry of the original model, paying the price of introducing non-unitary operators on the links of the lattice \cite{HORN1981,ORLAND1990,CHANDRASEKHARAN1997}. Due to the finiteness of the Hilbert space and the preservation of the local symmetry, they are suitable to be implemented and analysed as quantum simulators. While it is possible to recover the full, non-truncated, Wilson formulation from QLMs \cite{BrowerPRD1999,CiracPRL2012,ZoharPRD2015,Kasper_2017}, they provide an enriched playground where new phases are expected to appear, making them interesting also from this perspective
\cite{Banerjee_2013,Widmer_POS2014,banerjee2021,banerjee2021nematic,BanerjeePRL2021,banerjee2022}.

With respect to the quantum simulation of usual many-body quantum systems, there are additional features to be considered in the case of theories with gauge fields. The point is that the designed quantum platform should be consistent with the local symmetry, i.e. the \textit{gauge invariance}, of the theory. While in $d=1$ efficient ways to deal with the problem have been developed \cite{Martinez2016,Yang2019,Zhou2022,Dalmonte-Montangero}, in $d>1$ it seems that one necessarily needs to involve rather complicated many-body interaction terms to simulate the gauge fields dynamics. We mention that interesting physical phenomena can emerge in $d>1$ even in the absence of magnetic terms \cite{CardarelliPRL2017,OttPLB2020,GonzalezCuadraPRX2020,HashimuzeSciPost2022,Halimeh_arXiv_2022}. As a general consideration, one would like to have quantum simulation schemes which do not crucially depend on fine-tuning of the parameters of the systems, possibly not intrinsically perturbative, and extendable to higher dimensions.

For the $d>1$ case, both in the Wilson and quantum link formulations, different proposals have been put forward involving ultracold atoms in optical lattice and Rydberg atoms \cite{banuls_QTreview2020,aidelsburger_RSP2022}. Concerning the first platforms, in Ref. \cite{ZoharReznikPRA2013} the gauge invariance of the theory is obtained through angular momentum conservation for the gauge-matter interaction, while the dynamics of the gauge field emerges effectively in perturbation theory, employing the so-called ``loop method" in $d=2$ for the compact quantum electrodynamics (QED), realizing the plaquette term in terms of bosons. At variance, using the dual formulation \cite{Gras_Lewenstein2016} of the $U(1)$ spin$-1/2$ model in $d=2$,
plaquette interactions are mapped into single constrained hopping processes on the dual lattice. Ref. \cite{Celi2019} proposes to simulate this model through Rydberg configurable arrays, in which the physical states have a blockade character. While in \cite{ZoharReznikPRA2013} the plaquette terms are emerging at fourth-order of a strong coupling (cold atomic) expansion, in this proposal they are implemented directly, without the use of any perturbative expansion. At the same time, the approach of Ref. \cite{Celi2019} relies on the two-dimensional nature of the system, and does not seem to be easily generalizable to higher dimensions.

In this paper we propose a quantum simulator for the $U(1)$ spin$-1/2$ pure Abelian QLM using spinor dipolar Bose-Einstein condensates (BECs) loaded in a spin-dependent optical lattice. With respect to Ref. \cite{ZoharReznikPRA2013}, we use only bosonic atoms of spin-2, so that we have access to five internal states that, through angular momentum conservation in the various scattering channels, give rise to gauge invariance. As in Ref. \cite{ZoharReznikPRA2013}, the robustness of gauge invariance is tied to the one of angular momentum conservation, and in the present paper it is used to generate the plaquette term. The same principle can be achieved without conservation of angular momentum, provided the ultracold atom parameters are properly tuned in the strong coupling regime. The resulting effective Hamiltonian describes the dynamics of the gauge field at third-order in perturbation theory for a square lattice, or at second-order for a triangular lattice.

The paper is organized as follows. In Sec. \ref{halfspinLGT} we give a brief reminder about Abelian LGTs and then introduce the $U(1)$ spin-$1/2$ models in $d=2$, discussing both the bosonic and fermionic formulations. In Sec. \ref{spinordipolarBEC} we briefly review the theory about spinor dipolar BECs. In Sec. \ref{UCatomsrealization} we present our proposal using ultracold atomic platforms: we show how to construct our optical lattice, and derive the plaquette interctions using two approaches, the first one, non-perturbative (Subsec. \ref{non_pert_appr}), based entirely on angular momentum conservation, and the second one based on a perturbative expansion (Subsec. \ref{pert_appr}). In Subsec. \ref{gaugetheory_interpretation} we present the connection with the target gauge theory. In Sec. \ref{extensions} we discuss possible extensions and generalizations of our proposal. In Sec. \ref{conclusions} we summarize our results and present our conclusions.

\section{\label{halfspinLGT}$U(1)$ lattice gauge theories in two dimensions}
\begin{figure}[h]
    \centering	\includegraphics[width=0.8\linewidth]{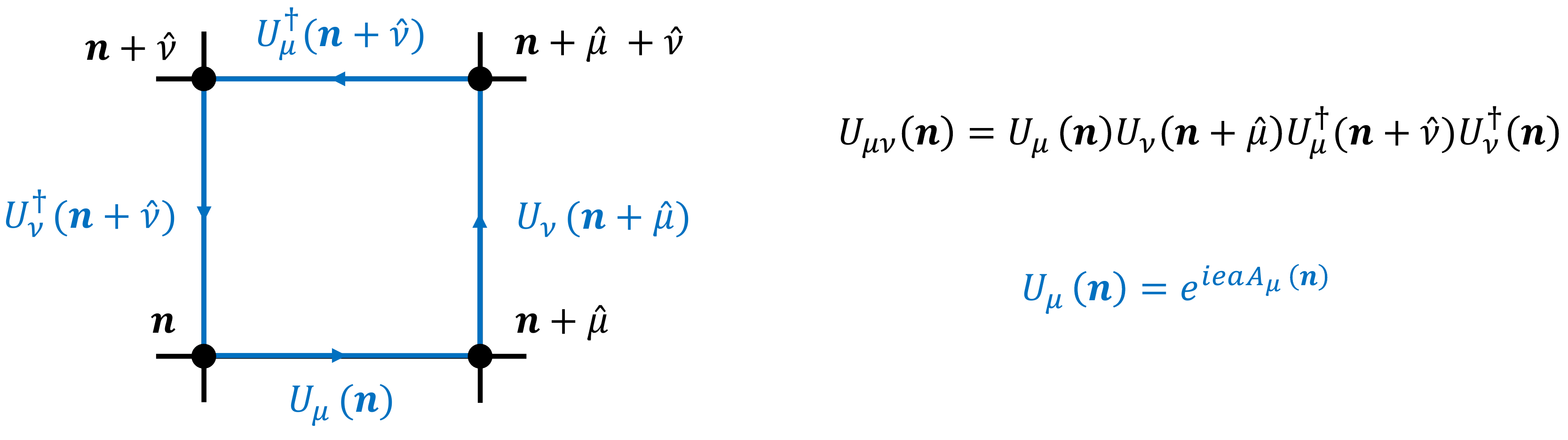}
    \caption{The gauge fields live on the links of the two-dimensional lattice, and are highlighted in blue in the elementary plaquette. The plaquette element $U_{\mu\nu}$ is the ordered product of the links, as presented in the right-hand side.}
    \label{link_plaquettes_plots}
\end{figure}
The purpose of this Section is to set the notation regarding the target theory, and provide a brief review review of the Hamiltonian formulation for $U\left(1\right)$ gauge theories. We will focus on the two dimensional case, later generalizing to higher dimensions. The matter fields live on the vertices of the lattice, here denoted by $\mathbf{n}=(n_1,n_2)$, while the gauge degrees of freedom live on the links, denoted by the site $\mathbf{n}$ and the direction towards which they point $\hat{\mu}=\{\hat{1},\hat{2}\}$. The electric field operator acting on the link, connecting the site $\mathbf{n}$ to the site $\mathbf{n}+\hat{\mu}$, is represented by $E_\mu(\mathbf{n})$ and commutes non-trivially with the Wilson operator  $U_\mu(\mathbf{n})$ on the same link
\begin{equation}
[U_\mu(\mathbf{n}),E_\nu(\mathbf{n}')]=-\delta_{\mu,\nu}\delta_{\mathbf{n},\mathbf{n}'}U_\mu(\mathbf{n}),\qquad[U^\dag_\mu(\mathbf{n}),E_\nu(\mathbf{n}')]=\delta_{\mu,\nu}\delta_{\mathbf{n},\mathbf{n}'}U^\dag_\mu(\mathbf{n}),
\label{commrel_Wilson_electricfield}
\end{equation}
with all remaining commutations set to zero. The Wilson and plaquette operators can be related to the gauge field $A_\mu$ and field strength $F_{\mu\nu}\equiv A_\nu(\mathbf{n}+\hat{\mu})-A_\nu(\mathbf{n})-A_\mu(\mathbf{n}+\hat{\nu})+A_\mu(\mathbf{n})$ by $U_\mu(\mathbf{n})=e^{ieaA_\mu(\mathbf{n})},\;U_{\mu\nu}(\mathbf{n})=e^{iea^2F_{\mu\nu}(\mathbf{n})}$. The Kogut-Susskind (KS) Hamiltonian is $H=H_g+H_m$, where
\begin{equation}
H_g=\frac{e^2}{2}\sum_{\mathbf{n},\mu}E^2_\mu(\mathbf{n})-\frac{1}{4a^2e^2}\sum_P (U_{\mu\nu}+U^\dag_{\mu\nu})
\label{gaugepart_KS_hamiltonian}
\end{equation}
is the pure gauge field contribution, while $H_m$ is the matter contribution, and depends on the employed discretization scheme for the fermions on the lattice. When matter is absent, Eq. \eqref{gaugepart_KS_hamiltonian} represent the KS Hamiltonian of a pure Abelian $U(1)$ LGTs in $2$D \cite{Kogut-Susskind,KogutRMP1979}. 
The KS Hamiltonian is gauge invariant, i.e. it commutes with the set of local operators
\begin{equation}
G(\mathbf{n})=\sum_{\mu}[E_\mu(\mathbf{n})-E_\mu(\mathbf{n}-\hat{\mu})],\qquad [H,G(\mathbf{n})]=0.
\end{equation}
Possible gauge invariant extensions can be added to the Hamiltonian, in the sense that the above symmetries are preserved. More about this point will be discussed below. In the absence of static charges, which is the case considered in the present work, the physical states $|\psi\rangle$ of the system are the ones satisfying the Gauss' law $G(\mathbf{n})|\psi\rangle=0,\;\forall\mathbf{n}$. 

\subsection{\label{bosonicQLM}Bosonic quantum link models}
Quantum link models \cite{HORN1981,ORLAND1990,CHANDRASEKHARAN1997} realize the commutation relations in Eq. \eqref{commrel_Wilson_electricfield} using quantum spin operators as
\begin{equation}
U_\mu(\mathbf{n})=S^+_\mu(\mathbf{n}),\qquad U^\dag_\mu(\mathbf{n})=S^-_\mu(\mathbf{n}),\qquad E_\mu(\mathbf{n})=S^z_\mu(\mathbf{n}).
\label{QLM_bosonic_spin_variables}
\end{equation}
In this framework, the operators $U_{\mu}$ and $U^\dagger_{\mu}$ are no longer unitary nor commuting, but rather satisfy
\begin{equation}
[U_\mu(\mathbf{n}),U^\dag_\nu(\mathbf{n}')]=2E_\mu(\mathbf{n})\delta_{\mu,\nu}\delta_{\mathbf{n},\mathbf{n}'},
\label{commrel_Wilson_Wilson}
\end{equation}
making the Hilbert space at each link finite. This difference gives rise to interesting physical phenomena \cite{Widmer_POS2014,banerjee2021nematic,BanerjeePRL2021}, while still providing a route to recover the Wilson discretization as one takes the spin representation $S$ to be large. In the particular case of $S=1/2$ there are only two states per link, associated with the values $E_\mu(\mathbf{n})=\pm 1/2$ of the electric field. The Hamiltonian gets simplified because $(S^z_\mu)^2=1/4$: the electric part is trivial and we are left with magnetic interactions only. The physics described by the Hamiltonian \eqref{gaugepart_KS_hamiltonian} can be enriched by introducing the Rokhsar-Kivelson (RK) term, with coupling $\lambda$, giving rise to the Hamiltonian\cite{RK1988}
\begin{equation}
H_{RK}=H_g+\lambda\sum_P(U_{\mu\nu}+U^\dag_{\mu\nu})^2,
\label{RK_Hamiltonian}
\end{equation}
which remains gauge invariant. Among all the possible $2^4$ states of the four links joining a vertex $\mathbf{n}$, only six satisfy the Gauss' law in $2$D. Despite the apparent simplicity of the model, its physics is very rich \cite{Wiese}, being closely related to the quantum dimer model \cite{Moessner2002}.

An alternative way to view this model is provided by mapping spins to hardcore bosons. There $+$ or $-$ signs of $E_\mu(\mathbf{n})$ label, respectively, the presence or absence of an hardcore boson in the link $\mathbf{n}\rightarrow\mathbf{n}+\hat{\mu}$ \cite{banerjee2022}. In terms of bosons, the gauge operators are written as
\begin{equation}
U_\mu(\mathbf{n})=b^\dag_\mu(\mathbf{n}),\qquad U^\dag_\mu(\mathbf{n})=b_\mu(\mathbf{n}),\qquad E_\mu(\mathbf{n})=n_\mu(\mathbf{n})-\frac{1}{2}.
\label{bosonic_QLMs_wilson_E_operators}
\end{equation}
The plaquette term becomes
\begin{equation}
U_{\mu\nu}(\mathbf{n})=b_\mu(\mathbf{n})b_\nu(\mathbf{n}+\hat{\nu})b^\dag_\mu(\mathbf{n}+\hat{\nu})b^\dag_\nu(\mathbf{n})
\label{plaquette_bosonic_QLMs}
\end{equation}
and can can be interpreted as a correlated hopping of two bosons. As will be discussed below, we will interpret the above terms as a particle at the link ($\mu,\mathbf{n}$) hopping to the link ($\nu,\mathbf{n}$) and one at ($\nu,\mathbf{n}+\hat{nu}$) hopping to ($\mu,\mathbf{n}+\hat{nu}$). The RK term can be written in this language as a sum of two-, three- and four-particles interactions. While this is simple to write, it does not arise as easily in an ultracold atomic setting. For this reason, we will focus only on the generation of the plaquette term in the present work. In this language, the generators take the form
\begin{equation}
G(\mathbf{n})=\sum_\mu[n_\mu(\mathbf{n})-n_\mu(\mathbf{n}-\hat{\mu})].
\label{gauss_fermionic_QLMs}
\end{equation}
and commute with the Hamiltonian by construction.

\subsection{\label{fermionicQLM}Fermionic quantum link models}
The particle representation opens the door to the construction of an alternative gauge theory: a gauge theory constructed with fermionic links \cite{banerjee2021,banerjee2022}, which is achieved by replacing the bosonic operators by fermionic ones. This is still gauge theory (there is still a set of local symmetries) but possibly hosting different physics due to the different commutation relations among the gauge field operators $U_\mu$, $U^\dagger_\mu$ and $E_{\mu}$. It turns out that for $2$D the theories are equivalent, while for $3$D they represent truly different models \cite{banerjee2021,banerjee2022}.

For concreteness, in the fermionic case, we can choose as a basis for the two-dimensional Hilbert space, the states $|0\rangle$ and $|1\rangle=c^\dag_\mu(\mathbf{n})|0\rangle$, and identify the Wilson and electric field through
\begin{equation}
U_\mu(\mathbf{n})=c^\dag_\mu(\mathbf{n}),\qquad U^\dag_\mu(\mathbf{n})=c_\mu(\mathbf{n}),\qquad E_\mu(\mathbf{n})=n_\mu(\mathbf{n})-\frac{1}{2},
\label{fermionic_QLMs_wilson_E_operators}
\end{equation}
where $n_\mu(\mathbf{n})\equiv c^\dag_\mu(\mathbf{n})c_\mu(\mathbf{n})$ is the number operator. It is straightforward to verify that Eq.s \eqref{commrel_Wilson_electricfield}
, \eqref{commrel_Wilson_Wilson} are satisfied with these definitions. As anticipated, the Wilson operators anticommute. 

Fermionic QLMs have been subjected to much less intense research when compared to their bosonic counterparts. Their analysis can lead, in principle, to the characterization of new phases of matter for LGTs. At the same time, quantum simulators of $2$D LGTs with ultracold atoms may profit from the fermionic interpretation of the plaquette interactions, as they provide an alternative equivalent way of realizing the same physics.

\section{\label{spinordipolarBEC}Spinor dipolar Bose-Einstein condensates}
Before addressing the details of our proposal, we give a brief reminder about spinor dipolar BECs, which are the basic tool that we need to build up the quantum simulator. Spinor BECs are degenerate Bose gases with spin internal degrees of freedom. With respect to usual (scalar) BECs, they present multicomponent order parameters and display richer physical phenomena, due to the interplay between superfluidity and magnetic effects. As a consequence, they provide a useful platform for the study of different physical aspects, such as the role of symmetry breaking and long-range order in quantum-ordered materials, quantum phase transitions and non-equilibrium quantum dynamics \cite{Lewenstein_book_2012,KAWAGUCHI2012,stapmerkurnRMP2013}.

The general atomic Hamiltonian of spinor BECs can be written on the basis of symmetry arguments, and, apart from the usual single-particle terms, it includes quantum number dependent interaction terms. For a spin-$f$ BEC we denote with $\phi_m(\mathbf{r})$ the bosonic field operators, satisfying the canonical commutation relations $[\phi_m(\mathbf{r}),\phi^\dag_{m'}(\mathbf{r}')]=\delta_{m,m'}\delta_{\mathbf{r},\mathbf{r}'}$, where $m=-f,-f+1,\ldots,f$ is the magnetic quantum number and $f$ is the hyperfine spin of the given atomic species. The microscopic Hamiltonian is
\begin{equation}
H=H_0+H^{(f)}_{int},\qquad H_0=\int\;d\mathbf{r}\;\sum_m\phi^\dag_m(\mathbf{r})\bigg[-\frac{\hbar^2\nabla^2}{2M}+U_{\text{trap}}(\mathbf{r})\bigg]\phi_m(\mathbf{r}),
\label{spinorBEC_generalH}
\end{equation}
\begin{equation}
H^{(f)}_{int}=\frac{1}{2}\int\;d\mathbf{r}\;\sum_{m_1,m_2,m_1',m_2'} C^{m_1m_2}_{m_1'm_2'}\;\phi^\dag_{m_1}(\mathbf{r})\phi^\dag_{m_2}(\mathbf{r})\phi_{m_1'}(\mathbf{r})\phi_{m_2'}(\mathbf{r}).
\label{spinorBEC_generalinteraction}
\end{equation}
The single-particle term, $H_0$, includes the possibility of having a trapping potential $U_{\text{trap}}(\mathbf{r})$. $H^{(f)}_{int}$ is the most general on-site interaction term for hyperfine spin $f$. For our purposes, it is enough to consider the $f=2$ case
\begin{equation}
H_{int}^{(2)}=\frac{1}{2}\int\;d\mathbf{r}\;[c_0:n^2(\mathbf{r}):+c_1:\mathbf{F}^2(\mathbf{r}):+c_2A^\dag_{00}(\mathbf{r})A_{00}(\mathbf{r})],
\label{spin2_spinorBEC_interactionterms}
\end{equation}
where $:\mathcal{O}:$ represents the normal order for the operator $\mathcal{O}$, $c_0,\;c_1$ and $c_2$ are numerical coefficients related to the scattering lengths $a_F$ in the various channels, and
\begin{equation}
n(\mathbf{r})=\sum_{m=-2}^2\phi^\dag_m\phi_m,\qquad A_{00}(\mathbf{r})=\frac{2\phi_2\phi_{-2}-2\phi_1\phi_{-1}+\phi_0^2}{\sqrt{5}},\qquad F_i(\mathbf{r})=\sum_{m,m'=-2}^2\phi^\dag_m(f_i)_{mm'}\phi_{m'}.
\label{spin2_spinorBEC_operators}
\end{equation}
The dependence on $\mathbf{r}$, on the right hand side, was ommitted for simplicity. The above defined quantities are $n(\mathbf{r})$ the total density operator, $A_{00}$ the amplitude of the spin singlet pair, and $F_i$ the spin density operators, with $f_i$ representing the spin-$2$ rotation matrices. Without further interactions, the spinor BECs in spin-independent optical lattices can be described by the Bose-Hubbard (BH) model \cite{jakschPRL1998,Eckert2007}. Expanding the field operators in terms of Wannier functions, and introducing the associated annihilation and creation operators $b_{\mathbf{i}m},\;b^\dag_{\mathbf{i}m}$, the BH Hamiltonian can be written as
\begin{equation}
H_{BH}=-t\sum_{\langle \mathbf{i,j}\rangle,m}(b^\dag_{\mathbf{i}m}b_{\mathbf{j}m}+\text{h.c.})+\frac{U_0}{2}\sum_{\mathbf{i}}n_\mathbf{i}(n_\mathbf{i}-1)+U_1\sum_\mathbf{i} (A_{00}^\dag)_\mathbf{i}(A_{00})_\mathbf{i}+\frac{U_2}{2}\sum_\mathbf{i}\mathbf{F}_\mathbf{i}^2-\mu\sum_\mathbf{i}n_\mathbf{i},
\label{spin2_BH_Hamiltonian}
\end{equation}
with $n_\mathbf{i}=\sum_m b^\dag_{\mathbf{i}m}b_{\mathbf{i}m}$, $F_{\mathbf{i}\alpha}=\sum_{m,m'}b^\dag_{\mathbf{i}m}(f_\alpha)_{mm'}b_{\mathbf{i}m'}$ and $A_{00}$ is the spin singlet amplitude written in terms of $b_{\mathbf{i}m},\;b^\dag_{\mathbf{i}m}$. The single site interactions are not enough to generate the desired plaquette terms within our proposal. However, this can be accomplished by including magnetic dipole-dipole interaction (MDDI) terms, and considering \textit{spinor dipolar BECs}. The MDDI couples the spin degrees of freedom with the orbital ones, conserving the total angular momentum. For spinor BECs, the MDDI can be relevant, as it is spin dependent and long-ranged. Its Hamiltonian in second quantization is given by
\begin{equation}
V_{dd}=\frac{c_{dd}}{2}\int\;d\mathbf{r}d\mathbf{r}'\;\sum_{\nu,\nu'}:F_{\nu}(\mathbf{r})Q_{\nu\nu'}(\mathbf{r}-\mathbf{r}')F_{\nu'}(\mathbf{r}'):,\qquad Q_{\nu\nu'}(\mathbf{r})=\frac{\delta_{\nu\nu'}-3\hat{r}_\nu\hat{r}_{\nu'}}{r^3},
\label{MDDI_interaction_hamiltonian}
\end{equation}
with the coefficient $c_{dd}\propto d^2$ related to the electric dipole moment. In the optical lattice Hamiltonian this generates a series of long-range terms
\begin{equation}
H_{dd}=\frac{1}{2}\sum_{\mathbf{i}\neq \mathbf{j}} U_{dd}^{\mathbf{ij}}n_\mathbf{i}n_\mathbf{j},\qquad U_{dd}^{\mathbf{ij}}\equiv c_{dd}\int\;d\mathbf{r}d\mathbf{r}'\;|w(\mathbf{r}-\mathbf{r}_i)|^2\frac{1-3\cos^2\theta}{|\mathbf{r}-\mathbf{r}'|^3}|w(\mathbf{r}'-\mathbf{r}_j)|^2,
\label{MDDI_interaction_extBH_Hamiltonian}
\end{equation}
where $\theta$ is the angle between the dipole moment and the vector $\mathbf{r}-\mathbf{r}_i$. The full Hamiltonian $H_{EBH}=H_{BH}+H_{dd}$ falls in the class of the so-called \textit{extended Bose-Hubbard models}. Depending on the values of $t,U_0,U_1$ and $U_{dd}$, that can be tuned independently, the extended BH has different quantum transitions and phases, including Mott insulator, superfluid and even supersolid phases, provided that more than nearest neighbors interaction terms are considered in the extended Hamiltonian \cite{vanotterlo1995,capogrossosansonePRL2010,Zhang2015}.

In principle, the dipolar interaction is dominant in gases of polar molecules when the application of a strong electric field is considered, due to their strong electric dipole moments. In this case, these are called spin-polarized dipolar BECs. The dipole-dipole interaction can be properly tuned through a rotating field \cite{giovinazzi2002}, allowing for the control of the interaction strength $c_{dd}$, that can be positive or negative according to the relative orientation of the dipoles. On the other hand, the MDDI can be neglected in several ultracold atomic systems, such as scalar alkali atoms, while they play an important role for other species, e.g. $\text{Cr}$ and $\text{Dy}$ \cite{Baranov_2002,Lahaye_2009}. We refer to the reviews \cite{KAWAGUCHI2012,stapmerkurnRMP2013}, and the references therein, for more details on the various physical properties of spinor (dipolar) BECs.

\section{\label{UCatomsrealization}Plaquette terms from angular momentum conservation}
In this Section we describe how the plaquette interactions in the $2$D Abelian spin-$1/2$ QLMs can be interpreted as a correlated hopping obtained through angular momentum conservation. The use of angular momentum conservation in scattering processes to ensure local gauge invariance was introduced, for the first time, in Ref. \cite{ZoharReznikPRA2013}. In that case, it guarantees that the gauge-matter interaction satisfies gauge invariance. By other side, plaquette terms are still obtained perturbatively. In contrast, our target model does not include matter and uses the conservation of angular momentum as a mean to obtain robust plaquette terms of the pure gauge theory. 

\begin{figure}[h]
		\centering
		\includegraphics[width=0.45\linewidth]{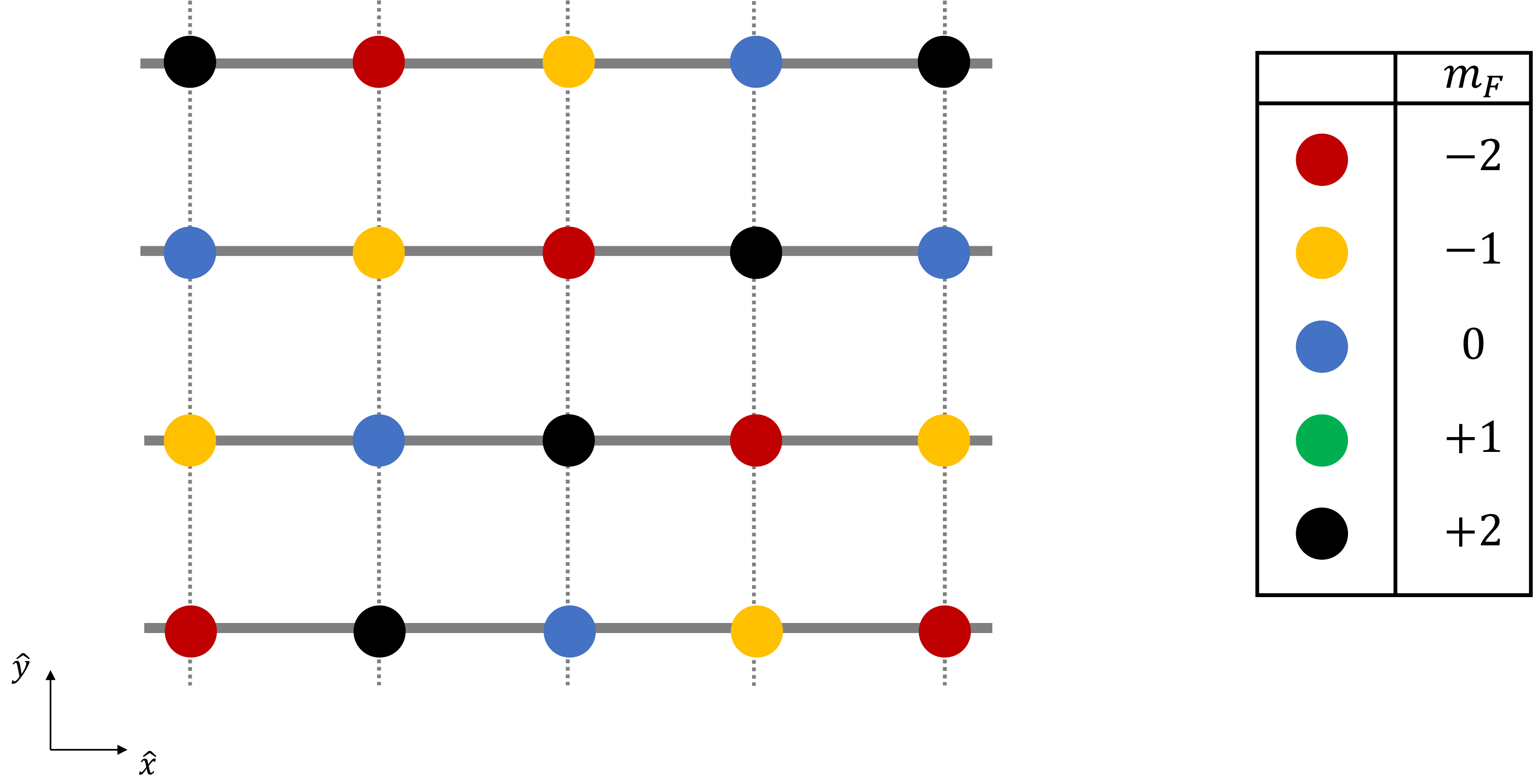}
		\qquad\qquad\qquad\qquad\qquad
		\includegraphics[width=0.28\linewidth]{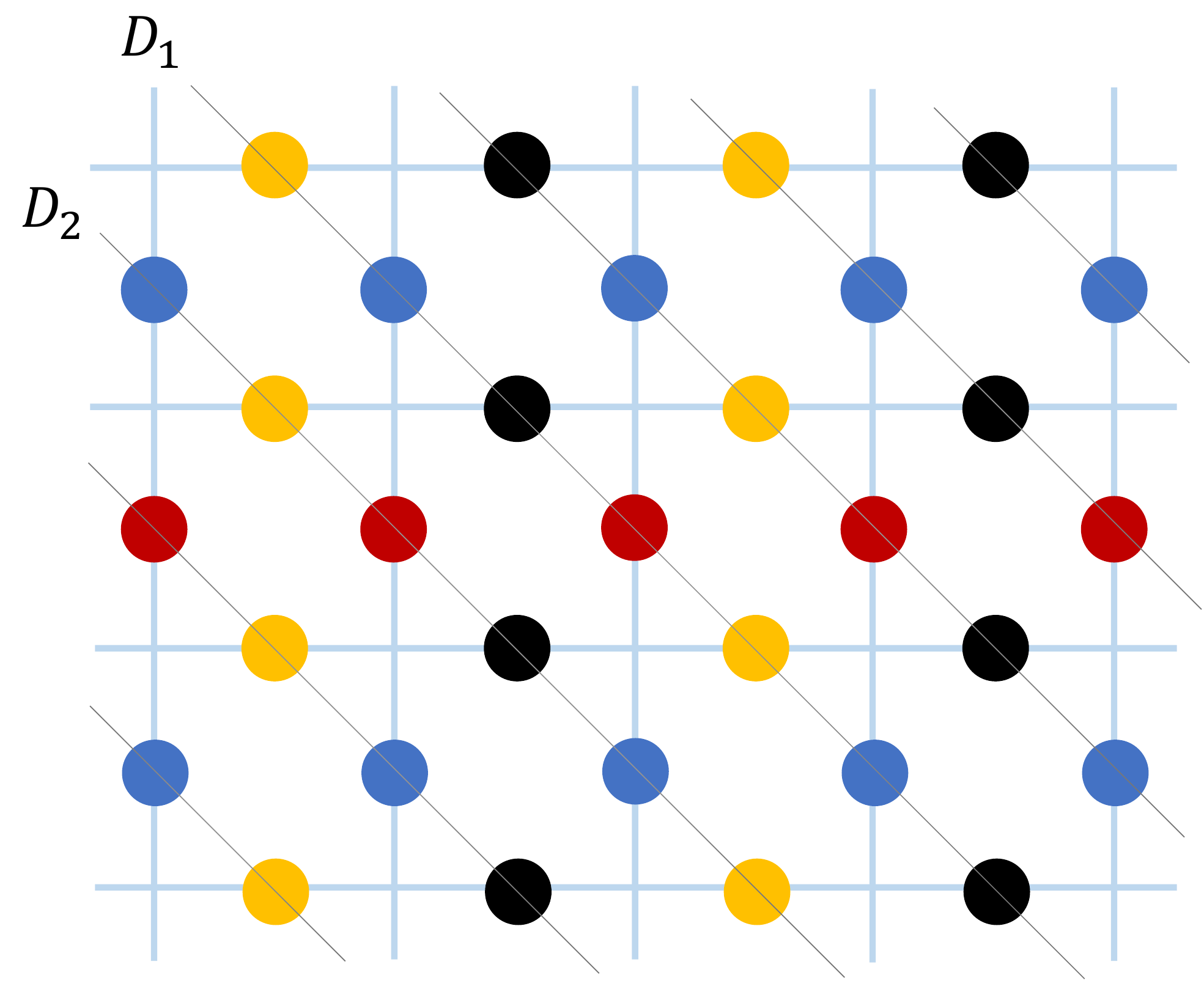}
		\caption{Left panel: spin-dependent optical lattice with atoms on the vertices in the five internal states $m_F=0,\pm1,\pm2$, whose color code is reported in the table. The dotted vertical lines represent avoided hoppings. Right panel: simulated lattice (light blue lines) alongside the allowed hoppings on the original spin-dependent optical lattice (inclined grey lines). While it is easier to visualize the simulator as it is represented in the left panel, the mapping to the gauge theory is more transparent when the lattice is rotated by $45^{\circ}$, as in the right panel.}
		\label{optical_lattices}
\end{figure}

In our proposal, we consider a spin-2 dipolar BEC loaded in a square optical lattice, whose structure is showed in the left panel of Fig. \ref{optical_lattices}. This figure has to be read as follows: the bosons are located on the different vertices of the lattice, and the lattice itself has a spin-dependent structure, so an atom can sit at a generic site $\mathbf{n}$ if it has the magnetic quantum number $m_F=0,\pm1,\pm2$ associated with that site. In other words, the color with which the site $\mathbf{n}$ is painted, in Fig. \ref{optical_lattices}, is associated with the magnetic quantum number of the atom that can sit here. This could be accomplished in two ways: {\it a) } by realizing a state-dependent optical superlattice, with different periods and minima \cite{optsuperlattice_yang_PRA2017}; or {\it b)} by introducing a one-site one-body term in the Hamiltonian penalizing or favoring, at the site $\mathbf{i}$, particles with different internal states. We will discuss in the next Subsections details concerning these approaches.

As a second condition, we require the presence of asymmetric hopping amplitudes within the lattice. With reference to the left panel Fig. \ref{optical_lattices}, we denote with $t_x$ and $t_y$ the horizontal and vertical hopping parameters and we assume that $t_x\gg t_y$, in a way that only horizontal hoppings processes are possible. In the left panel of Fig. \ref{optical_lattices}, this is represented by dotted ($\sim t_y$) and full ($\sim t_x$) lines, i.e. the dotted lines mean ``no hopping".  The simulated lattice, whose dynamics will be analyzed, is plotted in the right panel of Fig. \ref{optical_lattices}, rotated for convenience by $45^{\circ}$. Here we associate each well in which the spinor dipolar BEC lies to a single link of the simulated lattice, and we assume that the hopping of the atoms can happen only along its diagonals, due to the requirement on $t_{x}$ and $t_{y}$.

\begin{figure}[h]
	\centering
	\includegraphics[width=0.45\linewidth]{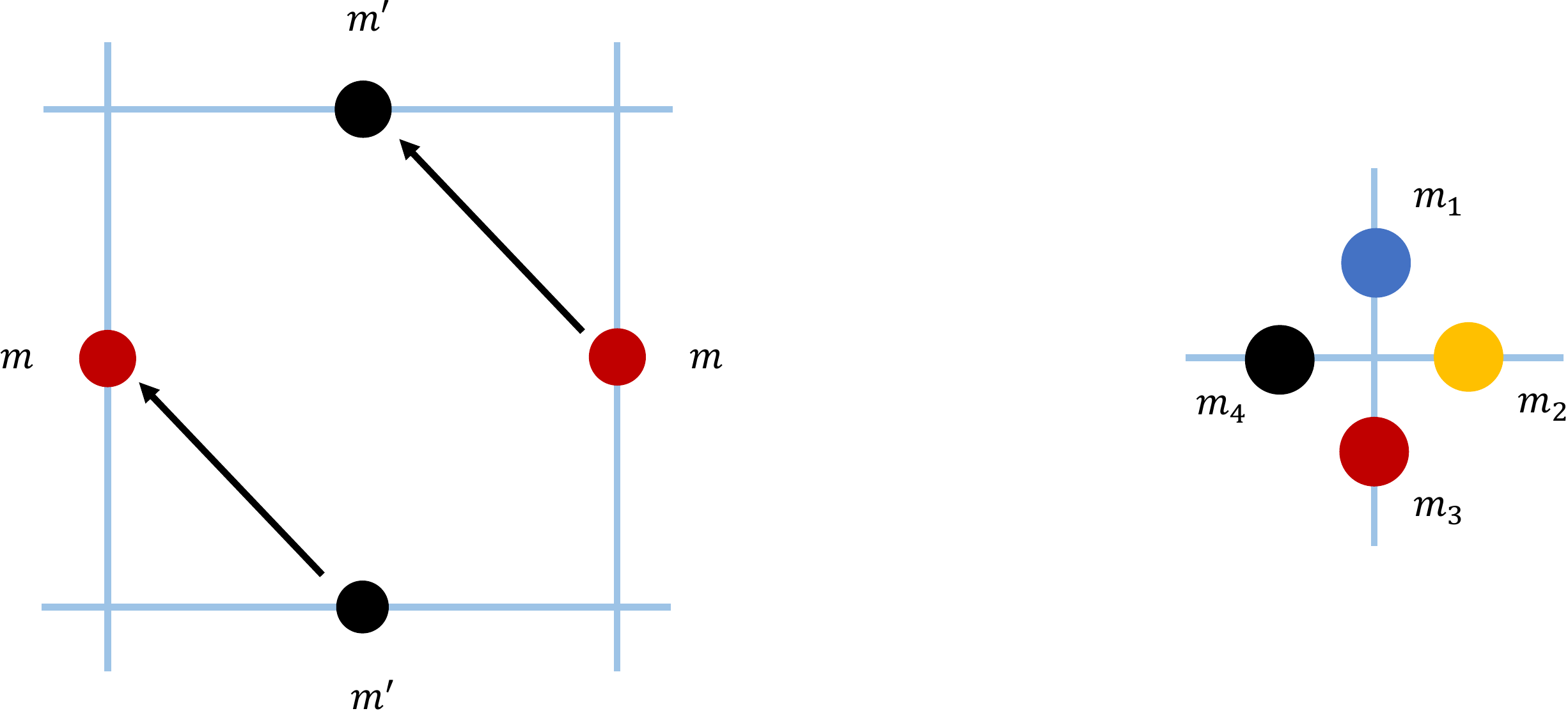}
	\caption{Left panel: generic plaquette structure in the simulated lattice with the same color appearing at opposite links. Right panel: generic vertex structure in the simulated lattice with all four different link colors meeting at a vertex.}
	\label{plaquette_vertex_structure}
\end{figure}

The crucial point, on which we are going to elaborate later, is that in this setup with the described requirements, a plaquette term is generated by correlated hoppings induced by angular momentum conservation. The geometric structure of the plaquette, as in left panel of Fig. \ref{plaquette_vertex_structure}, guarantees that the correspondent correlated hoppings generate the plaquette term. Additionally, by judicious choice of the four hyperfine levels, no correlated hoppings occur at vertices, as in left panel of Fig. \ref{plaquette_vertex_structure}. We anticipate that it is not enough to have four different colours meeting at a vertex, in order to forbid gauge symmetry breaking processes.

For these reasons, we consider two types of periodic sequences for the hyperfine levels along the diagonals in the right panel of Fig. \ref{optical_lattices}:

\begin{align}
    \nonumber
    &\text{D}_1:\;m_F=-1\;\rightarrow\;0\;\rightarrow\;2\;\rightarrow\;-2\;\rightarrow\;-1\;\rightarrow\;\ldots,
    \\\\
    \nonumber
    &\text{D}_2:\;m_F=0\;\rightarrow\;-1\;\rightarrow\;-2\;\rightarrow\;2\;\rightarrow\;0\;\rightarrow\;\ldots,
    \label{diagonal_cycles}
\end{align}
already depicted in Fig. \ref{optical_lattices}. Only four out of five possible hyperfine levels are used in our proposal for the spin-dependent superlattice. However, the use of total spin 2 is necessary in order to avoid correlated hopping processes at the vertices (see the Appendix \ref{species_number} for more details).

\begin{figure}[h]
	\centering
	\includegraphics[width=0.70\linewidth]{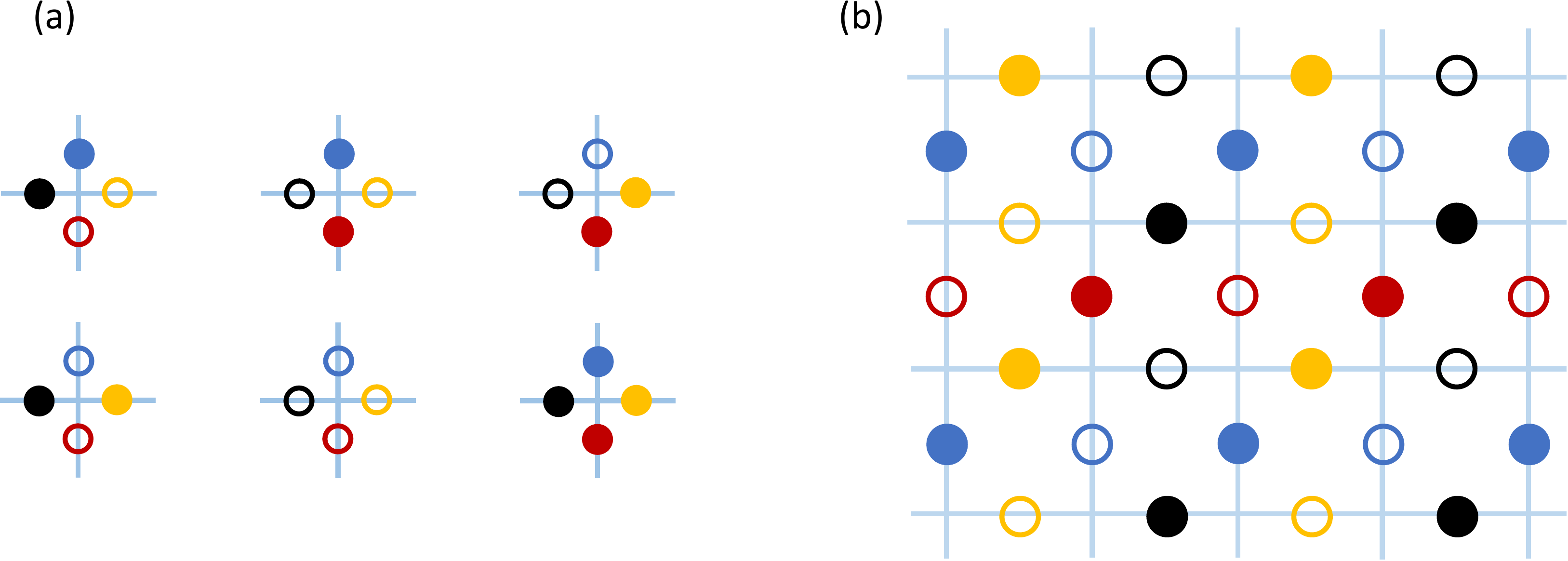}
	\caption{(a) The six vertices of the spin-dependent optical lattice compatible with Gauss' law. Filled circles represent sites occupied with a single particle, empty circles do not contain particles. (b) Fully flippable ground state of $H_0$, see Eq. \ref{H0_ourproposal}, made of disconnected flippable plaquettes.}
	\label{gauss_law_vertices}
\end{figure}

We are now ready to show how the plaquette term emerges in the ultracold atom dynamics. We are going to do it using the two approaches mentioned in the previous paragraphs. In the first scenario {\it a)}, the superlattice is realized by superimposing different state-dependent optical lattices. The plaquette interaction emerges directly from the optical lattice structure and the ultracold atom Hamiltonian, without the need of using perturbation theory in some large Bose-Hubbard parameter. For this reason, we refer to this as the \textit{non-perturbative scheme}. In the second scenario, dubbed before as approach {\it b)}, we enforce the structure of the optical lattice by means of large one-site one-body terms in the Hamiltonian, deriving the plaquette terms using a perturbative expansion on a proper gauge invariant manifold. We will refer to this case as the \textit{perturbative scheme}, following the same lines of the existing proposals in the literature \cite{ZoharPRL2011,CiracPRL2012,ZoharReznikPRA2013}.

\subsection{\label{non_pert_appr}Non-perturbative scheme}
To generate the one-dimensional chains depicted in the left panel of Fig. \ref{optical_lattices} we consider a state-dependent trapping potential of the form $U_{\text{trap},m_F}(\mathbf{r})=U_{x,m_F}(x)+U_{y,m_F}(y)+U_{z,m_F}(z)$. We require that the potentials along $y$ and $z$-directions are very deep, to prevent hoppings along the vertical direction and confine the particles into the two-dimensional plane. The form of the potential along the $x$-direction is
\begin{equation}
    U_{x,m_F}(x)=A_1\cos^2[q(x+\delta_{m_F})]+A_2\cos^2[2q(x+\delta_{m_F})],
    \label{x_potential}
\end{equation}
obtained by superimposing two laser beams with different amplitudes $A_1,\;A_2$ and momenta $q,\;2q$, respectively. This is to ensure the desired periodicity of the full optical superlattice, which is finally obtained by summing over all the selected internal states of Fig. \ref{optical_lattices} (see Appendix \ref{opticallattice_computations}).

We compute the optical lattice Hamiltonian by expanding the field operators in terms of the Wannier functions associated to $U_{\text{trap},m_F}(\mathbf{r})$, as described in Section \ref{spinordipolarBEC} for the spin-independent case. The final result is
\begin{equation}
    H_{\text{NP}}=\sum_{\mathbf{i},\mathbf{j}}\sum_m\;T^{m}_{\mathbf{ij}}(b^\dagger_{\mathbf{i}m}b_{\mathbf{j}m}+\text{h.c.})+\sum_{\substack{\mathbf{i,j}\\\mathbf{k,p}}}\sum_{\substack{m,m'\\ \ell,\ell'}}\;U^{mm'\ell\ell'}_{\mathbf{ijkp}}b^\dagger_{\mathbf{i}m}b^\dagger_{\mathbf{k}\ell}b_{\mathbf{j}m'}b_{\mathbf{p}\ell'},
    \label{fullH_optical_superlattice}
\end{equation}
and we refer to Appendix \ref{opticallattice_computations} for all the technical details of the computation and the explicit expression of the Hamiltonian amplitudes. 

In the hopping term we considered the only non-trivial contributions compatible with the conservation of angular momentum. We observe that, due to the structure reported in Fig. \ref{optical_lattices}, the case in which a boson can hop from $\mathbf{i}\rightarrow\mathbf{j}$ within the same sublattice involves fourth neighboring sites along the $\hat{x}$-direction. This term is safely negligible due to the small overlap of the Wannier functions, that are highly localized around the superlattice sites.

For the interaction terms, we require that the amplitudes satisfy some conditions in a way to generate the plaquette term. Firstly, we assume the hardcore boson limit for any internal state, i.e. a strong on-site repulsion $U^{mmmm}_{\mathbf{iiii}}\equiv U^m_\mathbf{i},$  for any value of $m$, much larger than any other energy scale in the system. We also have $U^{mm'\ell\ell'}_{\mathbf{ijkp}}=0$ for any combination of the internal indices not satisfying the angular momentum conservation. 

In light of these considerations, we can disregard the hopping terms in Eq. \eqref{fullH_optical_superlattice}, and the only relevant terms coming from the interactions are
\begin{equation}
    H_{\text{NP}}=\sum_{\substack{\langle\mathbf{i,j}\rangle\\ m,\ell}}\;U^{m\ell}_{\mathbf{ij}}n_{\mathbf{i}m}n_{\mathbf{j}\ell}+\sum_{\substack{\mathbf{i,j,k,p}\in\square\\m,\ell}}U^{m\ell}_{\square}b^\dag_{\mathbf{i}m}b^\dag_{\mathbf{k}\ell}b_{\mathbf{p}\ell}b_{\mathbf{j}m},
    \label{relevantH_optical_superlattice}
\end{equation}
i.e. nearest-neighbors extended interactions and the plaquette terms, with amplitudes depending on the overlaps of Wannier functions modulated by the presence of dipolar interactions.

This description guarantees that the plaquette terms are obtained directly, {\em without} the use of perturbation theory, as one of the most local processes allowed by conservation of angular momentum. By ``most local", we emphasize that there are other processes which preserve angular momentum. For example, a particle hopping to its fourth neighbor (same color) along the diagonal in the right panel of Fig. \ref{plaquette_vertex_structure} described above. This process is ``less local" in a rather concrete sense, and it is expected to be highly suppressed in the ultracold atom dynamics.

\subsection{\label{pert_appr}Perturbative scheme}
The purpose of this Subsection is to show how the plaquette terms can emerge in perturbation theory, even without conservation of angular momentum, if the lattice  parameters are properly tuned in a suitably defined strong coupling regime to enforce the superlattice structure of Fig. \ref{optical_lattices}. More precisely, in the scheme {\it b)} we favour the occupation of certain sites of the lattice by atoms with a certain $m_F$ by adding on-site energies to penalize the occupation with atoms carrying unwanted values of $m_F$. By doing the mapping of the system on a Bose-Hubbard Hamiltonian, there will be on-site energy terms of the form $\sim \epsilon_{\mathbf{i}m}b^\dagger_{\mathbf{i}m} b_{\mathbf{i}m}$. When the $\epsilon_{\mathbf{i}m}$ of the appropriate $m_F$'s are very large, then the corresponding site is occupied by the desired species. At this point, one has to do a perturbative expansion with large values of such on-site energies to obtain the plaquette term. Therefore the difference with the previous Subsection is that here we do perturbation theory in large values of certain Bose-Hubbard parameters, while in the previous scheme the conservation of the angular momentum enables us to avoid it, and obtain the plaquette term directly. The main point we want to make is that we retrieve the structure of Eq. \eqref{relevantH_optical_superlattice}, but now with extra terms that we are going to calculate explicitly.

We consider then an ultracold atom Hamiltonian that has regular hopping terms along a one-dimensional line, a one-body potential that promotes the superlattice structure, and angular momentum-preserving interactions among nearest neighbors. By performing a perturbative expansion for large amplitude values of the one-body potential, we can construct an effective Hamiltonian that exhibits gauge invariance at lowest orders, and contains all other angular momentum-conserving processes suppressed at higher orders. We also consider the hardcore bosons limit, so that we have, at most, one particle per site. In other words, there is a strong contact repulsion between bosons characterized by a parameter $U_0$ much larger than the relevant energy scales of the problem. Explicitly, the full Hamiltonian reads $H=H_0+H_1$ where
\begin{equation}
    H_1=H_{\text{hop}}+H_{\text{int}}\equiv-t_x\sum_{\langle \mathbf{i},\mathbf{j}\rangle_d,m}(b^\dag_{\mathbf{i}m}b_{\mathbf{j}m}+\text{h.c.})+\frac{1}{2}\sum_{\langle \mathbf{i},\mathbf{j}\rangle,m,m'}V^{\mathbf{ij}}_{mm'}b^\dag_{\mathbf{i}m}b^\dag_{\mathbf{j}m'}b_{\mathbf{i}m'}b_{\mathbf{j}m},
\label{H1_ourproposal}
\end{equation}
with the operators satisfying the hardcore bosons commutation relations $[b_{\mathbf{i}m},b^\dag_{\mathbf{j}m'}]=\delta_{mm'}\delta_{\mathbf{ij}}(1-2b^\dag_{\mathbf{i}m}b_{\mathbf{i}m})$. In addition to these terms we add a one-body term
\begin{equation}
    H_0=-h\sum_{\mathbf{i},m}\mathcal{\epsilon}_{\mathbf{i}m}b^\dag_{\mathbf{i}m}b_{\mathbf{i}m}
    \label{H0_ourproposal}
\end{equation}
with $h\gg t_x,V^{\mathbf{ij}}_{mm^\prime}$, and the function $\epsilon_{\mathbf{i}m}$ is such that it is equal to 1 if $m$ is the hyperfine state associated with site $\mathbf{i}$ (according to Fig. \ref{optical_lattices}) and 0 otherwise. This will enforce the desired lattice structure. Of course, this choice for $h$ implies that we have two large energy scales in the system ($h$ and $U_0$), and the further assumption that any effect of the on-site interaction is much beyond the scale we are interested in (see the Appendix \ref{pert_theory} for more details). 

We pause here to establish the connection with the non-perturbative scheme. The sum of the two Hamiltonians \eqref{H1_ourproposal} and \eqref{H0_ourproposal} does not correspond exactly to the scenario described in Subsection \ref{non_pert_appr}. In fact, the hopping terms presented in \eqref{H1_ourproposal} imply that particles with the same angular momentum can sit at nearest neighbor sites, in contrast to the situation described in Fig. \ref{optical_lattices}, for example. In turn, the Hamiltonian \eqref{H0_ourproposal} enforces this lattice structure through energy penalty. This allows the construction of the gauge theory as an effective theory at low energies, and the quantum numbers referred here could be different from angular momentum quantum numbers. However, the spin-dependent lattice structure described in Subsection \ref{non_pert_appr} represents a much more robust construction, as the plaquette terms rely on angular momentum conservation, and not on the large magnitude of $h$ with respect to the other parameters of the model.

In the same spirit of Ref. \cite{ZoharReznikPRA2013}, the idea is to prepare the system in a gauge invariant configuration that is a ground state of $H_0$. The dynamics generated by the full Hamiltonian $H=H_0+H_1$ is gauge invariant at low energies, and our aim is to construct an effective Hamiltonian in perturbation theory, using $h$ as large scale, giving rise to the plaquette interaction. 
We have then to characterize the ground states of $H_0$ that are compatible with the hardcore bosons constraint and with Gauss' law. Calling $N$ the number of lattice sites, and $N_p$ the number of particles in the lattice, we have two trivial cases, i.e. $N_p=0$ (empty lattice) and $N_p=N$ (full lattice), for which the dynamics is completely frozen. The other possibilities are represented by gluing different vertices compatible with Gauss' law, reported in Fig. \ref{gauss_law_vertices} (left panel), to form the full square lattice. The fully flippable ground state is composed by alternating filled anti-diagonals as showed in the right panel of Fig. \ref{gauss_law_vertices}.

\begin{figure}[h]
	\centering
	\includegraphics[width=0.70\linewidth]{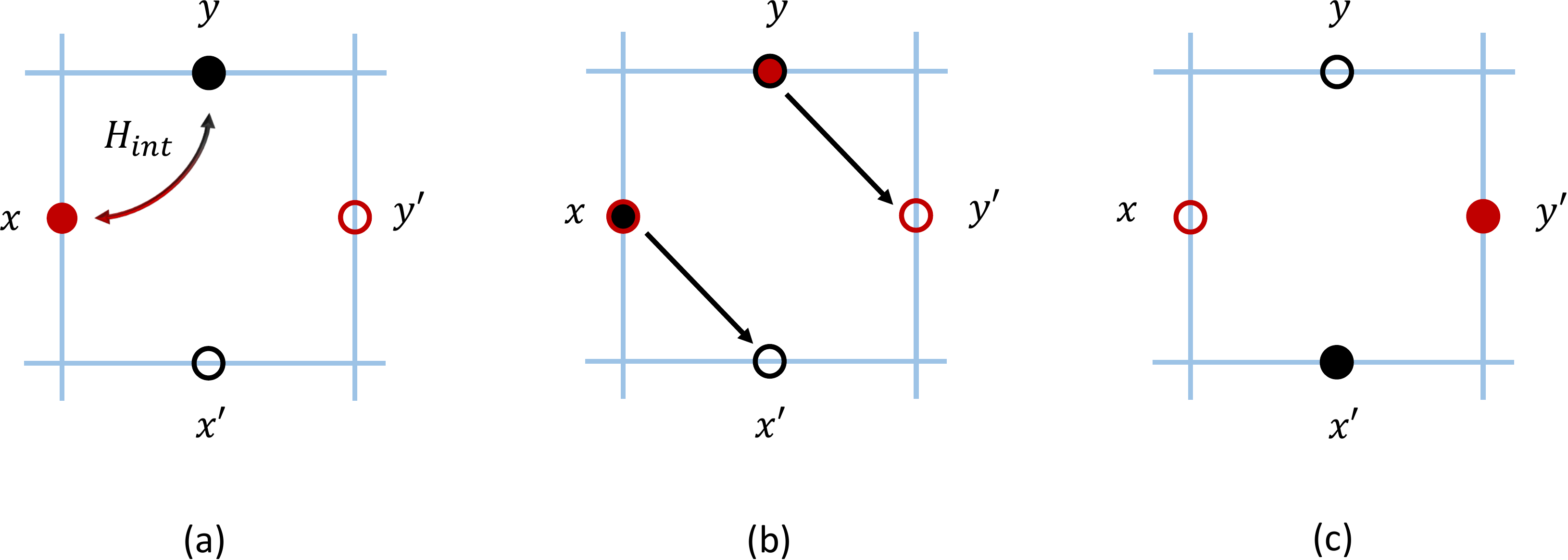}
	\caption{Virtual processes that build-up a plaquette flip, emerging at third-order in perturbation theory. (a) Spin-exchange interactions change the internal state of the atoms in the position $\mathbf{x},\;\mathbf{y}$; (b) two hoppings from $\mathbf{x}\rightarrow \mathbf{x}'$ and $\mathbf{y}\rightarrow \mathbf{y}'$; (c) final state after the whole process, with the flipped plaquette.}
	\label{3rdorder_processes_Heff}
\end{figure}

We denote with $\mathcal{M}_0$ the ground state manifold of $H_0$. The system must be prepared in a state $|\alpha\rangle\in\mathcal{M}_0$, and we work in a subspace $\mathcal{M}\subset\mathcal{M}_0$ which is gauge invariant. As $h$ is the largest scale in our system, we construct a low-energy Hamiltonian $H^{\text{(eff)}}$ within $\mathcal{M}_0$, that includes the plaquette interactions as correlated hoppings emerging from $H_1$. Up to third-order in perturbation theory, the effective Hamiltonian is
\begin{align}
    \nonumber
    H^{\text{(eff)}}&=\frac{t_x^2}{h}\sum_{\langle \mathbf{i},\mathbf{j}\rangle_d,m,m'}n_{\mathbf{i}m}n_{\mathbf{j}m'}-\frac{1}{h}\sum_{\langle \mathbf{i},\mathbf{j}\rangle,m,m'}(V^{\mathbf{ij}}_{mm'})^2n_{\mathbf{i}m}n_{\mathbf{j}m'}\\
    &+\frac{1}{h^2}\sum_{\substack{\mathbf{i}',\mathbf{j},\mathbf{j}'\in\;\square\\m,m'}}(V^{\mathbf{jj}'}_{mm'})^2V^{\mathbf{j}'\mathbf{i}'}_{mm}n_{\mathbf{j}'m}n_{\mathbf{i}'m}n_{\mathbf{j}m'}+ \frac{t_x^2}{h^2}\sum_{\substack{\mathbf{i},\mathbf{i}',\mathbf{j},\mathbf{j}'\in\;\square\\m,m'}}V^{\mathbf{ii}'}_{mm'}b_{\mathbf{j}'m}^\dag b_{\mathbf{j}m'}^\dag b_{\mathbf{i}'m'} b_{\mathbf{i}m}+\text{h.c.},
    \label{3rdorder_Heff}
\end{align}
and we refer to Appendix \ref{pert_theory} for details on the computation. We observe that the last term in the previous equation corresponds exactly to the plaquette interaction, as the effect of two correlated hoppings and a spin-exchange interaction, deriving from the dipolar term in $H_1$. The pictorial virtual processes are showed in Fig. \ref{3rdorder_processes_Heff}.

The other terms appearing in the first line of Eq. \eqref{3rdorder_Heff} arise from the second-order of the perturbative expansion, and are related to back and forth hoppings ($\sim t_x^2/h$) and double spin-exchange  ($\sim V^2/h$) between nearest neighbors. Similarly, the first term in the second line arises from the third-order of the expansion, and it is related to spin-exchange interactions ($\sim V^3/h^2$) within a given plaquette. These terms are not present in the initial model but are diagonal on the occupation number and, therefore, associated with products of the electric field at different links, in the gauge theory language. As a consequence, they are trivially gauge invariant.

\subsection{\label{gaugetheory_interpretation}Gauge theory interpretation}
The Hamiltonians in Eq.s \eqref{relevantH_optical_superlattice} and \eqref{3rdorder_Heff} can be interpreted in the language of QLMs. We can identify
\begin{equation}
    U_{\mathbf{i}m}=b^\dag_{\mathbf{i}m},\qquad U^{\dag}_{\mathbf{i}m}=b_{\mathbf{i}m}
    \label{link_operators_ourproposal}
\end{equation}
as the link operators of the associated LGT. In this way, the plaquette term has the desired form, and the mapping of the operators is such that the commutation relation of the QLMs are satisfied, using the hardcore bosonic commutation relations. Indeed, it is immediate to verify that
\begin{equation}
    [U_{\mathbf{i}m},U^\dag_{\mathbf{j}m'}]=\delta_{\mathbf{ij}}\delta_{mm'}(2n_{\mathbf{i}m}-1)=2\delta_{\mathbf{ij}}\delta_{mm'}\bigg(n_{\mathbf{i}m}-\frac{1}{2}\bigg)
\end{equation}
allowing for the identification of the electric field operator in terms of the particle number operator
\begin{equation}
    E_{\mathbf{i}m}\equiv n_{\mathbf{i}m}-\frac{1}{2}.
    \label{electric_field_operator_ourproposal}
\end{equation}
As anticipated $[U_{\mathbf{i}m},U^\dag_{\mathbf{j}m'}]=2\delta_{\mathbf{ij}}\delta_{mm'}E_{\mathbf{i}m}$. This is an explicit realization of the spin-$1/2$ QLMs, because, due to the hardcore boson constraint, the eigenvalues of $n_{\mathbf{i}m}\in\{0,1\}$, and therefore the possible values of the electric fields are $E_{\mathbf{i}m}=\pm1/2$. With this comparison, different particle sectors of the underlying bosonic theory are associated with different electric field sectors of the related QLM. While we have always assumed that the we were dealing with ultracold bosonic gases, the hardcore constraint makes the translation to fermionic link models trivial. As a consequence, and to the best of our knowledge, this constitutes the first proposal for the realization of the fermionic link models introduced in Ref.s \cite{banerjee2021,banerjee2022}.

With the mapping in Eq. \ref{link_operators_ourproposal}, the non-perturbative Hamiltonian can be written in the gauge theoretical language as
\begin{equation}
    H_{\text{NP}}=\sum_{\substack{\langle\mathbf{i,j}\rangle\\ m,\ell}}\;U^{m\ell}_{\mathbf{ij}}E_{\mathbf{i}m}E_{\mathbf{j}\ell}+\sum_{\substack{\mathbf{i,j,k,p}\in\square\\m,\ell}}J^{m\ell}_{\text{NP}}b^\dag_{\mathbf{i}m}(U_\square+U^\dagger_\square).
    \label{nonperturbativeH_gaugetheory_interpretation}
\end{equation}
In this Hamiltonian, the first term is the square of electric fields of nearest neighbor links in the target lattice, while the second term is the desired plaquette interaction with coupling given by $J^{m\ell}_{\text{NP}}\equiv U^{m\ell}_{\square}$. The explicit expressions for these amplitudes as functions of the optical superlattice components is reported in Appendix \ref{opticallattice_computations}.

On the other hand, the effective Hamiltonian derived in perturbation theory is
\begin{align}
    \nonumber
    H^{(\text{eff})}&=\lambda_1\sum_{\langle  \mathbf{i},\mathbf{j}\rangle_d,m,m'}E_{\mathbf{i}m}E_{\mathbf{j}m'}-\sum_{\langle \mathbf{i},\mathbf{j}\rangle,m,m'} \lambda_2^{(mm')} E_{\mathbf{i}m}E_{\mathbf{j}m'}\\
    &+\sum_{\substack{\mathbf{i}',\mathbf{j},\mathbf{j}'\in\;\square\\m,m'}} \lambda_3^{(mm')} E_{\mathbf{i}'m}E_{\mathbf{j}m'}E_{\mathbf{j}'m}-\sum_{\square}J^{(mm')}(U_\square+U^\dag_\square)
    \label{Heff_gaugetheory_interpretation}
\end{align}
with $\lambda_1,\;\lambda_2,\;\lambda_3,\;J>0$ defined in terms of the ultracold atom lattice parameters as
\begin{equation}
    \lambda_1=\frac{t_x^2}{h},\qquad\qquad\lambda_2^{(mm')}=\frac{(V^{\mathbf{ij}}_{mm'})^2}{h},\qquad\qquad\lambda_3^{(mm')}=\frac{(V^{\mathbf{jj}'}_{mm'})^2V^{\mathbf{j}'\mathbf{i}'}_{mm}}{h^2},\qquad\qquad J^{(mm')}=\frac{t_x^2V^{\mathbf{ii}'}_{mm'}}{h^2}.
\end{equation}

We observe, as already commented, that the plaquette term is properly generated within this perturbative scheme alongside asymmetric terms containing the square of the electric field. The asymmetry in the sums and the different coefficients $\lambda_1,\;\lambda_2,\;\lambda_3$ are directly related to the construction of the spin-dependent optical lattice. 
The physics of the model in Eq. \eqref{Heff_gaugetheory_interpretation} can be analysed for some values of the parameters. We consider $\lambda_3=0$, to discuss what happens in the parameters space spanned by $\lambda_1,\;\lambda_2$. We also restrict our analysis to the half-filling case, since in this sector there is the fully flippable ground state of $H_0$. Clearly, for $\lambda_1=\lambda_2=0$, our model is equivalent to the RK Hamiltonian with $\lambda=0$. When the two parameters are switched on, we can discuss two limiting cases: for $\lambda_1\gg\lambda_2$, the Hamiltonian favors the configurations in which the diagonals of the simulated lattice are independently filled as in the right panel of Fig. \ref{gauss_law_vertices}. In the opposite limit, i.e. for $\lambda_2\gg\lambda_1$, the spin-exchange interaction between nearest neighbors dominates. When instead both the parameters are such that $\lambda_1,\;\lambda_2\gg J$, this situation is exactly the one showed in Fig. \ref{gauss_law_vertices}. This is analogous to the $\lambda\rightarrow-\infty$ limit of the RK model, displaying therefore a Néel state \cite{Banerjee_2013,Wiese}. We point out these features to highlight the fact that, despite the low-energy properties of Eq. \eqref{Heff_gaugetheory_interpretation} and the RK model are very similar, further considerations about the specific phases of our effective model at intermediate couplings may not be easy to guess. In general, we may expect that the two phase diagrams should be different, based on the different symmetries of Eq.s \eqref{Heff_gaugetheory_interpretation} and \eqref{RK_Hamiltonian}.

\section{\label{extensions}Extensions and generalizations}
In this Section we discuss how to generalize our proposal to different cases. The most important is the extension to higher dimensions, and we discuss explicitly the $d=3$ case. We then address how the proposal can be generalized to other geometries, with the aim to generate $U(1)$ plaquette terms in lower order perturbation theory and to extend our analysis to discrete gauge groups. Finally, we comment about the higher spin QLMs, closely related to the relaxation of the hardcore bosons constraint in the underlying microscopic model.

\subsection{\label{higher_d}Higher dimensions: the $d=3$ case}
The extension to $d=3$ is challenging for all the schemes that have been proposed so far in $d=2$. In our proposal this is manifest by the increment of the number of atomic species. While the perturbative scheme will exhibit gauge breaking terms, we argue that, by using on the non-perturbative one based on angular momentum conservation, such terms are highly suppressed relatively to the gauge invariant dynamics.

To see this, we discuss how the extension to $d=3$ would work by using the same lattice and the same number of internal states employed in $d=2$, i.e. four out of five hyperfine levels of the spinor dipolar BEC. The principle that allowed the construction of the plaquette term in $d=2$ was to associate the increase of the electric field in one of the links of the plaquette, with the decrease of one of the other links. We can then associate the change of two links with a single hopping term, and the full plaquette term with a correlated hopping. The resulting particles move in single (diagonal) lines. The exact same principles apply in $d=3$. The extra dimension lifts the lines into planes, i.e., particles become confined in planes.
\begin{figure}[h]
	\centering
	\includegraphics[width=0.6\linewidth]{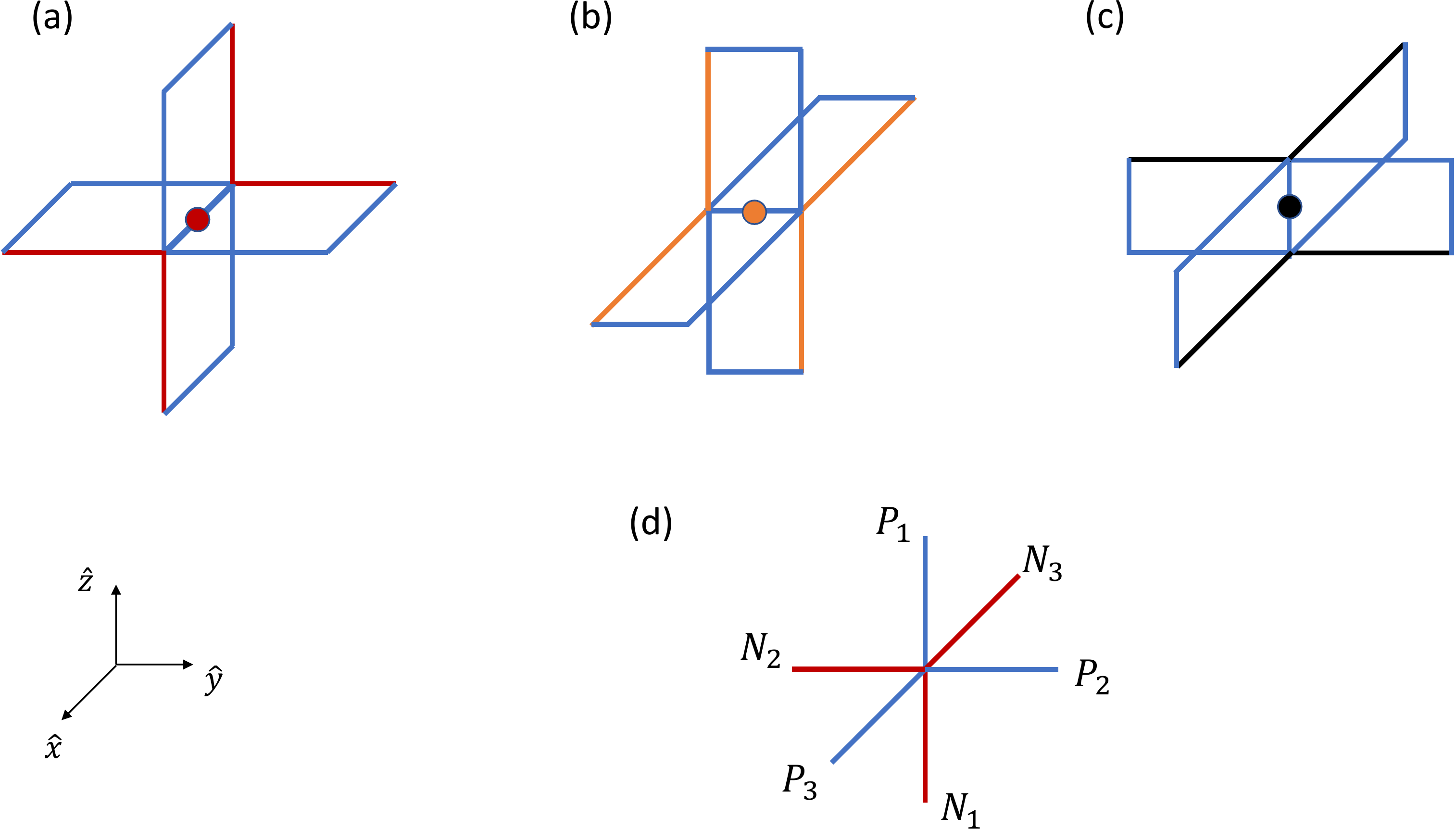}
	\caption{Structure of the staples for (a) $x$-links (red full dot), (b) $y$-links (orange full dot) and (c) $z$-links (black full dot) in $d=3$. As in $d=2$, we represent with blue links the target lattice. The full dots represent bosons. The colored links in (a)-(c) are the links that the particle can occupy after an hopping in the various possible directions. (d) Vertex structure in $d=3$. The blue and red links identify the sublattices in which the particles moving in different planes can hop. The $P_i$, $N_i$ are the possible internal states of the hardcore bosons at the given link.}
	\label{higherd_staples}
\end{figure}

To identify these planes, it is useful to consider each link separately and construct the relative staples that constitute the set of all links which are coupled to the central one, as showed in Fig. \ref{higherd_staples}. The centers of links of the lattice, where the particles reside, can be identified with the positions $\left(n_1,n_2,n_3\right)+\hat{\mu}/2$, with $\hat{\mu}$ one of the Cartesian unit vectors and $n_\nu$ integers. In Fig. \ref{higherd_staples} the cases $\mu=1,2,3$ are represented in (a), (b) and (c) respectively. The links to which the particle can hop are colored the same as the particle. It becomes clear that particles at position $\mathbf{r}=\left(n_1,n_2,n_3\right)+\hat{\mu}/2$ can (only) hop to positions $\mathbf{r}\pm\hat{\mu}/2\hat\mp\hat{\nu}/2$ with $\mu\neq\nu$. We can conclude that the planes along which the hardcore bosons move are described by the equation $\mathbf{r}\cdot\mathbf{n}=c$, where $\mathbf{n}=(1,1,1)$ and $c$ is a constant that distinguishes the different parallel planes.

\begin{figure}[h]
	\centering
	\includegraphics[width=0.8\linewidth]{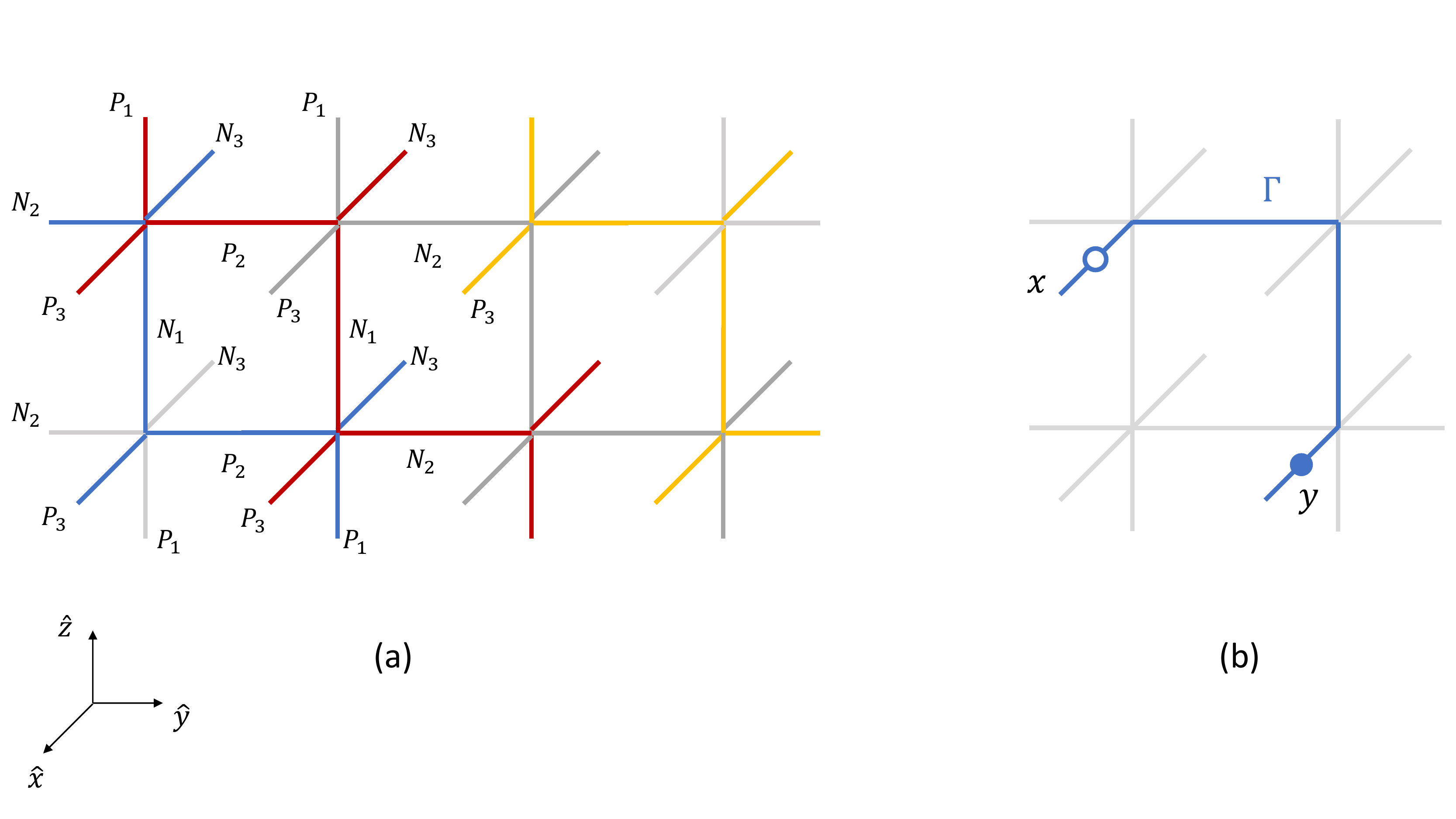}
	\caption{(a) Structure of the lattice in $d=3$ with four out of five internal states. For the explicit values of the levels $N_i,\;P_i$ we refer to Table \ref{table1}. We highlight here with different colors the sublattices in which the particles can hop, compatible with the planes identified in the main text. (b) Graphical structure of the gauge breaking terms that appear alongside the plaquettes, associated to a particle hopping from $\textbf{x}\rightarrow \textbf{y}$ along the blue path $\Gamma$.}
	\label{3d_lattice}
\end{figure}

Having identified the planes, the subsequent task consists in identifying the values of angular momentum associated to each link in a way that allows for the generation of plaquette terms, but still forbids hoppings at the vertices, i.e. gauge symmetry breaking terms. The first part is constructed in complete parallel with the $d=2$: by associating the same angular momentum to opposite links of the plaquette, we guarantee that the plaquette terms are such that conserve angular momentum. Guaranteeing that these are the only allowed processes is less straightforward. We refer to Fig. \ref{higherd_staples}(d) to denote a generic vertex and call $N_i$ and $P_i$ the internal states of the two sublattices that are associated with two different planes. The full lattice is constructed by reflecting this vertex relatively to the different planes and gluing them, as represented in Fig. \ref{3d_lattice}(a). The equations that must be imposed for $N_i$ and $P_i$ result from requiring two types of conditions. As a first point, differences of angular momentum across sublattices must be unequal so that such hoppings are removed from the dynamics. Secondly, within a sublattice, we must guarantee that each link does not have any neighbor with the same angular momentum in which it can hop to. These requirements lead to
	\begin{equation}
	|A_i-A_j|\neq|B_k-B_\ell|,\qquad\forall (i,j),(k,\ell)\in\{(1,2),(2,3),(3,1)\}\;\text{and}\;A,B\in\{P,N\},
	\label{ang_momenta_higherd}
	\end{equation}
	\begin{equation}
	N_i\neq N_j,\;P_i\neq P_j,\qquad N_i\neq P_j\qquad\forall i\neq j,
	\label{ang_momenta_higherd_1}
	\end{equation}
	where the first equation comes from the first type of condition and the other two from the second. 
\begin{table}[h]
	\setlength{\tabcolsep}{11pt}
	\begin{center}
		\begin{tabular}{cccccc}
			\hline 
			\hline
			$P_1$ & $P_2$ & $P_3$ & $N_1$ & $N_2$ & $N_3$\\  
			\hline
			-3/2 & 1/2 & 7/2 & -7/2 & -5/2 &	7/2\\
			-7/2 & 5/2 & 7/2 & -7/2 & -1/2 & 3/2\\
			1/2 & -3/2 & 7/2 & -5/2 & -7/2 & 7/2\\
			5/2 & -7/2 & 7/2 & -1/2 & -7/2 & 3/2\\
			\hline
			\hline
		\end{tabular}
	\end{center}
	\caption{Some solutions to the Eq.s \eqref{ang_momenta_higherd}. As commented in the main text, it is impossible to satisfy them with the same number of internal levels employed in $d=2$. To preserve the lattice structure and the scheme employed in the previous Section, we need five out of eight internal levels.}
	\label{table1}
\end{table}
The inspection of these equations show that they cannot be satisfied with four levels, see Table \ref{table1} and the discussion in the caption. We conclude that to extend our scheme to $d=3$ we have to increase the number of levels from four to {\em five} levels. 

We point out that, within the perturbative scheme, there is yet the problem that, even using five levels, we have the occurrence of gauge breaking terms of the form
\begin{equation}
H_{\text{GB}}=-\frac{t_xt_yt_z}{h^2}\sum_{\mathbf{x}.\mathbf{y}\in\Gamma}b^\dag_{\mathbf{y}m}b_{\mathbf{x}m}
\label{3d_breaking_terms}
\end{equation}
at the same order in perturbation theory as the plaquette term. If these terms are really competing with the plaquettes, the effective theory is not gauge invariant and this formulation needs further refinement. In this scheme we need to use {\em ad hoc} hopping terms, which by the way must have a tunneling coefficient of opposite sign with respect to the sign of the $t_{x,y,z}$ hopping parameters entering the microscopic Bose-Hubbard Hamiltonian of Eq. \eqref{H1_ourproposal}. In principle, this can be accomplished by adding further lattices, one for each internal level, suitably located to allow just the needed hoppings. These have to be shaken lattices \cite{ARIMONDO2012}, in a way to enforce the right sign for the hopping parameters. It is clear that this particular extension appears to be rather involved, and its realizability is far from being easy or conceivable. Finally, by modifying properly the superlattice structure, it could be possible to remove these third-order hoppings from the low-energy effective theory.

However, we remind that the presented perturbation theory scheme was introduced in order to offer a quantitative approach that shows how the plaquette term is the dominant dynamical process of the system, and that it can also emerge without conservation of angular momentum. If we follow the non-perturbative scheme, i.e. the construction of plaquette terms using the conservation of angular momentum, the process in Eq. \eqref{3d_breaking_terms} would correspond to tunneling to a third neighbor directly. In a real experimental setting, we expect that the atomic wave function can extend, even if slightly, to its nearest neighbors. Together with the dipolar interaction, which promotes exchange of angular momentum, the correlated hopping appears as the simplest dynamical process. We then argue that plaquette terms, which involve only nearest neighbors, are highly dominant over the above process in the non-perturbative framework.

In order to follow the perturbative scheme, we need to use {\em ad hoc} hopping terms, which by the way must have a tunneling coefficient of opposite sign with respect to the sign of the $t_{x,y,z}$ hopping parameters entering the microscopic Bose-Hubbard Hamiltonian of Eq. \eqref{H1_ourproposal}. In principle, this can be accomplished by adding further lattices, one for each internal level, suitably located to allow just the needed hoppings. These have to be shaken lattices \cite{ARIMONDO2012}, in a way to enforce the right sign for the hopping parameters. It is clear that this particular extension appears to be rather involved, and its realizability is far from being easy or conceivable. Finally, by modifying properly the superlattice structure, it could be possible to remove these third-order hoppings from the low-energy effective theory.

While it would be interesting to investigate if a suitable modification of the supperlatice would remove these terms in perturbation theory, we believe that the key advantage of the proposal is to tie the conservation of angular momentum to the correlated hopping that generates the plaquettes and, in that case, these processes should be irrelevant.

\subsection{\label{other_geometres}Triangular lattice}

The main purpose of this Section is to show that these principles are generalizable to other geometries and, inclusively, we can use fewer internal states to generate plaquette interactions in lower order in perturbation theory. In this specific case, we are able to use three consecutive internal states, i.e. a spin-$1$ dipolar BEC, to generate plaquettes at second-order in perturbation theory. Moreover, as we are going to discuss, the proposal looks arguably simpler in this geometry, if compared with the square lattice one. The phase diagram of the spin-$1/2$ QLMs on a triangular lattice, in the presence of the RK term, as been studied in \cite{moessnerPRL2001}.

\begin{figure}[h]
	\centering
	\includegraphics[width=0.7\linewidth]{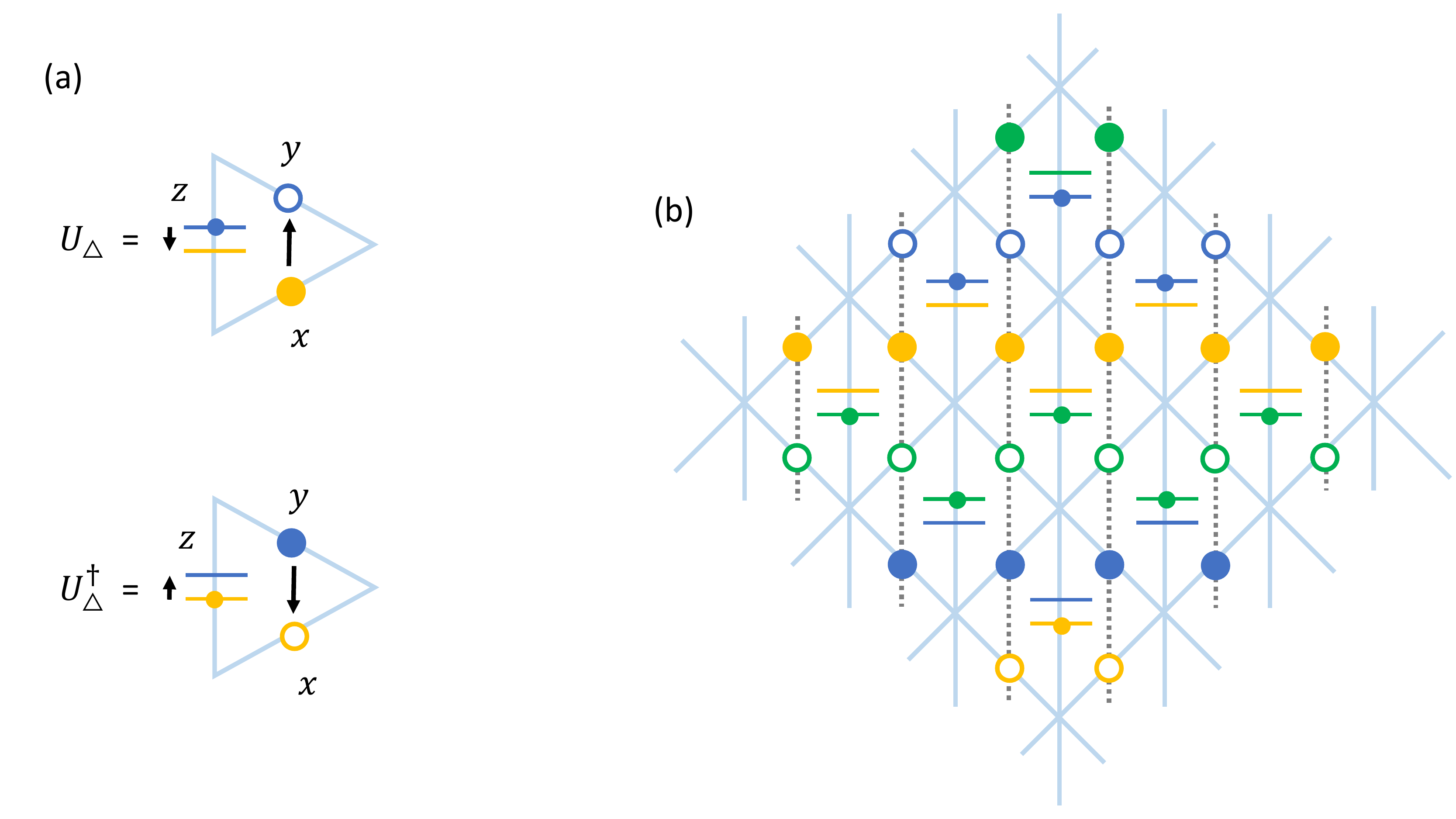}
	\caption{(a): Hopping processes generating the plaquette terms in the triangular lattice. (b): Structure of the triangular lattice using three internal states. The color code is the same of Fig. \ref{optical_lattices}, while the horizontal lines are the two-level systems described in the main text. The grey dotted lines represent the $d=1$ systems along which hopping processes can occur.}
	\label{triangular_proposal}
\end{figure}

The structure of the triangular lattice is presented in Fig. \ref{triangular_proposal}(b). The particles are constrained to hop along one-dimensional vertical lines, that we alternate with a set of two-level systems placed on the hypotenuses of the triangles. The sites of the spin-dependent optical lattice coincide with the sides of the target (triangular) lattice. According to the color code of Fig. \ref{optical_lattices}, we use here the three internal states $m_F=0,\pm1$, that can be realized by using a spin-$1$ dipolar BEC. With this geometry, the plaquettes of the lattice can be identified graphically as in Fig. \ref{triangular_proposal}(a). They are made of three links and we adopt the convention $U_\triangle=U_zU_xU_y$, alongside the corresponding one for $U^\dag_\triangle$. Here the Gauss' law takes a similar form, with the generator of gauge transformations being an oriented sum of the six links joining at each vertex of the triangular lattice. Explicitly $G(\mathbf{n})=\sum_\mu[E_\mu(\mathbf{n})-E_\mu(\mathbf{n}-\hat{\mu})]$, with $\mu$ representing the three directions $x,y$ and $z$ as in Fig. \ref{triangular_proposal}. By the judicious choice of the internal states for each site, in the non-perturbative scheme plaquette terms arise as one of the angular momentum conserving local processes.

Alternatively, using the perturbative scheme, this can be made precise by constructing an Hamiltonian along the same lines discussed before. If we denote by $t_y$ the coefficient of the hopping term along the vertical lines, in the hardcore bosons limit the Hamiltonian reads
\begin{equation}
H_\triangle=H_{\text{hop}}+H_{\text{int}}\equiv-t_y\sum_{\langle \mathbf{i},\mathbf{j}\rangle_\text{lines},m}(b^\dag_{\mathbf{i}m}b_{\mathbf{j}m}+\text{h.c.})+\frac{1}{2}\sum_{\langle \mathbf{i},\mathbf{j}\rangle,m,m'}V^{\mathbf{ij}}_{mm'}b^\dag_{\mathbf{i}m}b^\dag_{\mathbf{j}m'}b_{\mathbf{i}m'}b_{\mathbf{j}m},
\label{Htriangle_ourproposal}
\end{equation}
where the sum over nearest neighbors in the second term is extended to all the links of the triangular lattice, including the ones hosting the two-level systems. In turn, the hopping only occurs between neighbors along the line.

With this structure, we can proceed with perturbation theory exactly in the same logic of the square lattice, i.e. introducing $H_0$ as in Eq. \eqref{H0_ourproposal} and take $h$ as the largest scale in the system. While the first-order in the expansion vanishes, as in the square lattice, in this case the plaquette term emerges directly at second-order in perturbation theory. The effective Hamiltonian here contains two terms, i.e.
\begin{equation}
H^{\text{(eff)}}_\triangle=-\frac{t_y^2}{h}\sum_{\langle \mathbf{i},\mathbf{j}\rangle_\text{lines},m}n_{\mathbf{i}m}(1-n_{\mathbf{j}m})-\frac{t_y}{h}\sum_{\substack{\mathbf{i},\mathbf{j},\mathbf{k}\in\;\triangle\\m,m'}}V^{\textbf{jk}}_{mm'}b^\dag_{\textbf{k}m} b^\dag_{\textbf{j}m'} b_{\textbf{k}m'} b_{\textbf{i}m},
\label{Heff_triangularlattice}
\end{equation}
where the first one is associated to back and forth hoppings along the lines, and the second one is the plaquette term. To make concrete the mapping with the gauge theory, we identify the link operators on the different sides of the triangle as
\begin{equation}
U_{z}=b^\dag_{zm'}b_{zm},\qquad U_{y}=b^\dag_{ym},\qquad U_{x}=b_{xm},
\label{operators_triangularlattice}
\end{equation}
making reference to the notation of Fig. \ref{triangular_proposal}(a).
With this identification, the underlying LGT effective Hamiltonian has the form
\begin{equation}
H_\triangle^{(\text{eff})}=\lambda_{\triangle}\sum_{\langle \textit{i,j}\rangle_\text{lines},m}E_{\textbf{i}m}E_{\textbf{j}m}-\sum_{\triangle}J_\triangle^{(mm')}(U_\triangle+U^\dag_\triangle).
\label{Heff_triangular_gaugetheory}
\end{equation}
As in the square lattice case, we have the generation of the plaquette term plus an asymmetric term in the square of the electric field, due to the optical lattice structure. This is again different from the full RK model on the triangular lattice \cite{moessnerPRL2001}, and the various phases of the two models are in principle different.

\subsection{\label{higher_spin}Higher spin quantum link models}
In Section \ref{UCatomsrealization} we showed how to simulate spin-$1/2$ QLMs using ultracold gases of hardcore bosons, i.e. working in the limit $U_0\rightarrow\infty$. This allows for single particle occupations, and makes each link of the target lattice an effective spin-$1/2$ variable. Here we comment about the relaxation of the hardcore constraint on the ultracold bosons and about the quantum simulation of higher spin QLMs.

First of all, let us consider $U_0\nrightarrow\infty$: then one could resort to a Schrieffer-Wolff transformation \cite{SW_1966} to find the corresponding effective field theory and the terms generated by the finite $U_0$, i.e. the non-hardcore condition. Out of this procedure, one expects to obtain terms that can be incorporated at low energies in the parameters of the QLM: in other words, the QLM parameters are renormalized by $U_0$ (as an example see \cite{Giuliano} for the computation in the case of the extended BH model at half-filling). Of course, this expectation is valid as soon that $U_0$ is varied in a way such that other phases appear.

The implementation of higher spin QLMs is related to the previous comment. Indeed, suppose that one is able to allow for multiple particles occupations at different sites of the optical lattice, e.g. that one can impose that in each site there may be at most a certain number of bosons of the same level. By imposing that the number of particles within the lattice is fixed, the above mentioned procedure could be used to obtain the simulation of higher spin QLMs. However, it is not a finite $U_0$ the key ingredient to have such models, but rather higher-body terms to fix the maximum occupation of the lattice sites. 

\section{\label{conclusions}Conclusions}
In this paper we presented a proposal for the quantum simulation of Abelian spin-$1/2$ quantum link models (QLMs) using spinor dipolar Bose-Einstein condensates (BECs) loaded in spin-dependent optical lattices. We showed that plaquette interactions can be obtained, and are directly related to correlated hoppings of bosons within a spin-dependent lattice.

In order to derive the effective theory corresponding to a gauge theory, we considered two different approaches. The first one, that we referred to as the \textit{non-perturbative scheme}, does not employ perturbation theory as the analytical tool to derive the plaquette interaction.  We propose to load a spin-2 dipolar BECs in a spin-dependent superlattice obtained by the superimposition of several spin-dependent lattices. By requiring angular momentum conservation and expanding in the correct basis of lattice bosonic operators, we derived an optical superlattice Hamiltonian containing plaquette interactions as one of the local processes compatible with the symmetries of the system. In this approach, it is not needed to do a perturbative expansion in large parameters of the Bose-Hubbard Hamiltonian describing the system. The plaquette term $H_\square$ is straightforwardly obtained, and the corresponding Hamiltonian \eqref{relevantH_optical_superlattice} features the $H_\square$ term and a density-density interaction, having coefficients which can be, to a certain extent, tuned. A study of the phase diagram of Eq. \eqref{relevantH_optical_superlattice} when the its parameters are varied would be, in our opinion, very interesting.

In the second scheme, that we called the \textit{perturbative scheme}, we considered an extended Bose-Hubbard (BH) model with anisotropic hoppings and isotropic nearest neighbors interactions, with a further site-dependent energy penalty term to enforce the lattice structure. By doing perturbation theory using the on-site energies as large parameters, we derived an effective Hamiltonian, whose gauge theoretical interpretation includes the plaquette interactions at third-order in perturbation theory. This process is associated to correlated hoppings of bosons within elementary squares, with a subsequent spin-dependent interaction to make their internal states compatible with the spin lattice structure. We notice that in this second, perturbative, scheme, extra terms not present in Eq. \eqref{relevantH_optical_superlattice} are, at variance, present.

The use of angular momentum conservation in scattering processes to ensure gauge invariance was used in Ref. \cite{ZoharReznikPRA2013}. In that case, it guarantees that the gauge-matter interaction satisfies gauge invariance. By other side, plaquette terms are still obtained perturbatively. In contrast, our target model does not include matter and uses the conservation of angular momentum as a mean to obtain robust plaquette terms of the underlying pure gauge theory. The net result is that we must use at least four internal states of a spinor dipolar gas, moving in one-dimensional chains coupled by the dipolar interactions. Without angular momentum conservation, on-site energies superimposed via additional superlattices are needed to produce the target lattice. Conversely, by relying entirely on angular momentum conservation and building the superlattice with a superimposition of spin-dependent lattices, we obtained the plaquette interactions without any large parameter’s expansion, in parallel with gauge-matter coupling obtained in \cite{ZoharReznikPRA2013}. We think, for these reasons, that it is worth to ascertain whether this can be concretely an advantage to simulate QLMs with respect to the perturbative scheme, where the requirements related to perturbative arguments on the experimental control of the atoms are absent.

We further discussed possible extensions of our proposal to the triangular lattice, showing that it is possible to lower the perturbative order at which the plaquette terms appear, with the proper choice of the spin-dependent lattice, and to higher spin QLMs, allowing for multiple site occupations with a fixed number of bosons. In this last case it is not trivial to generalize our proposal, since we get non-local effective interactions in the perturbative expansions, and the plaquette term is coupled to the electric field. Finally, our proposal can be generalized to the three-dimensional case by increasing the number of internal states employed in $d=2$: with respect to the four levels used in $d=2$, we need to use five internal states in $d=3$. This is done by identifying the planes in which the particles move \cite{banerjee2021} and derive a set of equations for the internal states of the links in the third spatial directions. These equations can be satisfied at the cost of introducing extra atomic species. In the perturbative scheme, gauge breaking terms emerge, which should be removed by adding {\it ad hoc} terms. These complications show, in our particular example, the difficulties of extending a scheme from $d=2$ to $d=3$.

To put our results in the wider context of quantum simulations of higher dimensional lattice gauge theories (LGTs), we observe that these last are currently challenging to realize in the realm of quantum simulators, even if there have been a lot of recent technical progresses in the field \cite{banuls_QTreview2020,banulsRPP2020,aidelsburger_RSP2022,davoudi_QSforHEP2022}. This is mainly due to local symmetry of the model, and its direct implementation in controllable physical systems. Our proposal employs many-body interaction symmetries to achieve this target. The main advantage of our scheme is to relate the local conservation of angular momentum to the gauge symmetric plaquette terms. Even in the perturbative scheme, there is advantage coming from the order at which the plaquette terms come out with respect to Ref. \cite{ZoharReznikPRA2013}, where it appears at fourth-order, here it shows up at third-order, involving only two correlated hoppings between bosons. A disadvantage is provided by the complications to generalize the scheme to $d=3$, as discussed, anyway noticing that -- to the best of our knowledge -- the extension to $d=3$ is an issue for all the schemes present in literature so far.

There are open questions that are worth to pursue. First of all, it would be interesting to understand the role of the additional terms coming from the perturbative expansion regarding the phase diagram of the model. In the gauge theoretical interpretation they are anisotropic in the electric field, and may give rise to a different phase diagram if compared with the Roksha-Kivelson (RK) Hamiltonian \cite{RK1988}, based on symmetry arguments. From the condensed matter perspective, due to the extended nature of the Bose-Hubbard Hamiltonian, they could be related to supersolid phases \cite{vanotterlo1995,capogrossosansonePRL2010,Zhang2015}. 
Another important point regards the inclusion of matter. There is no straightforward way of doing this within this proposal but may be possible through the inclusion of ancillary particles. This difficulty is also present in the proposal of \cite{Celi2019}, where it is proposed a way to include static charges, but not dynamical ones. In the present proposal, static charges can be easily included by violating Gauss' law at the desired sites in the initial state. In the proposal of \cite{ZoharReznikPRA2013}, instead, they include dynamical matter by increasing the number of fermionic species, in addition to the ancillary ones. Finally, it would be interesting to generalize the presented scheme to non-Abelian gauge theories. However, the advantage gained in encoding plaquettes in correlated hopping is quickly lost by the increasing complicated substructure of the superlattice required by non-Abelian symmetries. A more reasonable goal may consist on considering discrete groups, in order to explore different physics and, possibly, simplify the superlattice structure. 

\begin{acknowledgements}
	We thank D. Banerjee, L. Santos and F. M. Surace  for very useful discussions. P.F. thanks the Pauli Center for Theoretical Studies and the Institute for Theoretical Physics of ETH Zurich for the kind hospitality, where the final part of the work has been performed. 
\end{acknowledgements}

\widetext

\appendix

\section{\label{species_number}Determination of the number of species}

\begin{figure}[h]
	\centering
	\includegraphics[width=0.60\linewidth]{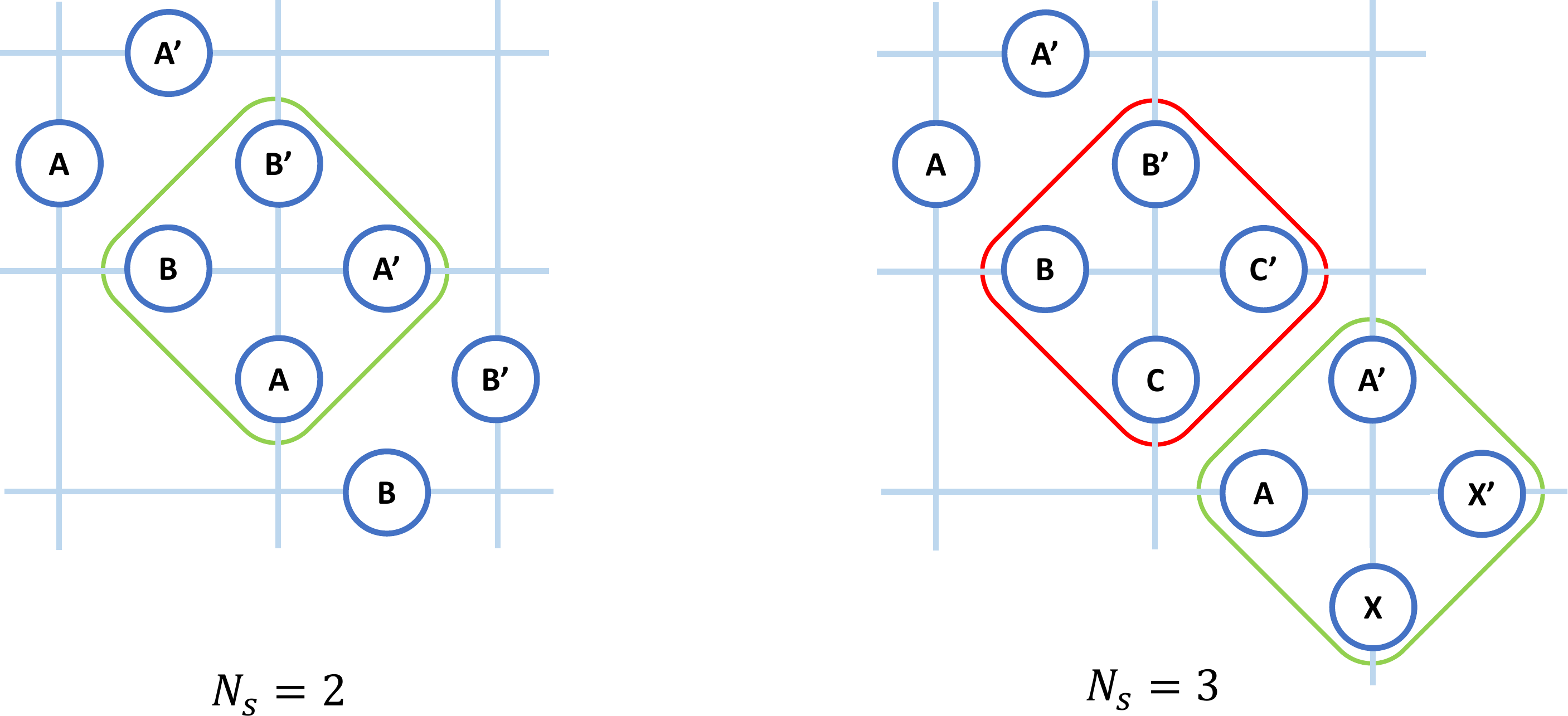}
	\caption{Lattice structure with $N_s=2$ (left panel) and $N_s=3$ (right panel) internal states. The quantum numbers are denoted respectively with $A,\;B$ and $C$. The primed letters denote a different permutations of the internal states along the two lattice diagonals. The green (red) square encloses the vertices that are problematic (correct) in the spirit of our proposal.}
	\label{ns2_ns3_lattices}
\end{figure}
We motivate here the choice of four internal states for our proposal on the square lattice. To do this, we show what are the inconsistencies when less internal states are considered in the construction of the spin-dependent optical lattice.

Let us firstly discuss the case of $N_s=2$ internal states (left panel of Fig. \ref{ns2_ns3_lattices}). We consider two diagonals of the lattice, with generic internal levels $A,\;B$ and $A',B'$ that can assume only two integer values. If we want to generate the plaquette term through angular momentum conservation, we have to impose the condition
\begin{equation}
\Delta_1=B'-A',\qquad\Delta_2=B-A\qquad\Rightarrow\qquad\Delta_1+\Delta_2=0.
\end{equation}
These conditions will inevitably lead to unwanted hoppings at the vertices (see the green square in the left panel of Fig. \ref{ns2_ns3_lattices}), since they break gauge symmetry. 

If we try to repeat the same reasoning with $N_s=3$ internal levels (right panel of Fig. \ref{ns2_ns3_lattices}), we can solve this inconsistency in the first vertex (red square in Fig. \ref{ns2_ns3_lattices}) by choosing $C$ and $C'$ such that $\Delta_1+\Delta_2=0$, and at the same time $\Delta_3+\Delta_2\neq0$, with $\Delta_3=C'-C$. We have then two different types of plaquettes in this scheme. However, inconsistencies arise again when we go to the subsequent vertex (green square in the right panel of Fig. \ref{ns2_ns3_lattices}), and try to combine $A,\;A'$ with $X,X'$ that could be either $B$ ($B'$) or $C$ ($C'$) internal states. Also in this case, for any $X,\;X'$
there are unwanted hopping processes at this vertex.

The solution to this is to use four internal states, but this is still not enough. To avoid hopping at the vertices throughout all the lattice, there must be two internal states $m,\;m'$ such that their difference is $|m-m'|>1$, as in our proposal of Fig. \ref{optical_lattices}. The minimal requirement is then to use four out of five hyperfine levels, as developed in the main text.

\section{\label{opticallattice_computations}Derivation of the Hamiltonian \eqref{relevantH_optical_superlattice}}
\begin{figure}[h]
		\centering
		\includegraphics[width=0.45\linewidth]{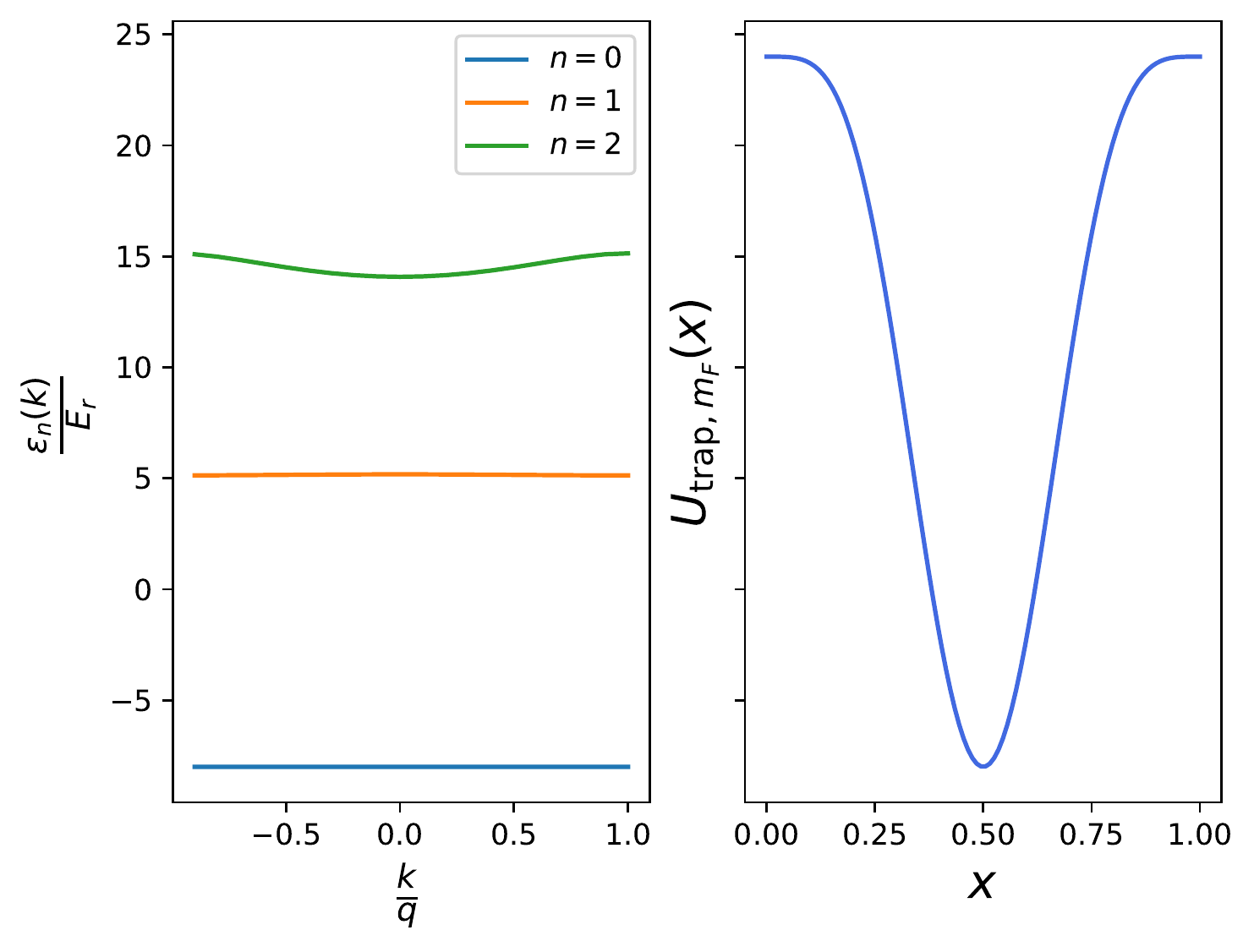}
		\qquad\qquad\qquad\includegraphics[width=0.43\linewidth]{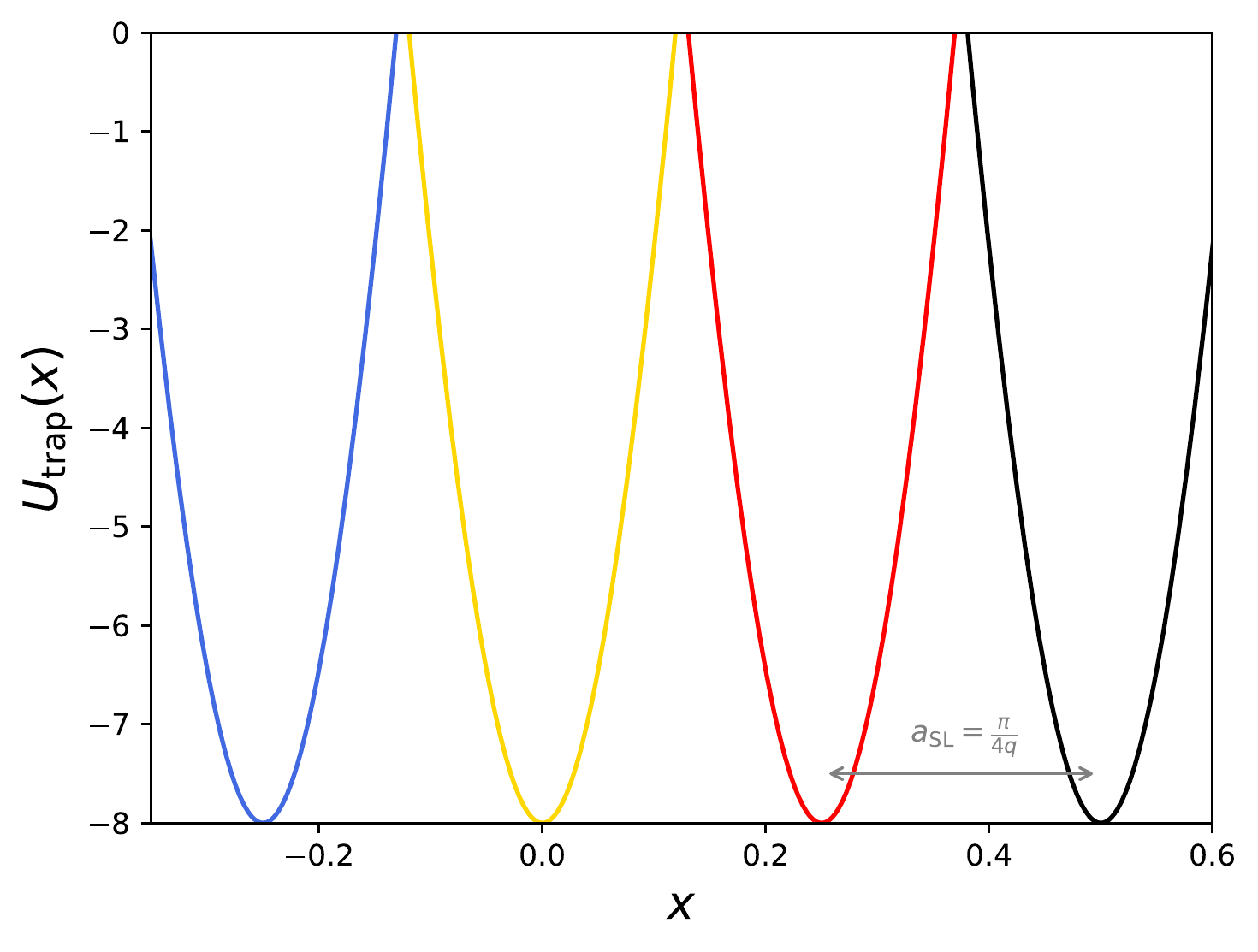}
		\caption{Left panel: band structure of $U_{\text{trap},m_F}(x)$ for a fixed internal level and in units of the recoil energy $E_r=\hbar^2q^2/2M$. Right panel: superimposition of all the potential along the $x$-axis to generate the periodic structure in Fig. \ref{optical_lattices}, with lattice spacing given by $a_{\text{SL}}$ (grey arrow). In both these plots we fixed $A_1=32$, $A_2=-8$ and $q=\pi$.}
		\label{bands_fulllattice_x}
\end{figure}

We show here the computations leading to the Hamiltonian presented in Eq. \eqref{relevantH_optical_superlattice}. The initial point is the solution of the single-particle Hamiltonian $H_0$ with the spin-dependent optical potential $U_{\text{trap},m_F}(\mathbf{r})$, defined in Section \ref{non_pert_appr}. In this Appendix we fix $\delta_{m_F}=n(m_F) \pi/4q$, with $n(m_F)$ such that $n(-2)=0,\;n(-1)=1,\;n(0)=2,\;n(2)=3$, to reproduce the superlattice structure of Fig. \ref{bands_fulllattice_x} (right panel). With this choice, the lattice spacing of the optical superlattice is $a_{\text{SL}}=\pi/4q$, with $q=2\pi/\lambda_L$ the laser wave-vector.

Due to the periodicity of the potential, we can apply the Bloch theorem and obtain the eigenfunctions of the single-particle problem, which will be factorized in the spatial components. If we assume that the particle can be in any of the considered internal states, the single particle Hamiltonian density $\mathcal{H}(\mathbf{r})$ looks like
\begin{equation}
    \mathcal{H}(\mathbf{r})=\sum_{m_F\in\{-2,-1,0,2\}}\phi^\dag_{m_F}(\mathbf{r})\bigg[-\frac{\hbar^2\nabla^2}{2M}+U_{\text{trap},m_F}(\mathbf{r})\bigg]\phi_{m_F}(\mathbf{r})\equiv\mathbf{\Phi}^\dagger(\mathbf{r})\mathcal{K}\mathbf{\Phi}(\mathbf{r}),
\end{equation}
where we defined the column vector $\mathbf{\Phi}(\mathbf{r})=(\phi_{-2}(\mathbf{r}),\phi_{-1}(\mathbf{r}),\phi_{0}(\mathbf{r}),\phi_2(\mathbf{r}))^T$ and a block diagonal matrix $\mathcal{K}$. This can be seen as a single-particle problem with four levels, and it is enough to analyse it along the $\hat{x}$-direction, due to the separability of the potential. The lowest energy bands of Eq. \eqref{x_potential} are showed in Fig. \ref{bands_fulllattice_x} (left panel). Since we assumed that all the potentials along $\hat{x}$ are the same apart from a shift, we see that the energies of the first four bands are equal. For the same reason, the Bloch functions $u^{(m_F)}_\mathbf{k}(\mathbf{r})$ will be the same for each internal states, but shifted by the superlattice spacing $a_{\text{SL}}$. 

Summarizing we have
\begin{equation}
    \mathbf{u}_\mathbf{k}(\mathbf{r})=(u^{(-2)}_\mathbf{k}(\mathbf{r}),u^{(-1)}_\mathbf{k}(\mathbf{r}+a_{\text{SL}}\hat{x}),u^{(0)}_\mathbf{k}(\mathbf{r}+2a_{\text{SL}}\hat{x}),u^{(2)}_\mathbf{k}(\mathbf{r}+3a_{\text{SL}}\hat{x}))^T,\qquad E_{0,\mathbf{k}}=\sum_{m_F\in\{-2,-1,0,2\}}\epsilon^{(m_F)}_{0,\mathbf{k}}.
    \label{generalized_blochwaves}
\end{equation}
Given this form of the Bloch functions, the procedure to derive the optical lattice Hamiltonian is the standard one \cite{Lewenstein_book_2012}, with the difference that we have to employ the Wannier functions in the expansion of the field operator $\mathbf{\Phi}(\mathbf{r})$, i.e.
\begin{equation}
    w_{0,m_F}(\mathbf{r}-\mathbf{r}_\mathbf{i})\equiv\frac{1}{\mathcal{N}}\sum_{\mathbf{k}}e^{-i\mathbf{k}\cdot(\mathbf{r}-\mathbf{r}_\mathbf{i}-na_{\text{SL}}\hat{x})}u^{(m_F)}_\mathbf{k}(\mathbf{r}+n(m_F)a_{\text{SL}}\hat{x})\qquad\Rightarrow\qquad\phi_m(\mathbf{r})=\sum_\mathbf{i}w_{0,m_F}(\mathbf{r}-\mathbf{r}_\mathbf{i})b_{\mathbf{i}m}.
    \label{generalized_wannier_fieldexp}
\end{equation}
By inserting this expansion in Eq.s \eqref{spinorBEC_generalH} and \eqref{MDDI_interaction_hamiltonian} we end up in Eq. \eqref{fullH_optical_superlattice} with amplitudes
\begin{equation}
    T^m_{\mathbf{ij}}=\int\;d\mathbf{r}d\mathbf{r}'\;\sum_{\nu,\nu'}\sum_{\ell,\mathbf{k}}\;\mathcal{W}^{m\ell\ell m}_{\mathbf{ikkj}}(\mathbf{r},\mathbf{r'})\mathcal{F}^{\nu\nu'}_{m\ell\ell m}Q_{\nu\nu'}(\mathbf{r}-\mathbf{r}'),
    \label{2body_NP_amplitude}
\end{equation}
\begin{equation}
    U^{mm'\ell\ell'}_{\mathbf{ijkp}}=\int\;d\mathbf{r}d\mathbf{r}'\;\sum_{\nu,\nu'}\mathcal{W}^{m\ell m'\ell'}_{\mathbf{ijkp}}(\mathbf{r},\mathbf{r'})\mathcal{F}^{\nu\nu'}_{m m'\ell\ell' }Q_{\nu\nu'}(\mathbf{r}-\mathbf{r}'),
    \label{4body_NP_amplitude}
\end{equation}
where we defined
\begin{equation}
    \mathcal{W}^{m\ell m'\ell'}_{\mathbf{i}\mathbf{k}\mathbf{j}\mathbf{p}}(\mathbf{r},\mathbf{r'})\equiv w^*_{0,m}(\mathbf{r}-\mathbf{r}_i)w^*_{0,\ell}(\mathbf{r}'-\mathbf{r}_k)w_{0,m'}(\mathbf{r}-\mathbf{r}_j)w_{0,\ell'}(\mathbf{r}'-\mathbf{r}_p),
\end{equation}
\begin{equation}
    \mathcal{F}^{\nu\nu'}_{mm'\ell\ell'}\equiv[f_\nu]_{mm'}[f_{\nu'}]_{\ell\ell'}.
\end{equation}
The plaquette terms are directly contained in Eq. \eqref{4body_NP_amplitude} as one of the most local processes in the optical superlattice, leading to a non-trivial contribution to the Hamiltonian without the need of using perturbation theory.

Indeed, by looking at the target lattice in Fig. \ref{optical_lattices} (right panel), the values of the internal states on the opposite links composing the plaquettes are the same. This means to consider $m=\ell'$ and $\ell=m'$ in Eq. \eqref{4body_NP_amplitude}. Moreover, we can define $\mathbf{R}\equiv\mathbf{r}-\mathbf{r}'$ and simplify the structure of $\mathcal{F}\cdot Q$ as
\begin{equation}
	\sum_{\nu,\nu'}[f_\nu]_{m\ell}[f_{\nu'}]_{\ell m}(\delta_{\nu\nu'}-3\hat{R}_\nu\hat{R}_{\nu'})=\mathbf{f}_{m\ell}\cdot\mathbf{f}_{\ell m}-3[\mathbf{f}_{m\ell}\cdot\hat{\mathbf{R}}][\mathbf{f}_{\ell m}\cdot\hat{\mathbf{R}}],
\end{equation}
with the spatial scalar product defined as $\mathbf{A}\cdot\mathbf{B}\equiv\sum_\nu A_\nu B_\nu$. We have therefore to evaluate how these contributions extend throughout the lattice.

On the basis of the considered optical superlattice structure, we argue the most significant extended terms arise when the lattice sites are such that $\langle\mathbf{i,j}\rangle$ and $\langle\mathbf{k,p}\rangle$, i.e. when the plaquette process is considered, for two reasons. 
The first one is angular momentum conservation: the only correlated hopping processes emerging in the optical superlattice are generating the plaquettes, the others being forbidden as explained in the main text. The second reason is based on the localized feature of Wannier's
functions, which is preserved even in the presence of the dipolar potential. Actually, the effective dipole-dipole interactions in optical lattices are modified with respect to the continuum ones. The modifications related to the overlap of the Wannier's functions, i.e. the computation of the integral in Eq. \eqref{4body_NP_amplitude}, affect the exchange channel: the power-law decay of the dipolar interaction gets modulated by an exponential decay for short (or moderate) separations on the optical lattice \cite{Wall_2013}.

\section{\label{pert_theory}Perturbation theory contributions}
\begin{figure}[h]
	\centering
	\includegraphics[width=0.8\linewidth]{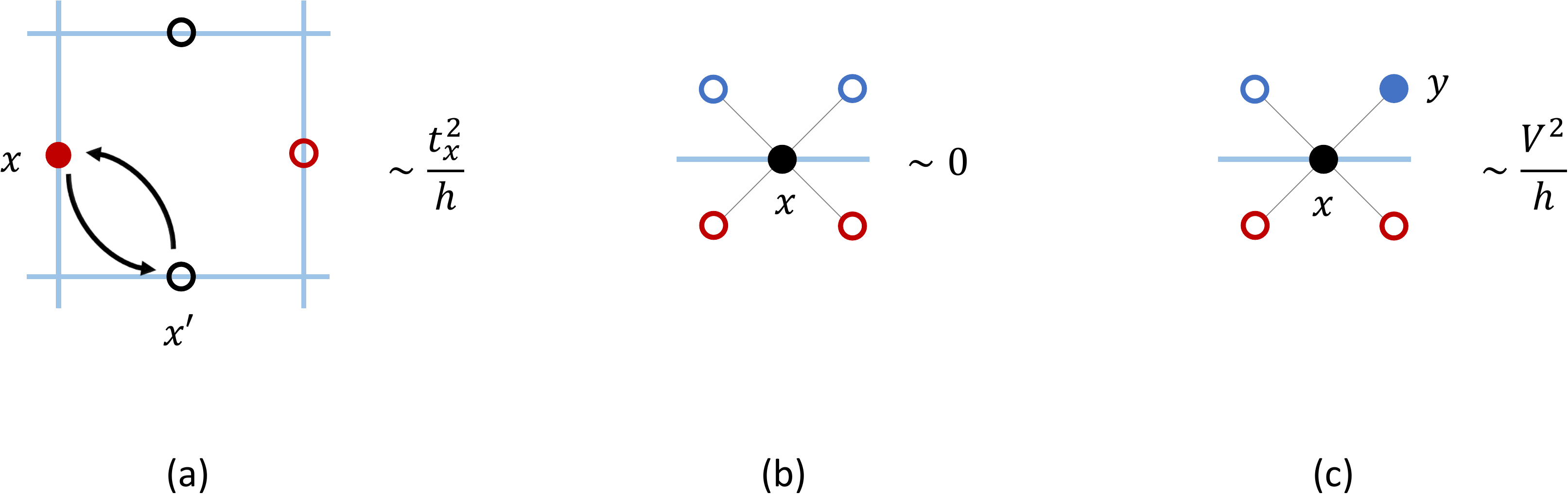}
	\caption{Virtual processes at second-order in perturbation theory. In all the panels, light blue lines are representing the target lattice, while grey lines the optical one, according to the convention of the main text. (a) Back and forth hopping process $\textbf{x}\rightarrow \textbf{x}'\rightarrow \textbf{x}$, generating the term in Eq. \eqref{H2_eff_hophop}. This process can happen if $\textbf{x}$ is occupied and $\textbf{x}'$ is empty, independently of the other two sites of the plaquette. (b)-(c): Processes associated to the $H_{\text{int}}-H_{\text{int}}$ contribution of Eq. \eqref{H2_eff_hophop}. The only trivial situation is when all of the nearest neighbors of $\textbf{x}$ are empty.}
	\label{2ndorder_processes_Heff}
\end{figure}
We present here the computations in perturbation theory that give rise to the effective model in Eq. \eqref{3rdorder_Heff}. The starting point is the full Hamiltonian $H=H_1+H_0$, with $h\gg t_x,\;V^{\textbf{ij}}_{mm'}$. Making reference to the definition of $\mathcal{M}_0$ given in the main text, we define respectively the projector on $\mathcal{M}_0$ and the restricted inverse of $(H_0-E_0)$ as
\begin{equation}
\mathcal{P}_{0}=\sum_{\alpha\in\mathcal{M}_0}|\alpha\rangle\langle\alpha|,\qquad\qquad\mathcal{K}=\sum_{\alpha\notin\mathcal{M}_0}\frac{|\alpha\rangle\langle\alpha|}{E_0(\alpha)-E_0(0)},
\end{equation}
where $E_0$ is the eigenvalue of $H_0$, and proceed with the computations order by order \cite{soliverez1969}. At first-order the contribution is $H^{\text{(eff)}}_1=\mathcal{P}_0H_1\mathcal{P}_0$, and it is always zero. This is because, both for the hopping and interaction term, with a single application of $H_1$ we move out of the ground state manifold. Therefore, once $\mathcal{P}_0$ acts again, the projection gives zero. At second-order we have $H^{\text{(eff)}}_2=-\mathcal{P}_0H_1\mathcal{K}H_1\mathcal{P}_0$, and the two non-trivial contributions are 
\begin{equation}
\mathcal{P}_0H_{\text{hop}}\mathcal{K}H_{\text{hop}}\mathcal{P}_0=-\frac{t_x^2}{h}\sum_{\langle \textbf{x,x'}\rangle_d,m,m'}n_{\textbf{x}m}(1-n_{\textbf{x}'m'}),
\label{H2_eff_intint}
\end{equation}
\begin{equation}
\mathcal{P}_0H_{\text{int}}\mathcal{K}H_{\text{int}}\mathcal{P}_0=-\frac{1}{h}\sum_{\langle \textbf{x,y}\rangle,m,m'}(V^{\textbf{xy}}_{mm'})^2n_{\textbf{x}m}n_{\textbf{y}m'}.
\label{H2_eff_hophop}
\end{equation}

The graphical representation of the virtual processes associated to these two terms is showed in Fig. \ref{2ndorder_processes_Heff}. Concerning the pure hopping contribution of the first equation, it is associated to back and forth hopping processes happening along the diagonals; for the interaction part, in the second line, we have that two generic nearest neighbors in the optical lattice interact mutually. These are the only processes that do not vanish when projected back to the ground state manifold $\mathcal{M}_0$.

\begin{figure}[h]
	\centering
	\includegraphics[width=0.8\linewidth]{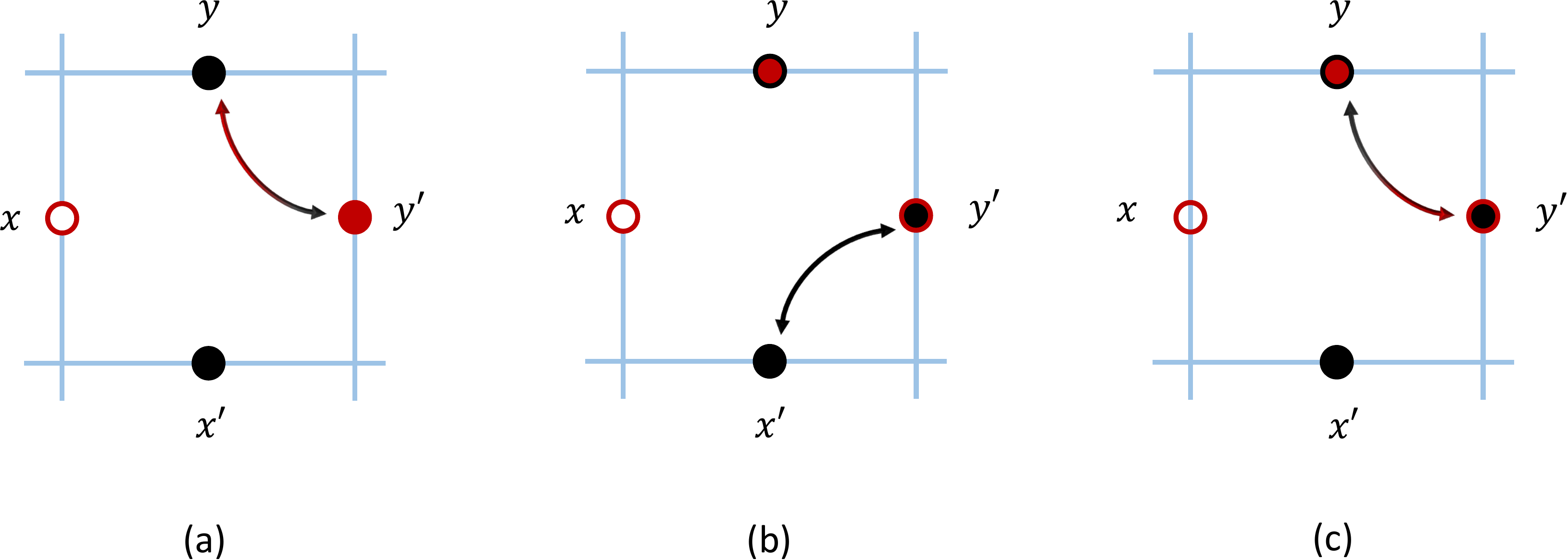}
	\caption{Virtual processes due to three spin-exchange interactions at third-order in perturbation theory. We have the spin-exchange between $\textbf{y}-\textbf{y}'$ (a), followed by the ones involving $\textbf{y}'-\textbf{x}'$ (b) and again $\textbf{y}'-\textbf{y}$ (c). The final state here coincides with the initial one. This process can happen if all the three involved sites contain the bosons in the proper initial states, according to the lattice structure. In all the panels, light blue lines are representing the target lattice, while grey lines the optical one, according to the convention of the main text}
	\label{3rdorder_extraprocesses_Heff}
\end{figure}

At third-order we get two non-trivial contribution, including the plaquette term already commented in the main text. The structure of the effective Hamiltonian is
\begin{equation}
H^{\text{(eff)}}_3=\mathcal{P}_0H_1\mathcal{K}H_1\mathcal{K}H_1\mathcal{P}_0-\frac{1}{2}\{\mathcal{P}_0H_1\mathcal{K}^2H_1\mathcal{P}_0,\mathcal{P}_0H_1\mathcal{P}_0\}\equiv \tilde{H}_3+\tilde{H}_2,
\end{equation}
where $\tilde{H}_2$ is a combination of lower order terms, and $\tilde{H}_3$ is the real third-order contribution. Due to the fact that 
\begin{equation}
\tilde{H}_2=-\frac{1}{2}\{\mathcal{P}_0H_1\mathcal{K}^2H_1,H^{\text{(eff)}}_1\}
\end{equation}
and since $H^{\text{(eff)}}_1=0$, this term vanishes. We are left only with $\tilde{H}_3$, and among the various contributions the only non-trivial ones are $\mathcal{P}_0H_{\text{int}}\mathcal{K}H_{\text{hop}}\mathcal{K}H_{\text{hop}}\mathcal{P}_0$, generating the plaquette term, and $\mathcal{P}_0H_{\text{int}}\mathcal{K}H_{\text{int}}\mathcal{K}H_{\text{int}}\mathcal{P}_0$, due to the spin-exchange interactions within a given plaquette. We make reference to Fig. \ref{3rdorder_extraprocesses_Heff} for the representation of the virtual process in this last case, which is constrained to have the sites $\textbf{x}',\textbf{y},\textbf{y}'$ in the plaquettes occupied by particles compatible with the spin-dependent lattice structure. The total third-order contribution is therefore
\begin{equation}
\tilde{H}_3=\frac{t_x^2}{h^2}\sum_{\substack{\textbf{x},\textbf{x}',\textbf{y},\textbf{y}'\in\;\square\\m,m'}}V^{\textbf{xx}'}_{mm'}b_{\textbf{y}'m}^\dag b_{\textbf{y}m'}^\dag b_{\textbf{x}'m'} b_{\textbf{x}m}+\text{h.c.}+\frac{1}{h^2}\sum_{\substack{\textbf{x}',\textbf{y},\textbf{y}'\in\;\square\\m,m'}}(V^{\textbf{yy}'}_{mm'})^2V^{\textbf{y}'\textbf{x}'}_{mm}n_{\textbf{y}'m}n_{\textbf{x}'m}n_{\textbf{y}m'}.
\end{equation}

\bibliography{biblio}

\begin{thebibliography}{78}%
\makeatletter
\providecommand \@ifxundefined [1]{%
 \@ifx{#1\undefined}
}%
\providecommand \@ifnum [1]{%
 \ifnum #1\expandafter \@firstoftwo
 \else \expandafter \@secondoftwo
 \fi
}%
\providecommand \@ifx [1]{%
 \ifx #1\expandafter \@firstoftwo
 \else \expandafter \@secondoftwo
 \fi
}%
\providecommand \natexlab [1]{#1}%
\providecommand \enquote  [1]{``#1''}%
\providecommand \bibnamefont  [1]{#1}%
\providecommand \bibfnamefont [1]{#1}%
\providecommand \citenamefont [1]{#1}%
\providecommand \href@noop [0]{\@secondoftwo}%
\providecommand \href [0]{\begingroup \@sanitize@url \@href}%
\providecommand \@href[1]{\@@startlink{#1}\@@href}%
\providecommand \@@href[1]{\endgroup#1\@@endlink}%
\providecommand \@sanitize@url [0]{\catcode `\\12\catcode `\$12\catcode
  `\&12\catcode `\#12\catcode `\^12\catcode `\_12\catcode `\%12\relax}%
\providecommand \@@startlink[1]{}%
\providecommand \@@endlink[0]{}%
\providecommand \url  [0]{\begingroup\@sanitize@url \@url }%
\providecommand \@url [1]{\endgroup\@href {#1}{\urlprefix }}%
\providecommand \urlprefix  [0]{URL }%
\providecommand \Eprint [0]{\href }%
\providecommand \doibase [0]{https://doi.org/}%
\providecommand \selectlanguage [0]{\@gobble}%
\providecommand \bibinfo  [0]{\@secondoftwo}%
\providecommand \bibfield  [0]{\@secondoftwo}%
\providecommand \translation [1]{[#1]}%
\providecommand \BibitemOpen [0]{}%
\providecommand \bibitemStop [0]{}%
\providecommand \bibitemNoStop [0]{.\EOS\space}%
\providecommand \EOS [0]{\spacefactor3000\relax}%
\providecommand \BibitemShut  [1]{\csname bibitem#1\endcsname}%
\let\auto@bib@innerbib\@empty
\bibitem [{\citenamefont {Feynman}(1982)}]{feynmanQS1982}%
  \BibitemOpen
  \bibfield  {author} {\bibinfo {author} {\bibfnamefont {R.~P.}\ \bibnamefont
  {Feynman}},\ }\href@noop {} {\bibfield  {journal} {\bibinfo  {journal}
  {\href{https://doi.org/10.1007/BF02650179}{International Journal of
  Theoretical Physics}}\ }\textbf {\bibinfo {volume} {21}},\ \bibinfo {pages}
  {467} (\bibinfo {year} {1982})}\BibitemShut {NoStop}%
\bibitem [{\citenamefont {Trabesinger}(2012)}]{Trabesinger2012}%
  \BibitemOpen
  \bibfield  {author} {\bibinfo {author} {\bibfnamefont {A.}~\bibnamefont
  {Trabesinger}},\ }\href@noop {} {\bibfield  {journal} {\bibinfo  {journal}
  {\href{https://doi.org/10.1038/nphys2258}{Nature Physics}}\ }\textbf
  {\bibinfo {volume} {8}},\ \bibinfo {pages} {263} (\bibinfo {year}
  {2012})}\BibitemShut {NoStop}%
\bibitem [{\citenamefont {Blatt}\ and\ \citenamefont
  {Roos}(2012)}]{Blatt_trappedions2012}%
  \BibitemOpen
  \bibfield  {author} {\bibinfo {author} {\bibfnamefont {R.}~\bibnamefont
  {Blatt}}\ and\ \bibinfo {author} {\bibfnamefont {C.}~\bibnamefont {Roos}},\
  }\href@noop {} {\bibfield  {journal} {\bibinfo  {journal}
  {\href{https://doi.org/10.1038/nphys2252}{Nature Physics}}\ }\textbf
  {\bibinfo {volume} {8}},\ \bibinfo {pages} {277} (\bibinfo {year}
  {2012})}\BibitemShut {NoStop}%
\bibitem [{\citenamefont {Houck}\ \emph {et~al.}(2012)\citenamefont {Houck},
  \citenamefont {Türeci},\ and\ \citenamefont {Koch}}]{Houck_SCcircuits2012}%
  \BibitemOpen
  \bibfield  {author} {\bibinfo {author} {\bibfnamefont {A.}~\bibnamefont
  {Houck}}, \bibinfo {author} {\bibfnamefont {H.}~\bibnamefont {Türeci}},\
  and\ \bibinfo {author} {\bibfnamefont {J.}~\bibnamefont {Koch}},\ }\href@noop
  {} {\bibfield  {journal} {\bibinfo  {journal}
  {\href{https://doi.org/10.1038/nphys2251}{Nature Physics}}\ }\textbf
  {\bibinfo {volume} {8}},\ \bibinfo {pages} {292} (\bibinfo {year}
  {2012})}\BibitemShut {NoStop}%
\bibitem [{\citenamefont {Wu}\ \emph {et~al.}(2021)\citenamefont {Wu},
  \citenamefont {Liang}, \citenamefont {Tian}, \citenamefont {Yang},
  \citenamefont {Chen}, \citenamefont {Liu}, \citenamefont {Tey},\ and\
  \citenamefont {You}}]{Wu_2021}%
  \BibitemOpen
  \bibfield  {author} {\bibinfo {author} {\bibfnamefont {X.}~\bibnamefont
  {Wu}}, \bibinfo {author} {\bibfnamefont {L.}~\bibnamefont {Liang}}, \bibinfo
  {author} {\bibfnamefont {T.}~\bibnamefont {Tian}}, \bibinfo {author}
  {\bibfnamefont {F.}~\bibnamefont {Yang}}, \bibinfo {author} {\bibfnamefont
  {C.}~\bibnamefont {Chen}}, \bibinfo {author} {\bibfnamefont {Y.-C.}\
  \bibnamefont {Liu}}, \bibinfo {author} {\bibfnamefont {M.~K.}\ \bibnamefont
  {Tey}},\ and\ \bibinfo {author} {\bibfnamefont {L.}~\bibnamefont {You}},\
  }\href@noop {} {\bibfield  {journal} {\bibinfo  {journal}
  {\href{https://dx.doi.org/10.1088/1674-1056/abd76f}{Chinese Physics B}}\
  }\textbf {\bibinfo {volume} {30}},\ \bibinfo {pages} {020305} (\bibinfo
  {year} {2021})}\BibitemShut {NoStop}%
\bibitem [{\citenamefont {Bloch}\ \emph {et~al.}(2012)\citenamefont {Bloch},
  \citenamefont {Dalibard},\ and\ \citenamefont
  {Nascimbène}}]{Dalibard_UCatoms2012}%
  \BibitemOpen
  \bibfield  {author} {\bibinfo {author} {\bibfnamefont {I.}~\bibnamefont
  {Bloch}}, \bibinfo {author} {\bibfnamefont {J.}~\bibnamefont {Dalibard}},\
  and\ \bibinfo {author} {\bibfnamefont {S.}~\bibnamefont {Nascimbène}},\
  }\href@noop {} {\bibfield  {journal} {\bibinfo  {journal}
  {\href{https://doi.org/10.1038/nphys2259}{Nature Physics}}\ }\textbf
  {\bibinfo {volume} {8}},\ \bibinfo {pages} {267} (\bibinfo {year}
  {2012})}\BibitemShut {NoStop}%
\bibitem [{\citenamefont {Wiese}(2013)}]{Wiese}%
  \BibitemOpen
  \bibfield  {author} {\bibinfo {author} {\bibfnamefont {U.-J.}\ \bibnamefont
  {Wiese}},\ }\href@noop {} {\bibfield  {journal} {\bibinfo  {journal}
  {\href{https://onlinelibrary.wiley.com/doi/abs/10.1002/andp.201300104}{Annalen
  der Physik}}\ }\textbf {\bibinfo {volume} {525}},\ \bibinfo {pages} {777}
  (\bibinfo {year} {2013})}\BibitemShut {NoStop}%
\bibitem [{\citenamefont {Dalmonte}\ and\ \citenamefont
  {Montangero}(2016)}]{Dalmonte-Montangero}%
  \BibitemOpen
  \bibfield  {author} {\bibinfo {author} {\bibfnamefont {M.}~\bibnamefont
  {Dalmonte}}\ and\ \bibinfo {author} {\bibfnamefont {S.}~\bibnamefont
  {Montangero}},\ }\href@noop {} {\bibfield  {journal} {\bibinfo  {journal}
  {\href{https://doi.org/10.1080/00107514.2016.1151199}{Contemporary Physics}}\
  }\textbf {\bibinfo {volume} {57}},\ \bibinfo {pages} {388} (\bibinfo {year}
  {2016})}\BibitemShut {NoStop}%
\bibitem [{\citenamefont {Zohar}\ \emph {et~al.}(2016)\citenamefont {Zohar},
  \citenamefont {Cirac},\ and\ \citenamefont {Reznik}}]{Zohar-Cirac-Reznik}%
  \BibitemOpen
  \bibfield  {author} {\bibinfo {author} {\bibfnamefont {E.}~\bibnamefont
  {Zohar}}, \bibinfo {author} {\bibfnamefont {J.~I.}\ \bibnamefont {Cirac}},\
  and\ \bibinfo {author} {\bibfnamefont {B.}~\bibnamefont {Reznik}},\
  }\href@noop {} {\bibfield  {journal} {\bibinfo  {journal}
  {\href{https://iopscience.iop.org/article/10.1088/0034-4885/79/1/014401}{Report
  on Progress in Physics}}\ }\textbf {\bibinfo {volume} {79}} (\bibinfo {year}
  {2016})}\BibitemShut {NoStop}%
\bibitem [{\citenamefont {Preskill}(2018)}]{Preskill2018}%
  \BibitemOpen
  \bibfield  {author} {\bibinfo {author} {\bibfnamefont {J.}~\bibnamefont
  {Preskill}},\ }\href@noop {} {\bibfield  {journal} {\bibinfo  {journal}
  {\href{https://doi.org/10.22331/q-2018-08-06-79}{{Quantum}}}\ }\textbf
  {\bibinfo {volume} {2}},\ \bibinfo {pages} {79} (\bibinfo {year}
  {2018})}\BibitemShut {NoStop}%
\bibitem [{\citenamefont {Bañuls}\ \emph {et~al.}(2020)\citenamefont
  {Bañuls}, \citenamefont {Blatt}, \citenamefont {Catani},\ and\ \citenamefont
  {et~al.}}]{banuls_QTreview2020}%
  \BibitemOpen
  \bibfield  {author} {\bibinfo {author} {\bibfnamefont {M.}~\bibnamefont
  {Bañuls}}, \bibinfo {author} {\bibfnamefont {R.}~\bibnamefont {Blatt}},
  \bibinfo {author} {\bibfnamefont {J.}~\bibnamefont {Catani}},\ and\ \bibinfo
  {author} {\bibnamefont {et~al.}},\ }\href@noop {} {\bibfield  {journal}
  {\bibinfo  {journal} {\href{https://doi.org/10.1140/epjd/e2020-100571-8}{The
  European Physical Journal D}}\ }\textbf {\bibinfo {volume} {74}},\ \bibinfo
  {pages} {165} (\bibinfo {year} {2020})}\BibitemShut {NoStop}%
\bibitem [{\citenamefont {Aidelsburger}\ \emph {et~al.}(2022)\citenamefont
  {Aidelsburger}, \citenamefont {Barbiero}, \citenamefont {Bermudez},\ and\
  \citenamefont {et~al.}}]{aidelsburger_RSP2022}%
  \BibitemOpen
  \bibfield  {author} {\bibinfo {author} {\bibfnamefont {M.}~\bibnamefont
  {Aidelsburger}}, \bibinfo {author} {\bibfnamefont {L.}~\bibnamefont
  {Barbiero}}, \bibinfo {author} {\bibfnamefont {A.}~\bibnamefont {Bermudez}},\
  and\ \bibinfo {author} {\bibnamefont {et~al.}},\ }\href@noop {} {\bibfield
  {journal} {\bibinfo  {journal}
  {\href{https://royalsocietypublishing.org/doi/abs/10.1098/rsta.2021.0064}{Philosophical
  Transactions of the Royal Society A: Mathematical, Physical and Engineering
  Sciences}}\ }\textbf {\bibinfo {volume} {380}},\ \bibinfo {pages} {20210064}
  (\bibinfo {year} {2022})}\BibitemShut {NoStop}%
\bibitem [{\citenamefont {Zohar}(2022)}]{Zohar2022}%
  \BibitemOpen
  \bibfield  {author} {\bibinfo {author} {\bibfnamefont {E.}~\bibnamefont
  {Zohar}},\ }\href@noop {} {\bibfield  {journal} {\bibinfo  {journal}
  {\href{https://royalsocietypublishing.org/doi/abs/10.1098/rsta.2021.0069}{Philosophical
  Transactions of the Royal Society A: Mathematical, Physical and Engineering
  Sciences}}\ }\textbf {\bibinfo {volume} {380}},\ \bibinfo {pages} {20210069}
  (\bibinfo {year} {2022})}\BibitemShut {NoStop}%
\bibitem [{\citenamefont {Klco}\ \emph {et~al.}(2022)\citenamefont {Klco},
  \citenamefont {Roggero},\ and\ \citenamefont {Savage}}]{Klco2022}%
  \BibitemOpen
  \bibfield  {author} {\bibinfo {author} {\bibfnamefont {N.}~\bibnamefont
  {Klco}}, \bibinfo {author} {\bibfnamefont {A.}~\bibnamefont {Roggero}},\ and\
  \bibinfo {author} {\bibfnamefont {M.~J.}\ \bibnamefont {Savage}},\
  }\href@noop {} {\bibfield  {journal} {\bibinfo  {journal}
  {\href{https://dx.doi.org/10.1088/1361-6633/ac58a4}{Reports on Progress in
  Physics}}\ }\textbf {\bibinfo {volume} {85}},\ \bibinfo {pages} {064301}
  (\bibinfo {year} {2022})}\BibitemShut {NoStop}%
\bibitem [{\citenamefont {Bauer}\ \emph {et~al.}(2022)\citenamefont {Bauer},
  \citenamefont {Davoudi}, \citenamefont {Balantekin},\ and\ \citenamefont
  {et~al.}}]{davoudi_QSforHEP2022}%
  \BibitemOpen
  \bibfield  {author} {\bibinfo {author} {\bibfnamefont {C.~W.}\ \bibnamefont
  {Bauer}}, \bibinfo {author} {\bibfnamefont {Z.}~\bibnamefont {Davoudi}},
  \bibinfo {author} {\bibfnamefont {A.~B.}\ \bibnamefont {Balantekin}},\ and\
  \bibinfo {author} {\bibnamefont {et~al.}},\ }\href@noop {} {\bibfield
  {journal} {\bibinfo  {journal} {\href{https://arxiv.org/abs/2204.03381}{arXiv
  preprint arXiv:2204.03381}}\ } (\bibinfo {year} {2022})}\BibitemShut
  {NoStop}%
\bibitem [{\citenamefont {Martinez}\ \emph {et~al.}(2016)\citenamefont
  {Martinez}, \citenamefont {Muschik}, \citenamefont {Schindler},\ and\
  \citenamefont {et~al.}}]{Martinez2016}%
  \BibitemOpen
  \bibfield  {author} {\bibinfo {author} {\bibfnamefont {E.}~\bibnamefont
  {Martinez}}, \bibinfo {author} {\bibfnamefont {C.}~\bibnamefont {Muschik}},
  \bibinfo {author} {\bibfnamefont {P.}~\bibnamefont {Schindler}},\ and\
  \bibinfo {author} {\bibnamefont {et~al.}},\ }\href@noop {} {\bibfield
  {journal} {\bibinfo  {journal}
  {\href{https://doi.org/10.1038/nature18318}{Nature}}\ }\textbf {\bibinfo
  {volume} {534}},\ \bibinfo {pages} {516} (\bibinfo {year}
  {2016})}\BibitemShut {NoStop}%
\bibitem [{\citenamefont {Kokail}\ \emph {et~al.}(2019)\citenamefont {Kokail},
  \citenamefont {Maier}, \citenamefont {van Bijnen},\ and\ \citenamefont
  {et~al.}}]{Kokail2019}%
  \BibitemOpen
  \bibfield  {author} {\bibinfo {author} {\bibfnamefont {C.}~\bibnamefont
  {Kokail}}, \bibinfo {author} {\bibfnamefont {C.}~\bibnamefont {Maier}},
  \bibinfo {author} {\bibfnamefont {R.}~\bibnamefont {van Bijnen}},\ and\
  \bibinfo {author} {\bibnamefont {et~al.}},\ }\href@noop {} {\bibfield
  {journal} {\bibinfo  {journal}
  {\href{https://doi.org/10.1038/s41586-019-1177-4}{Nature}}\ }\textbf
  {\bibinfo {volume} {569}},\ \bibinfo {pages} {355} (\bibinfo {year}
  {2019})}\BibitemShut {NoStop}%
\bibitem [{\citenamefont {Schweizer}\ \emph {et~al.}(2019)\citenamefont
  {Schweizer}, \citenamefont {Grusdt}, \citenamefont {Bernburger},\ and\
  \citenamefont {et~al.}}]{Schweizer2019}%
  \BibitemOpen
  \bibfield  {author} {\bibinfo {author} {\bibfnamefont {C.}~\bibnamefont
  {Schweizer}}, \bibinfo {author} {\bibfnamefont {F.}~\bibnamefont {Grusdt}},
  \bibinfo {author} {\bibfnamefont {M.}~\bibnamefont {Bernburger}},\ and\
  \bibinfo {author} {\bibnamefont {et~al.}},\ }\href@noop {} {\bibfield
  {journal} {\bibinfo  {journal}
  {\href{https://doi.org/10.1038/s41567-019-0649-7}{Nature}}\ }\textbf
  {\bibinfo {volume} {15}},\ \bibinfo {pages} {1168} (\bibinfo {year}
  {2019})}\BibitemShut {NoStop}%
\bibitem [{\citenamefont {Mil}\ \emph {et~al.}(2020)\citenamefont {Mil},
  \citenamefont {Zache}, \citenamefont {Apoorva~Hegde},\ and\ \citenamefont
  {et~al.}}]{Mil2020}%
  \BibitemOpen
  \bibfield  {author} {\bibinfo {author} {\bibfnamefont {A.}~\bibnamefont
  {Mil}}, \bibinfo {author} {\bibfnamefont {T.}~\bibnamefont {Zache}}, \bibinfo
  {author} {\bibfnamefont {A.}~\bibnamefont {Apoorva~Hegde}},\ and\ \bibinfo
  {author} {\bibnamefont {et~al.}},\ }\href@noop {} {\bibfield  {journal}
  {\bibinfo  {journal}
  {\href{https://www.science.org/doi/abs/10.1126/science.aaz5312}{Science}}\
  }\textbf {\bibinfo {volume} {367}},\ \bibinfo {pages} {1128} (\bibinfo {year}
  {2020})}\BibitemShut {NoStop}%
\bibitem [{\citenamefont {Yang}\ \emph {et~al.}(2020)\citenamefont {Yang},
  \citenamefont {Sun}, \citenamefont {Ott},\ and\ \citenamefont
  {et~al.}}]{Yang2019}%
  \BibitemOpen
  \bibfield  {author} {\bibinfo {author} {\bibfnamefont {B.}~\bibnamefont
  {Yang}}, \bibinfo {author} {\bibfnamefont {H.}~\bibnamefont {Sun}}, \bibinfo
  {author} {\bibfnamefont {R.}~\bibnamefont {Ott}},\ and\ \bibinfo {author}
  {\bibnamefont {et~al.}},\ }\href@noop {} {\bibfield  {journal} {\bibinfo
  {journal} {\href{https://doi.org/10.1038/s41586-020-2910-8}{Nature}}\
  }\textbf {\bibinfo {volume} {587}},\ \bibinfo {pages} {392} (\bibinfo {year}
  {2020})}\BibitemShut {NoStop}%
\bibitem [{\citenamefont {Semeghini}\ \emph {et~al.}(2021)\citenamefont
  {Semeghini}, \citenamefont {Levine}, \citenamefont {Keesling},\ and\
  \citenamefont {et~al.}}]{Semeghini2021}%
  \BibitemOpen
  \bibfield  {author} {\bibinfo {author} {\bibfnamefont {G.}~\bibnamefont
  {Semeghini}}, \bibinfo {author} {\bibfnamefont {H.}~\bibnamefont {Levine}},
  \bibinfo {author} {\bibfnamefont {A.}~\bibnamefont {Keesling}},\ and\
  \bibinfo {author} {\bibnamefont {et~al.}},\ }\href@noop {} {\bibfield
  {journal} {\bibinfo  {journal}
  {\href{https://www.science.org/doi/abs/10.1126/science.abi8794}{Science}}\
  }\textbf {\bibinfo {volume} {374}},\ \bibinfo {pages} {1242} (\bibinfo {year}
  {2021})}\BibitemShut {NoStop}%
\bibitem [{\citenamefont {Zhou}\ \emph {et~al.}(2022)\citenamefont {Zhou},
  \citenamefont {Guo-Xian~Su}, \citenamefont {rammeleh},\ and\ \citenamefont
  {et~al.}}]{Zhou2022}%
  \BibitemOpen
  \bibfield  {author} {\bibinfo {author} {\bibfnamefont {Z.-Y.}\ \bibnamefont
  {Zhou}}, \bibinfo {author} {\bibfnamefont {G.-X.}\ \bibnamefont
  {Guo-Xian~Su}}, \bibinfo {author} {\bibfnamefont {J.}~\bibnamefont
  {rammeleh}},\ and\ \bibinfo {author} {\bibnamefont {et~al.}},\ }\href@noop {}
  {\bibfield  {journal} {\bibinfo  {journal}
  {\href{https://www.science.org/doi/abs/10.1126/science.abl6277}{Science}}\
  }\textbf {\bibinfo {volume} {377}},\ \bibinfo {pages} {311} (\bibinfo {year}
  {2022})}\BibitemShut {NoStop}%
\bibitem [{\citenamefont {Riechert}\ \emph {et~al.}(2022)\citenamefont
  {Riechert}, \citenamefont {Halimeh}, \citenamefont {Kasper}, \citenamefont
  {Bretheau}, \citenamefont {Zohar}, \citenamefont {Hauke},\ and\ \citenamefont
  {Jendrzejewski}}]{RiechertPRB2022}%
  \BibitemOpen
  \bibfield  {author} {\bibinfo {author} {\bibfnamefont {H.}~\bibnamefont
  {Riechert}}, \bibinfo {author} {\bibfnamefont {J.~C.}\ \bibnamefont
  {Halimeh}}, \bibinfo {author} {\bibfnamefont {V.}~\bibnamefont {Kasper}},
  \bibinfo {author} {\bibfnamefont {L.}~\bibnamefont {Bretheau}}, \bibinfo
  {author} {\bibfnamefont {E.}~\bibnamefont {Zohar}}, \bibinfo {author}
  {\bibfnamefont {P.}~\bibnamefont {Hauke}},\ and\ \bibinfo {author}
  {\bibfnamefont {F.}~\bibnamefont {Jendrzejewski}},\ }\href@noop {} {\bibfield
   {journal} {\bibinfo  {journal}
  {\href{https://link.aps.org/doi/10.1103/PhysRevB.105.205141}{Phys. Rev. B}}\
  }\textbf {\bibinfo {volume} {105}},\ \bibinfo {pages} {205141} (\bibinfo
  {year} {2022})}\BibitemShut {NoStop}%
\bibitem [{\citenamefont {Kalinowski}\ \emph {et~al.}(2022)\citenamefont
  {Kalinowski}, \citenamefont {Maskara},\ and\ \citenamefont
  {Lukin}}]{Kalonowski2022}%
  \BibitemOpen
  \bibfield  {author} {\bibinfo {author} {\bibfnamefont {M.}~\bibnamefont
  {Kalinowski}}, \bibinfo {author} {\bibfnamefont {N.}~\bibnamefont
  {Maskara}},\ and\ \bibinfo {author} {\bibfnamefont {M.~D.}\ \bibnamefont
  {Lukin}},\ }\href@noop {} {\bibfield  {journal} {\bibinfo  {journal}
  {\href{https://doi.org/10.48550/arxiv.2211.00017}{arXiv preprint
  arXiv:2211.00017}}\ } (\bibinfo {year} {2022})}\BibitemShut {NoStop}%
\bibitem [{\citenamefont {Lewenstein}\ \emph {et~al.}(2012)\citenamefont
  {Lewenstein}, \citenamefont {Sanpera},\ and\ \citenamefont
  {Ahufinger}}]{Lewenstein_book_2012}%
  \BibitemOpen
  \bibfield  {author} {\bibinfo {author} {\bibfnamefont {M.}~\bibnamefont
  {Lewenstein}}, \bibinfo {author} {\bibfnamefont {A.}~\bibnamefont
  {Sanpera}},\ and\ \bibinfo {author} {\bibfnamefont {V.}~\bibnamefont
  {Ahufinger}},\ }\href@noop {} {\emph {\bibinfo {title} {Ultracold Atoms in
  Optical Lattices: Simulating quantum many-body systems}}}\ (\bibinfo
  {publisher} {Oxford University Press},\ \bibinfo {year} {2012})\BibitemShut
  {NoStop}%
\bibitem [{\citenamefont {Peskin}\ and\ \citenamefont
  {Schroeder}(1995)}]{Peskin}%
  \BibitemOpen
  \bibfield  {author} {\bibinfo {author} {\bibfnamefont {M.}~\bibnamefont
  {Peskin}}\ and\ \bibinfo {author} {\bibfnamefont {D.}~\bibnamefont
  {Schroeder}},\ }\href@noop {} {\emph {\bibinfo {title} {An introduction to
  Quantum Field Theory}}}\ (\bibinfo  {publisher} {Addison-Wesley Pub. Co.},\
  \bibinfo {address} {Reading, Mass},\ \bibinfo {year} {1995})\BibitemShut
  {NoStop}%
\bibitem [{\citenamefont {Schwartz}(2013)}]{Scwhartz}%
  \BibitemOpen
  \bibfield  {author} {\bibinfo {author} {\bibfnamefont {M.~D.}\ \bibnamefont
  {Schwartz}},\ }\href@noop {} {\emph {\bibinfo {title} {Quantum Field Theory
  and the Standard Model}}}\ (\bibinfo  {publisher} {Cambridge University
  Press},\ \bibinfo {year} {2013})\BibitemShut {NoStop}%
\bibitem [{\citenamefont {Maggiore}(2005)}]{Maggiore}%
  \BibitemOpen
  \bibfield  {author} {\bibinfo {author} {\bibfnamefont {M.}~\bibnamefont
  {Maggiore}},\ }\href@noop {} {\emph {\bibinfo {title} {A Modern Introduction
  to Quantum Field Theory}}}\ (\bibinfo  {publisher} {OUP Oxford},\ \bibinfo
  {year} {2005})\BibitemShut {NoStop}%
\bibitem [{\citenamefont {Wen}(2004)}]{Wen}%
  \BibitemOpen
  \bibfield  {author} {\bibinfo {author} {\bibfnamefont {X.-G.}\ \bibnamefont
  {Wen}},\ }\href@noop {} {\emph {\bibinfo {title} {Quantum Field Theory of
  Many-body Systems}}}\ (\bibinfo  {publisher} {OUP Oxford},\ \bibinfo {year}
  {2004})\BibitemShut {NoStop}%
\bibitem [{\citenamefont {Troyer}\ and\ \citenamefont
  {Wiese}(2005)}]{troyer_wiese2005}%
  \BibitemOpen
  \bibfield  {author} {\bibinfo {author} {\bibfnamefont {M.}~\bibnamefont
  {Troyer}}\ and\ \bibinfo {author} {\bibfnamefont {U.-J.}\ \bibnamefont
  {Wiese}},\ }\href@noop {} {\bibfield  {journal} {\bibinfo  {journal}
  {\href{https://link.aps.org/doi/10.1103/PhysRevLett.94.170201}{Phys. Rev.
  Lett.}}\ }\textbf {\bibinfo {volume} {94}},\ \bibinfo {pages} {170201}
  (\bibinfo {year} {2005})}\BibitemShut {NoStop}%
\bibitem [{\citenamefont {Fukushima}\ and\ \citenamefont
  {Hatsuda}(2010)}]{fukushima2010}%
  \BibitemOpen
  \bibfield  {author} {\bibinfo {author} {\bibfnamefont {K.}~\bibnamefont
  {Fukushima}}\ and\ \bibinfo {author} {\bibfnamefont {T.}~\bibnamefont
  {Hatsuda}},\ }\href@noop {} {\bibfield  {journal} {\bibinfo  {journal}
  {\href{https://doi.org/10.1088/0034-4885/74/1/014001}{Reports on Progress in
  Physics}}\ }\textbf {\bibinfo {volume} {74}},\ \bibinfo {pages} {014001}
  (\bibinfo {year} {2010})}\BibitemShut {NoStop}%
\bibitem [{\citenamefont {Wilson}(1974)}]{Wilson}%
  \BibitemOpen
  \bibfield  {author} {\bibinfo {author} {\bibfnamefont {K.~G.}\ \bibnamefont
  {Wilson}},\ }\href@noop {} {\bibfield  {journal} {\bibinfo  {journal}
  {\href{http://link.aps.org/doi/10.1103/PhysRevD.10.2445}{Phys. Rev. D}}\
  }\textbf {\bibinfo {volume} {10}},\ \bibinfo {pages} {2445} (\bibinfo {year}
  {1974})}\BibitemShut {NoStop}%
\bibitem [{\citenamefont {K\"uhn}\ \emph {et~al.}(2014)\citenamefont {K\"uhn},
  \citenamefont {Cirac},\ and\ \citenamefont
  {Ba\~nuls}}]{Cirac_Banuls_PhysRevA}%
  \BibitemOpen
  \bibfield  {author} {\bibinfo {author} {\bibfnamefont {S.}~\bibnamefont
  {K\"uhn}}, \bibinfo {author} {\bibfnamefont {J.~I.}\ \bibnamefont {Cirac}},\
  and\ \bibinfo {author} {\bibfnamefont {M.-C.}\ \bibnamefont {Ba\~nuls}},\
  }\href@noop {} {\bibfield  {journal} {\bibinfo  {journal}
  {\href{https://link.aps.org/doi/10.1103/PhysRevA.90.042305}{Phys. Rev. A}}\
  }\textbf {\bibinfo {volume} {90}},\ \bibinfo {pages} {042305} (\bibinfo
  {year} {2014})}\BibitemShut {NoStop}%
\bibitem [{\citenamefont {Ercolessi}\ \emph {et~al.}(2018)\citenamefont
  {Ercolessi}, \citenamefont {Facchi}, \citenamefont {Magnifico}, \citenamefont
  {Pascazio},\ and\ \citenamefont {Pepe}}]{Ercolessi_PhysRevD}%
  \BibitemOpen
  \bibfield  {author} {\bibinfo {author} {\bibfnamefont {E.}~\bibnamefont
  {Ercolessi}}, \bibinfo {author} {\bibfnamefont {P.}~\bibnamefont {Facchi}},
  \bibinfo {author} {\bibfnamefont {G.}~\bibnamefont {Magnifico}}, \bibinfo
  {author} {\bibfnamefont {S.}~\bibnamefont {Pascazio}},\ and\ \bibinfo
  {author} {\bibfnamefont {F.~V.}\ \bibnamefont {Pepe}},\ }\href@noop {}
  {\bibfield  {journal} {\bibinfo  {journal}
  {\href{https://link.aps.org/doi/10.1103/PhysRevD.98.074503}{Phys. Rev. D}}\
  }\textbf {\bibinfo {volume} {98}},\ \bibinfo {pages} {074503} (\bibinfo
  {year} {2018})}\BibitemShut {NoStop}%
\bibitem [{\citenamefont {Horn}(1981)}]{HORN1981}%
  \BibitemOpen
  \bibfield  {author} {\bibinfo {author} {\bibfnamefont {D.}~\bibnamefont
  {Horn}},\ }\href@noop {} {\bibfield  {journal} {\bibinfo  {journal}
  {\href{https://doi.org/10.1016/0370-2693(81)90763-2}{Physics Letters B}}\
  }\textbf {\bibinfo {volume} {100}},\ \bibinfo {pages} {149} (\bibinfo {year}
  {1981})}\BibitemShut {NoStop}%
\bibitem [{\citenamefont {Orland}\ and\ \citenamefont
  {Rohrlich}(1990)}]{ORLAND1990}%
  \BibitemOpen
  \bibfield  {author} {\bibinfo {author} {\bibfnamefont {P.}~\bibnamefont
  {Orland}}\ and\ \bibinfo {author} {\bibfnamefont {D.}~\bibnamefont
  {Rohrlich}},\ }\href@noop {} {\bibfield  {journal} {\bibinfo  {journal}
  {\href{https://doi.org/10.1016/0550-3213(90)90646-U}{Nuclear Physics B}}\
  }\textbf {\bibinfo {volume} {338}},\ \bibinfo {pages} {647} (\bibinfo {year}
  {1990})}\BibitemShut {NoStop}%
\bibitem [{\citenamefont {Chandrasekharan}\ and\ \citenamefont
  {Wiese}(1997)}]{CHANDRASEKHARAN1997}%
  \BibitemOpen
  \bibfield  {author} {\bibinfo {author} {\bibfnamefont {S.}~\bibnamefont
  {Chandrasekharan}}\ and\ \bibinfo {author} {\bibfnamefont {U.~J.}\
  \bibnamefont {Wiese}},\ }\href@noop {} {\bibfield  {journal} {\bibinfo
  {journal} {\href{{https://doi.org/10.1016/S0550-3213(97)80041-7}}{Nuclear
  Physics B}}\ }\textbf {\bibinfo {volume} {492}},\ \bibinfo {pages} {455}
  (\bibinfo {year} {1997})}\BibitemShut {NoStop}%
\bibitem [{\citenamefont {Brower}\ \emph {et~al.}(1999)\citenamefont {Brower},
  \citenamefont {Chandrasekharan},\ and\ \citenamefont
  {Wiese}}]{BrowerPRD1999}%
  \BibitemOpen
  \bibfield  {author} {\bibinfo {author} {\bibfnamefont {R.}~\bibnamefont
  {Brower}}, \bibinfo {author} {\bibfnamefont {S.}~\bibnamefont
  {Chandrasekharan}},\ and\ \bibinfo {author} {\bibfnamefont {U.-J.}\
  \bibnamefont {Wiese}},\ }\href@noop {} {\bibfield  {journal} {\bibinfo
  {journal} {\href{https://link.aps.org/doi/10.1103/PhysRevD.60.094502}{Phys.
  Rev. D}}\ }\textbf {\bibinfo {volume} {60}},\ \bibinfo {pages} {094502}
  (\bibinfo {year} {1999})}\BibitemShut {NoStop}%
\bibitem [{\citenamefont {Zohar}\ \emph {et~al.}(2012)\citenamefont {Zohar},
  \citenamefont {Cirac},\ and\ \citenamefont {Reznik}}]{CiracPRL2012}%
  \BibitemOpen
  \bibfield  {author} {\bibinfo {author} {\bibfnamefont {E.}~\bibnamefont
  {Zohar}}, \bibinfo {author} {\bibfnamefont {J.~I.}\ \bibnamefont {Cirac}},\
  and\ \bibinfo {author} {\bibfnamefont {B.}~\bibnamefont {Reznik}},\
  }\href@noop {} {\bibfield  {journal} {\bibinfo  {journal}
  {\href{https://link.aps.org/doi/10.1103/PhysRevLett.109.125302}{Phys. Rev.
  Lett.}}\ }\textbf {\bibinfo {volume} {109}},\ \bibinfo {pages} {125302}
  (\bibinfo {year} {2012})}\BibitemShut {NoStop}%
\bibitem [{\citenamefont {Zohar}\ and\ \citenamefont
  {Burrello}(2015)}]{ZoharPRD2015}%
  \BibitemOpen
  \bibfield  {author} {\bibinfo {author} {\bibfnamefont {E.}~\bibnamefont
  {Zohar}}\ and\ \bibinfo {author} {\bibfnamefont {M.}~\bibnamefont
  {Burrello}},\ }\href@noop {} {\bibfield  {journal} {\bibinfo  {journal}
  {\href{https://link.aps.org/doi/10.1103/PhysRevD.91.054506}{Phys. Rev. D}}\
  }\textbf {\bibinfo {volume} {91}},\ \bibinfo {pages} {054506} (\bibinfo
  {year} {2015})}\BibitemShut {NoStop}%
\bibitem [{\citenamefont {Kasper}\ \emph {et~al.}(2017)\citenamefont {Kasper},
  \citenamefont {Hebenstreit}, \citenamefont {Jendrzejewski}, \citenamefont
  {Oberthaler},\ and\ \citenamefont {Berges}}]{Kasper_2017}%
  \BibitemOpen
  \bibfield  {author} {\bibinfo {author} {\bibfnamefont {V.}~\bibnamefont
  {Kasper}}, \bibinfo {author} {\bibfnamefont {F.}~\bibnamefont {Hebenstreit}},
  \bibinfo {author} {\bibfnamefont {F.}~\bibnamefont {Jendrzejewski}}, \bibinfo
  {author} {\bibfnamefont {M.~K.}\ \bibnamefont {Oberthaler}},\ and\ \bibinfo
  {author} {\bibfnamefont {J.}~\bibnamefont {Berges}},\ }\href@noop {}
  {\bibfield  {journal} {\bibinfo  {journal}
  {\href{https://doi.org/10.1088/1367-2630/aa54e0}{New Journal of Physics}}\
  }\textbf {\bibinfo {volume} {19}},\ \bibinfo {pages} {023030} (\bibinfo
  {year} {2017})}\BibitemShut {NoStop}%
\bibitem [{\citenamefont {Banerjee}\ \emph {et~al.}(2013)\citenamefont
  {Banerjee}, \citenamefont {Jiang}, \citenamefont {Widmer},\ and\
  \citenamefont {Wiese}}]{Banerjee_2013}%
  \BibitemOpen
  \bibfield  {author} {\bibinfo {author} {\bibfnamefont {D.}~\bibnamefont
  {Banerjee}}, \bibinfo {author} {\bibfnamefont {F.-J.}\ \bibnamefont {Jiang}},
  \bibinfo {author} {\bibfnamefont {P.}~\bibnamefont {Widmer}},\ and\ \bibinfo
  {author} {\bibfnamefont {U.-J.}\ \bibnamefont {Wiese}},\ }\href@noop {}
  {\bibfield  {journal} {\bibinfo  {journal}
  {\href{https://iopscience.iop.org/article/10.1088/1742-5468/2013/12/P12010}{Journal
  of Statistical Mechanics: Theory and Experiment}}\ }\textbf {\bibinfo
  {volume} {2013}},\ \bibinfo {pages} {P12010} (\bibinfo {year}
  {2013})}\BibitemShut {NoStop}%
\bibitem [{\citenamefont {Widmer}\ \emph {et~al.}(2014)\citenamefont {Widmer},
  \citenamefont {Banerjee}, \citenamefont {Jiang},\ and\ \citenamefont
  {Wiese}}]{Widmer_POS2014}%
  \BibitemOpen
  \bibfield  {author} {\bibinfo {author} {\bibfnamefont {P.}~\bibnamefont
  {Widmer}}, \bibinfo {author} {\bibfnamefont {D.}~\bibnamefont {Banerjee}},
  \bibinfo {author} {\bibfnamefont {F.-J.}\ \bibnamefont {Jiang}},\ and\
  \bibinfo {author} {\bibfnamefont {U.-J.}\ \bibnamefont {Wiese}},\ }\href@noop
  {} {\bibfield  {journal} {\bibinfo  {journal}
  {\href{https://doi.org/10.22323/1.187.0333}{PoS}}\ }\textbf {\bibinfo
  {volume} {LATTICE 2013}},\ \bibinfo {pages} {333} (\bibinfo {year}
  {2014})}\BibitemShut {NoStop}%
\bibitem [{\citenamefont {Banerjee}\ \emph
  {et~al.}(2021{\natexlab{a}})\citenamefont {Banerjee}, \citenamefont
  {Huffman},\ and\ \citenamefont {Rammelmüller}}]{banerjee2021}%
  \BibitemOpen
  \bibfield  {author} {\bibinfo {author} {\bibfnamefont {D.}~\bibnamefont
  {Banerjee}}, \bibinfo {author} {\bibfnamefont {E.}~\bibnamefont {Huffman}},\
  and\ \bibinfo {author} {\bibfnamefont {L.}~\bibnamefont {Rammelmüller}},\
  }\href@noop {} {\bibfield  {journal} {\bibinfo  {journal}
  {\href{https://arxiv.org/abs/2111.00300}{\textnormal{arXiv preprint
  arXiv:2111.00300 [hep-lat]}}}\ } (\bibinfo {year}
  {2021}{\natexlab{a}})}\BibitemShut {NoStop}%
\bibitem [{\citenamefont {Banerjee}\ \emph
  {et~al.}(2021{\natexlab{b}})\citenamefont {Banerjee}, \citenamefont {Caspar},
  \citenamefont {Jiang}, \citenamefont {Peng},\ and\ \citenamefont
  {Wiese}}]{banerjee2021nematic}%
  \BibitemOpen
  \bibfield  {author} {\bibinfo {author} {\bibfnamefont {D.}~\bibnamefont
  {Banerjee}}, \bibinfo {author} {\bibfnamefont {S.}~\bibnamefont {Caspar}},
  \bibinfo {author} {\bibfnamefont {F.-J.}\ \bibnamefont {Jiang}}, \bibinfo
  {author} {\bibfnamefont {J.-H.}\ \bibnamefont {Peng}},\ and\ \bibinfo
  {author} {\bibfnamefont {U.-J.}\ \bibnamefont {Wiese}},\ }\href@noop {}
  {\bibfield  {journal} {\bibinfo  {journal}
  {\href{https://arxiv.org/abs/2107.01283}{arXiv preprint arXiv:2107.01283}}\ }
  (\bibinfo {year} {2021}{\natexlab{b}})}\BibitemShut {NoStop}%
\bibitem [{\citenamefont {Banerjee}\ and\ \citenamefont
  {Sen}(2021)}]{BanerjeePRL2021}%
  \BibitemOpen
  \bibfield  {author} {\bibinfo {author} {\bibfnamefont {D.}~\bibnamefont
  {Banerjee}}\ and\ \bibinfo {author} {\bibfnamefont {A.}~\bibnamefont {Sen}},\
  }\href@noop {} {\bibfield  {journal} {\bibinfo  {journal}
  {\href{{https://link.aps.org/doi/10.1103/PhysRevLett.126.220601}}{Phys. Rev.
  Lett.}}\ }\textbf {\bibinfo {volume} {126}},\ \bibinfo {pages} {220601}
  (\bibinfo {year} {2021})}\BibitemShut {NoStop}%
\bibitem [{\citenamefont {Banerjee}\ \emph {et~al.}(2022)\citenamefont
  {Banerjee}, \citenamefont {Huffman},\ and\ \citenamefont
  {Rammelm{\"u}ller}}]{banerjee2022}%
  \BibitemOpen
  \bibfield  {author} {\bibinfo {author} {\bibfnamefont {D.}~\bibnamefont
  {Banerjee}}, \bibinfo {author} {\bibfnamefont {E.}~\bibnamefont {Huffman}},\
  and\ \bibinfo {author} {\bibfnamefont {L.}~\bibnamefont {Rammelm{\"u}ller}},\
  }\href@noop {} {\bibfield  {journal} {\bibinfo  {journal}
  {\href{https://arxiv.org/abs/2201.07171}{arXiv preprint arXiv:2201.07171}}\ }
  (\bibinfo {year} {2022})}\BibitemShut {NoStop}%
\bibitem [{\citenamefont {Cardarelli}\ \emph {et~al.}(2017)\citenamefont
  {Cardarelli}, \citenamefont {Greschner},\ and\ \citenamefont
  {Santos}}]{CardarelliPRL2017}%
  \BibitemOpen
  \bibfield  {author} {\bibinfo {author} {\bibfnamefont {L.}~\bibnamefont
  {Cardarelli}}, \bibinfo {author} {\bibfnamefont {S.}~\bibnamefont
  {Greschner}},\ and\ \bibinfo {author} {\bibfnamefont {L.}~\bibnamefont
  {Santos}},\ }\href@noop {} {\bibfield  {journal} {\bibinfo  {journal}
  {\href{https://link.aps.org/doi/10.1103/PhysRevLett.119.180402}{Phys. Rev.
  Lett.}}\ }\textbf {\bibinfo {volume} {119}},\ \bibinfo {pages} {180402}
  (\bibinfo {year} {2017})}\BibitemShut {NoStop}%
\bibitem [{\citenamefont {Ott}\ \emph {et~al.}(2020)\citenamefont {Ott},
  \citenamefont {Zache}, \citenamefont {Mueller},\ and\ \citenamefont
  {Berges}}]{OttPLB2020}%
  \BibitemOpen
  \bibfield  {author} {\bibinfo {author} {\bibfnamefont {R.}~\bibnamefont
  {Ott}}, \bibinfo {author} {\bibfnamefont {T.}~\bibnamefont {Zache}}, \bibinfo
  {author} {\bibfnamefont {N.}~\bibnamefont {Mueller}},\ and\ \bibinfo {author}
  {\bibfnamefont {J.}~\bibnamefont {Berges}},\ }\href@noop {} {\bibfield
  {journal} {\bibinfo  {journal}
  {\href{https://www.sciencedirect.com/science/article/pii/S037026932030263X}{Physics
  Letters B}}\ }\textbf {\bibinfo {volume} {805}},\ \bibinfo {pages} {135459}
  (\bibinfo {year} {2020})}\BibitemShut {NoStop}%
\bibitem [{\citenamefont {Gonz\'alez-Cuadra}\ \emph {et~al.}(2020)\citenamefont
  {Gonz\'alez-Cuadra}, \citenamefont {Tagliacozzo}, \citenamefont
  {Lewenstein},\ and\ \citenamefont {Bermudez}}]{GonzalezCuadraPRX2020}%
  \BibitemOpen
  \bibfield  {author} {\bibinfo {author} {\bibfnamefont {D.}~\bibnamefont
  {Gonz\'alez-Cuadra}}, \bibinfo {author} {\bibfnamefont {L.}~\bibnamefont
  {Tagliacozzo}}, \bibinfo {author} {\bibfnamefont {M.}~\bibnamefont
  {Lewenstein}},\ and\ \bibinfo {author} {\bibfnamefont {A.}~\bibnamefont
  {Bermudez}},\ }\href@noop {} {\bibfield  {journal} {\bibinfo  {journal}
  {\href{https://link.aps.org/doi/10.1103/PhysRevX.10.041007}{Phys. Rev. X}}\
  }\textbf {\bibinfo {volume} {10}},\ \bibinfo {pages} {041007} (\bibinfo
  {year} {2020})}\BibitemShut {NoStop}%
\bibitem [{\citenamefont {Hashizume}\ \emph {et~al.}(2022)\citenamefont
  {Hashizume}, \citenamefont {Halimeh}, \citenamefont {Hauke},\ and\
  \citenamefont {Banerjee}}]{HashimuzeSciPost2022}%
  \BibitemOpen
  \bibfield  {author} {\bibinfo {author} {\bibfnamefont {T.}~\bibnamefont
  {Hashizume}}, \bibinfo {author} {\bibfnamefont {J.~C.}\ \bibnamefont
  {Halimeh}}, \bibinfo {author} {\bibfnamefont {P.}~\bibnamefont {Hauke}},\
  and\ \bibinfo {author} {\bibfnamefont {D.}~\bibnamefont {Banerjee}},\
  }\href@noop {} {\bibfield  {journal} {\bibinfo  {journal}
  {\href{https://scipost.org/10.21468/SciPostPhys.13.2.017}{SciPost Phys.}}\
  }\textbf {\bibinfo {volume} {13}},\ \bibinfo {pages} {017} (\bibinfo {year}
  {2022})}\BibitemShut {NoStop}%
\bibitem [{\citenamefont {Osborne}\ \emph {et~al.}(2022)\citenamefont
  {Osborne}, \citenamefont {McCulloch}, \citenamefont {Yang}, \citenamefont
  {Hauke},\ and\ \citenamefont {Halimeh}}]{Halimeh_arXiv_2022}%
  \BibitemOpen
  \bibfield  {author} {\bibinfo {author} {\bibfnamefont {J.}~\bibnamefont
  {Osborne}}, \bibinfo {author} {\bibfnamefont {I.~P.}\ \bibnamefont
  {McCulloch}}, \bibinfo {author} {\bibfnamefont {B.}~\bibnamefont {Yang}},
  \bibinfo {author} {\bibfnamefont {P.}~\bibnamefont {Hauke}},\ and\ \bibinfo
  {author} {\bibfnamefont {J.~C.}\ \bibnamefont {Halimeh}},\ }\href@noop {}
  {\bibfield  {journal} {\bibinfo  {journal}
  {\href{https://arxiv.org/abs/2211.01380}{\textnormal{arXiv preprint
  arxiv:2211.01380v2 [cond-mat.quant-gas]}}}\ } (\bibinfo {year}
  {2022})}\BibitemShut {NoStop}%
\bibitem [{\citenamefont {Zohar}\ \emph {et~al.}(2013)\citenamefont {Zohar},
  \citenamefont {Cirac},\ and\ \citenamefont {Reznik}}]{ZoharReznikPRA2013}%
  \BibitemOpen
  \bibfield  {author} {\bibinfo {author} {\bibfnamefont {E.}~\bibnamefont
  {Zohar}}, \bibinfo {author} {\bibfnamefont {J.~I.}\ \bibnamefont {Cirac}},\
  and\ \bibinfo {author} {\bibfnamefont {B.}~\bibnamefont {Reznik}},\
  }\href@noop {} {\bibfield  {journal} {\bibinfo  {journal}
  {\href{https://link.aps.org/doi/10.1103/PhysRevA.88.023617}{Phys. Rev. A}}\
  }\textbf {\bibinfo {volume} {88}},\ \bibinfo {pages} {023617} (\bibinfo
  {year} {2013})}\BibitemShut {NoStop}%
\bibitem [{\citenamefont {Gra{\ss}}\ \emph {et~al.}(2016)\citenamefont
  {Gra{\ss}}, \citenamefont {Lewenstein},\ and\ \citenamefont
  {Bermudez}}]{Gras_Lewenstein2016}%
  \BibitemOpen
  \bibfield  {author} {\bibinfo {author} {\bibfnamefont {T.}~\bibnamefont
  {Gra{\ss}}}, \bibinfo {author} {\bibfnamefont {M.}~\bibnamefont
  {Lewenstein}},\ and\ \bibinfo {author} {\bibfnamefont {A.}~\bibnamefont
  {Bermudez}},\ }\href@noop {} {\bibfield  {journal} {\bibinfo  {journal}
  {\href{https://doi.org/10.1088/1367-2630/18/3/033011}{New Journal of
  Physics}}\ }\textbf {\bibinfo {volume} {18}},\ \bibinfo {pages} {033011}
  (\bibinfo {year} {2016})}\BibitemShut {NoStop}%
\bibitem [{\citenamefont {Celi}\ \emph {et~al.}(2020)\citenamefont {Celi},
  \citenamefont {Vermersch}, \citenamefont {Viyuela}, \citenamefont {Pichler},
  \citenamefont {Lukin},\ and\ \citenamefont {Zoller}}]{Celi2019}%
  \BibitemOpen
  \bibfield  {author} {\bibinfo {author} {\bibfnamefont {A.}~\bibnamefont
  {Celi}}, \bibinfo {author} {\bibfnamefont {B.}~\bibnamefont {Vermersch}},
  \bibinfo {author} {\bibfnamefont {O.}~\bibnamefont {Viyuela}}, \bibinfo
  {author} {\bibfnamefont {H.}~\bibnamefont {Pichler}}, \bibinfo {author}
  {\bibfnamefont {M.~D.}\ \bibnamefont {Lukin}},\ and\ \bibinfo {author}
  {\bibfnamefont {P.}~\bibnamefont {Zoller}},\ }\href@noop {} {\bibfield
  {journal} {\bibinfo  {journal}
  {\href{https://journals.aps.org/prx/abstract/10.1103/PhysRevX.10.021057}{Phys.
  Rev. X}}\ }\textbf {\bibinfo {volume} {10}},\ \bibinfo {pages} {021057}
  (\bibinfo {year} {2020})}\BibitemShut {NoStop}%
\bibitem [{\citenamefont {Kogut}\ and\ \citenamefont
  {Susskind}(1975)}]{Kogut-Susskind}%
  \BibitemOpen
  \bibfield  {author} {\bibinfo {author} {\bibfnamefont {J.}~\bibnamefont
  {Kogut}}\ and\ \bibinfo {author} {\bibfnamefont {L.}~\bibnamefont
  {Susskind}},\ }\href@noop {} {\bibfield  {journal} {\bibinfo  {journal}
  {\href{https://link.aps.org/doi/10.1103/PhysRevD.11.395}{Phys. Rev. D}}\
  }\textbf {\bibinfo {volume} {11}},\ \bibinfo {pages} {395} (\bibinfo {year}
  {1975})}\BibitemShut {NoStop}%
\bibitem [{\citenamefont {Kogut}(1979)}]{KogutRMP1979}%
  \BibitemOpen
  \bibfield  {author} {\bibinfo {author} {\bibfnamefont {J.~B.}\ \bibnamefont
  {Kogut}},\ }\href@noop {} {\bibfield  {journal} {\bibinfo  {journal}
  {\href{https://link.aps.org/doi/10.1103/RevModPhys.51.659}{Rev. Mod. Phys.}}\
  }\textbf {\bibinfo {volume} {51}},\ \bibinfo {pages} {659} (\bibinfo {year}
  {1979})}\BibitemShut {NoStop}%
\bibitem [{\citenamefont {Rokhsar}\ and\ \citenamefont
  {Kivelson}(1988)}]{RK1988}%
  \BibitemOpen
  \bibfield  {author} {\bibinfo {author} {\bibfnamefont {D.~S.}\ \bibnamefont
  {Rokhsar}}\ and\ \bibinfo {author} {\bibfnamefont {S.~A.}\ \bibnamefont
  {Kivelson}},\ }\href@noop {} {\bibfield  {journal} {\bibinfo  {journal}
  {\href{https://link.aps.org/doi/10.1103/PhysRevLett.61.2376}{Phys. Rev.
  Lett.}}\ }\textbf {\bibinfo {volume} {61}},\ \bibinfo {pages} {2376}
  (\bibinfo {year} {1988})}\BibitemShut {NoStop}%
\bibitem [{\citenamefont {Moessner}\ \emph {et~al.}(2001)\citenamefont
  {Moessner}, \citenamefont {Sondhi},\ and\ \citenamefont
  {Fradkin}}]{Moessner2002}%
  \BibitemOpen
  \bibfield  {author} {\bibinfo {author} {\bibfnamefont {R.}~\bibnamefont
  {Moessner}}, \bibinfo {author} {\bibfnamefont {S.~L.}\ \bibnamefont
  {Sondhi}},\ and\ \bibinfo {author} {\bibfnamefont {E.}~\bibnamefont
  {Fradkin}},\ }\href@noop {} {\bibfield  {journal} {\bibinfo  {journal}
  {\href{https://link.aps.org/doi/10.1103/PhysRevB.65.024504}{Phys. Rev. B}}\
  }\textbf {\bibinfo {volume} {65}},\ \bibinfo {pages} {024504} (\bibinfo
  {year} {2001})}\BibitemShut {NoStop}%
\bibitem [{\citenamefont {Kawaguchi}\ and\ \citenamefont
  {Ueda}(2012)}]{KAWAGUCHI2012}%
  \BibitemOpen
  \bibfield  {author} {\bibinfo {author} {\bibfnamefont {Y.}~\bibnamefont
  {Kawaguchi}}\ and\ \bibinfo {author} {\bibfnamefont {M.}~\bibnamefont
  {Ueda}},\ }\href@noop {} {\bibfield  {journal} {\bibinfo  {journal}
  {\href{https://doi.org/10.1016/j.physrep.2012.07.005}{Physics Reports}}\
  }\textbf {\bibinfo {volume} {520}},\ \bibinfo {pages} {253} (\bibinfo {year}
  {2012})}\BibitemShut {NoStop}%
\bibitem [{\citenamefont {Stamper-Kurn}\ and\ \citenamefont
  {Ueda}(2013)}]{stapmerkurnRMP2013}%
  \BibitemOpen
  \bibfield  {author} {\bibinfo {author} {\bibfnamefont {D.~M.}\ \bibnamefont
  {Stamper-Kurn}}\ and\ \bibinfo {author} {\bibfnamefont {M.}~\bibnamefont
  {Ueda}},\ }\href@noop {} {\bibfield  {journal} {\bibinfo  {journal}
  {\href{https://link.aps.org/doi/10.1103/RevModPhys.85.1191}{Rev. Mod.
  Phys.}}\ }\textbf {\bibinfo {volume} {85}},\ \bibinfo {pages} {1191}
  (\bibinfo {year} {2013})}\BibitemShut {NoStop}%
\bibitem [{\citenamefont {Jaksch}\ \emph {et~al.}(1998)\citenamefont {Jaksch},
  \citenamefont {Bruder}, \citenamefont {Cirac}, \citenamefont {Gardiner},\
  and\ \citenamefont {Zoller}}]{jakschPRL1998}%
  \BibitemOpen
  \bibfield  {author} {\bibinfo {author} {\bibfnamefont {D.}~\bibnamefont
  {Jaksch}}, \bibinfo {author} {\bibfnamefont {C.}~\bibnamefont {Bruder}},
  \bibinfo {author} {\bibfnamefont {J.~I.}\ \bibnamefont {Cirac}}, \bibinfo
  {author} {\bibfnamefont {C.~W.}\ \bibnamefont {Gardiner}},\ and\ \bibinfo
  {author} {\bibfnamefont {P.}~\bibnamefont {Zoller}},\ }\href@noop {}
  {\bibfield  {journal} {\bibinfo  {journal}
  {\href{https://link.aps.org/doi/10.1103/PhysRevLett.81.3108}{Phys. Rev.
  Lett.}}\ }\textbf {\bibinfo {volume} {81}},\ \bibinfo {pages} {3108}
  (\bibinfo {year} {1998})}\BibitemShut {NoStop}%
\bibitem [{\citenamefont {Eckert}\ \emph {et~al.}(2007)\citenamefont {Eckert},
  \citenamefont {Zawitkowski}, \citenamefont {Leskinen}, \citenamefont
  {Sanpera},\ and\ \citenamefont {Lewenstein}}]{Eckert2007}%
  \BibitemOpen
  \bibfield  {author} {\bibinfo {author} {\bibfnamefont {K.}~\bibnamefont
  {Eckert}}, \bibinfo {author} {\bibfnamefont {L.}~\bibnamefont {Zawitkowski}},
  \bibinfo {author} {\bibfnamefont {M.~J.}\ \bibnamefont {Leskinen}}, \bibinfo
  {author} {\bibfnamefont {A.}~\bibnamefont {Sanpera}},\ and\ \bibinfo {author}
  {\bibfnamefont {M.}~\bibnamefont {Lewenstein}},\ }\href@noop {} {\bibfield
  {journal} {\bibinfo  {journal}
  {\href{https://doi.org/10.1088/1367-2630/9/5/133}{New Journal of Physics}}\
  }\textbf {\bibinfo {volume} {9}},\ \bibinfo {pages} {133} (\bibinfo {year}
  {2007})}\BibitemShut {NoStop}%
\bibitem [{\citenamefont {van Otterlo}\ \emph {et~al.}(1995)\citenamefont {van
  Otterlo}, \citenamefont {Wagenblast}, \citenamefont {Baltin}, \citenamefont
  {Bruder}, \citenamefont {Fazio},\ and\ \citenamefont
  {Sch{\"o}n}}]{vanotterlo1995}%
  \BibitemOpen
  \bibfield  {author} {\bibinfo {author} {\bibfnamefont {A.}~\bibnamefont {van
  Otterlo}}, \bibinfo {author} {\bibfnamefont {K.-H.}\ \bibnamefont
  {Wagenblast}}, \bibinfo {author} {\bibfnamefont {R.}~\bibnamefont {Baltin}},
  \bibinfo {author} {\bibfnamefont {C.}~\bibnamefont {Bruder}}, \bibinfo
  {author} {\bibfnamefont {R.}~\bibnamefont {Fazio}},\ and\ \bibinfo {author}
  {\bibfnamefont {G.}~\bibnamefont {Sch{\"o}n}},\ }\href@noop {} {\bibfield
  {journal} {\bibinfo  {journal}
  {\href{https://journals.aps.org/prb/pdf/10.1103/PhysRevB.52.16176}{Phys. Rev.
  B}}\ }\textbf {\bibinfo {volume} {52}},\ \bibinfo {pages} {16176} (\bibinfo
  {year} {1995})}\BibitemShut {NoStop}%
\bibitem [{\citenamefont {Capogrosso-Sansone}\ \emph
  {et~al.}(2010)\citenamefont {Capogrosso-Sansone}, \citenamefont {Trefzger},
  \citenamefont {Lewenstein}, \citenamefont {Zoller},\ and\ \citenamefont
  {Pupillo}}]{capogrossosansonePRL2010}%
  \BibitemOpen
  \bibfield  {author} {\bibinfo {author} {\bibfnamefont {B.}~\bibnamefont
  {Capogrosso-Sansone}}, \bibinfo {author} {\bibfnamefont {C.}~\bibnamefont
  {Trefzger}}, \bibinfo {author} {\bibfnamefont {M.}~\bibnamefont
  {Lewenstein}}, \bibinfo {author} {\bibfnamefont {P.}~\bibnamefont {Zoller}},\
  and\ \bibinfo {author} {\bibfnamefont {G.}~\bibnamefont {Pupillo}},\
  }\href@noop {} {\bibfield  {journal} {\bibinfo  {journal}
  {\href{https://link.aps.org/doi/10.1103/PhysRevLett.104.125301}{Phys. Rev.
  Lett.}}\ }\textbf {\bibinfo {volume} {104}},\ \bibinfo {pages} {125301}
  (\bibinfo {year} {2010})}\BibitemShut {NoStop}%
\bibitem [{\citenamefont {Zhang}\ \emph {et~al.}(2015)\citenamefont {Zhang},
  \citenamefont {Safavi-Naini}, \citenamefont {Rey},\ and\ \citenamefont
  {Capogrosso-Sansone}}]{Zhang2015}%
  \BibitemOpen
  \bibfield  {author} {\bibinfo {author} {\bibfnamefont {C.}~\bibnamefont
  {Zhang}}, \bibinfo {author} {\bibfnamefont {A.}~\bibnamefont {Safavi-Naini}},
  \bibinfo {author} {\bibfnamefont {A.~M.}\ \bibnamefont {Rey}},\ and\ \bibinfo
  {author} {\bibfnamefont {B.}~\bibnamefont {Capogrosso-Sansone}},\ }\href@noop
  {} {\bibfield  {journal} {\bibinfo  {journal}
  {\href{https://doi.org/10.1088/1367-2630/17/12/123014}{New Journal of
  Physics}}\ }\textbf {\bibinfo {volume} {17}},\ \bibinfo {pages} {123014}
  (\bibinfo {year} {2015})}\BibitemShut {NoStop}%
\bibitem [{\citenamefont {Giovanazzi}\ \emph {et~al.}(2002)\citenamefont
  {Giovanazzi}, \citenamefont {G\"orlitz},\ and\ \citenamefont
  {Pfau}}]{giovinazzi2002}%
  \BibitemOpen
  \bibfield  {author} {\bibinfo {author} {\bibfnamefont {S.}~\bibnamefont
  {Giovanazzi}}, \bibinfo {author} {\bibfnamefont {A.}~\bibnamefont
  {G\"orlitz}},\ and\ \bibinfo {author} {\bibfnamefont {T.}~\bibnamefont
  {Pfau}},\ }\href@noop {} {\bibfield  {journal} {\bibinfo  {journal}
  {\href{https://link.aps.org/doi/10.1103/PhysRevLett.89.130401}{Phys. Rev.
  Lett.}}\ }\textbf {\bibinfo {volume} {89}},\ \bibinfo {pages} {130401}
  (\bibinfo {year} {2002})}\BibitemShut {NoStop}%
\bibitem [{\citenamefont {Baranov}\ \emph {et~al.}(2002)\citenamefont
  {Baranov}, \citenamefont {Dobrek}, \citenamefont {Goral}, \citenamefont
  {Santos},\ and\ \citenamefont {Lewenstein}}]{Baranov_2002}%
  \BibitemOpen
  \bibfield  {author} {\bibinfo {author} {\bibfnamefont {M.}~\bibnamefont
  {Baranov}}, \bibinfo {author} {\bibfnamefont {L.}~\bibnamefont {Dobrek}},
  \bibinfo {author} {\bibfnamefont {K.}~\bibnamefont {Goral}}, \bibinfo
  {author} {\bibfnamefont {L.}~\bibnamefont {Santos}},\ and\ \bibinfo {author}
  {\bibfnamefont {M.}~\bibnamefont {Lewenstein}},\ }\href@noop {} {\bibfield
  {journal} {\bibinfo  {journal}
  {\href{https://doi.org/10.1238/physica.topical.102a00074}{Physica Scripta}}\
  }\textbf {\bibinfo {volume} {T102}},\ \bibinfo {pages} {74} (\bibinfo {year}
  {2002})}\BibitemShut {NoStop}%
\bibitem [{\citenamefont {Lahaye}\ \emph {et~al.}(2009)\citenamefont {Lahaye},
  \citenamefont {Menotti}, \citenamefont {Santos}, \citenamefont {Lewenstein},\
  and\ \citenamefont {Pfau}}]{Lahaye_2009}%
  \BibitemOpen
  \bibfield  {author} {\bibinfo {author} {\bibfnamefont {T.}~\bibnamefont
  {Lahaye}}, \bibinfo {author} {\bibfnamefont {C.}~\bibnamefont {Menotti}},
  \bibinfo {author} {\bibfnamefont {L.}~\bibnamefont {Santos}}, \bibinfo
  {author} {\bibfnamefont {M.}~\bibnamefont {Lewenstein}},\ and\ \bibinfo
  {author} {\bibfnamefont {T.}~\bibnamefont {Pfau}},\ }\href@noop {} {\bibfield
   {journal} {\bibinfo  {journal}
  {\href{https://doi.org/10.1088/0034-4885/72/12/126401}{Reports on Progress in
  Physics}}\ }\textbf {\bibinfo {volume} {72}},\ \bibinfo {pages} {126401}
  (\bibinfo {year} {2009})}\BibitemShut {NoStop}%
\bibitem [{\citenamefont {Yang}\ \emph {et~al.}(2017)\citenamefont {Yang},
  \citenamefont {Dai}, \citenamefont {Sun}, \citenamefont {Reingruber},
  \citenamefont {Yuan},\ and\ \citenamefont
  {Pan}}]{optsuperlattice_yang_PRA2017}%
  \BibitemOpen
  \bibfield  {author} {\bibinfo {author} {\bibfnamefont {B.}~\bibnamefont
  {Yang}}, \bibinfo {author} {\bibfnamefont {H.-N.}\ \bibnamefont {Dai}},
  \bibinfo {author} {\bibfnamefont {H.}~\bibnamefont {Sun}}, \bibinfo {author}
  {\bibfnamefont {A.}~\bibnamefont {Reingruber}}, \bibinfo {author}
  {\bibfnamefont {Z.-S.}\ \bibnamefont {Yuan}},\ and\ \bibinfo {author}
  {\bibfnamefont {J.-W.}\ \bibnamefont {Pan}},\ }\href@noop {} {\bibfield
  {journal} {\bibinfo  {journal}
  {\href{https://link.aps.org/doi/10.1103/PhysRevA.96.011602}{Phys. Rev. A}}\
  }\textbf {\bibinfo {volume} {96}},\ \bibinfo {pages} {011602(R)} (\bibinfo
  {year} {2017})}\BibitemShut {NoStop}%
\bibitem [{\citenamefont {Zohar}\ and\ \citenamefont
  {Reznik}(2011)}]{ZoharPRL2011}%
  \BibitemOpen
  \bibfield  {author} {\bibinfo {author} {\bibfnamefont {E.}~\bibnamefont
  {Zohar}}\ and\ \bibinfo {author} {\bibfnamefont {B.}~\bibnamefont {Reznik}},\
  }\href@noop {} {\bibfield  {journal} {\bibinfo  {journal}
  {\href{https://link.aps.org/doi/10.1103/PhysRevLett.107.275301}{Phys. Rev.
  Lett.}}\ }\textbf {\bibinfo {volume} {107}},\ \bibinfo {pages} {275301}
  (\bibinfo {year} {2011})}\BibitemShut {NoStop}%
\bibitem [{\citenamefont {Arimondo}\ \emph {et~al.}(2012)\citenamefont
  {Arimondo}, \citenamefont {Ciampini}, \citenamefont {Eckardt}, \citenamefont
  {Holthaus},\ and\ \citenamefont {Morsch}}]{ARIMONDO2012}%
  \BibitemOpen
  \bibfield  {author} {\bibinfo {author} {\bibfnamefont {E.}~\bibnamefont
  {Arimondo}}, \bibinfo {author} {\bibfnamefont {D.}~\bibnamefont {Ciampini}},
  \bibinfo {author} {\bibfnamefont {A.}~\bibnamefont {Eckardt}}, \bibinfo
  {author} {\bibfnamefont {M.}~\bibnamefont {Holthaus}},\ and\ \bibinfo
  {author} {\bibfnamefont {O.}~\bibnamefont {Morsch}},\ }in\ \href@noop {}
  {\emph {\bibinfo {booktitle} {Advances in Atomic, Molecular, and Optical
  Physics}}},\ \bibinfo {series}
  {\href{https://www.sciencedirect.com/science/article/pii/B9780123964823000107}{Advances
  In Atomic, Molecular, and Optical Physics}}, Vol.~\bibinfo {volume} {61},\
  \bibinfo {editor} {edited by\ \bibinfo {editor} {\bibfnamefont
  {P.}~\bibnamefont {Berman}}, \bibinfo {editor} {\bibfnamefont
  {E.}~\bibnamefont {Arimondo}},\ and\ \bibinfo {editor} {\bibfnamefont
  {C.}~\bibnamefont {Lin}}}\ (\bibinfo  {publisher} {Academic Press},\ \bibinfo
  {year} {2012})\ pp.\ \bibinfo {pages} {515--547}\BibitemShut {NoStop}%
\bibitem [{\citenamefont {Moessner}\ and\ \citenamefont
  {Sondhi}(2001)}]{moessnerPRL2001}%
  \BibitemOpen
  \bibfield  {author} {\bibinfo {author} {\bibfnamefont {R.}~\bibnamefont
  {Moessner}}\ and\ \bibinfo {author} {\bibfnamefont {S.~L.}\ \bibnamefont
  {Sondhi}},\ }\href@noop {} {\bibfield  {journal} {\bibinfo  {journal}
  {\href{https://link.aps.org/doi/10.1103/PhysRevLett.86.1881}{Phys. Rev.
  Lett.}}\ }\textbf {\bibinfo {volume} {86}},\ \bibinfo {pages} {1881}
  (\bibinfo {year} {2001})}\BibitemShut {NoStop}%
\bibitem [{\citenamefont {Schrieffer}\ and\ \citenamefont
  {Wolff}(1966)}]{SW_1966}%
  \BibitemOpen
  \bibfield  {author} {\bibinfo {author} {\bibfnamefont {J.~R.}\ \bibnamefont
  {Schrieffer}}\ and\ \bibinfo {author} {\bibfnamefont {P.~A.}\ \bibnamefont
  {Wolff}},\ }\href@noop {} {\bibfield  {journal} {\bibinfo  {journal}
  {\href{https://link.aps.org/doi/10.1103/PhysRev.149.491}{Phys. Rev.}}\
  }\textbf {\bibinfo {volume} {149}},\ \bibinfo {pages} {491} (\bibinfo {year}
  {1966})}\BibitemShut {NoStop}%
\bibitem [{\citenamefont {Giuliano}\ \emph {et~al.}(2013)\citenamefont
  {Giuliano}, \citenamefont {Rossini}, \citenamefont {Sodano},\ and\
  \citenamefont {Trombettoni}}]{Giuliano}%
  \BibitemOpen
  \bibfield  {author} {\bibinfo {author} {\bibfnamefont {D.}~\bibnamefont
  {Giuliano}}, \bibinfo {author} {\bibfnamefont {D.}~\bibnamefont {Rossini}},
  \bibinfo {author} {\bibfnamefont {P.}~\bibnamefont {Sodano}},\ and\ \bibinfo
  {author} {\bibfnamefont {A.}~\bibnamefont {Trombettoni}},\ }\href
  {https://doi.org/10.1103/PhysRevB.87.035104} {\bibfield  {journal} {\bibinfo
  {journal} {Phys. Rev. B}\ }\textbf {\bibinfo {volume} {87}},\ \bibinfo
  {pages} {035104} (\bibinfo {year} {2013})}\BibitemShut {NoStop}%
\bibitem [{\citenamefont {Bañuls}\ and\ \citenamefont
  {Cichy}(2020)}]{banulsRPP2020}%
  \BibitemOpen
  \bibfield  {author} {\bibinfo {author} {\bibfnamefont {M.~C.}\ \bibnamefont
  {Bañuls}}\ and\ \bibinfo {author} {\bibfnamefont {K.}~\bibnamefont
  {Cichy}},\ }\href@noop {} {\bibfield  {journal} {\bibinfo  {journal}
  {\href{https://doi.org/10.1088/1361-6633/ab6311}{Rep. Prog. Phys.}}\ }\textbf
  {\bibinfo {volume} {83}} (\bibinfo {year} {2020})}\BibitemShut {NoStop}%
\bibitem [{\citenamefont {Wall}\ and\ \citenamefont {Carr}(2013)}]{Wall_2013}%
  \BibitemOpen
  \bibfield  {author} {\bibinfo {author} {\bibfnamefont {M.~L.}\ \bibnamefont
  {Wall}}\ and\ \bibinfo {author} {\bibfnamefont {L.~D.}\ \bibnamefont
  {Carr}},\ }\href@noop {} {\bibfield  {journal} {\bibinfo  {journal}
  {\href{https://dx.doi.org/10.1088/1367-2630/15/12/123005}{New Journal of
  Physics}}\ }\textbf {\bibinfo {volume} {15}},\ \bibinfo {pages} {123005}
  (\bibinfo {year} {2013})}\BibitemShut {NoStop}%
\bibitem [{\citenamefont {Soliverez}(1969)}]{soliverez1969}%
  \BibitemOpen
  \bibfield  {author} {\bibinfo {author} {\bibfnamefont {C.~E.}\ \bibnamefont
  {Soliverez}},\ }\href@noop {} {\bibfield  {journal} {\bibinfo  {journal}
  {\href{https://doi.org/10.1088/0022-3719/2/12/301}{Journal of Physics C:
  Solid State Physics}}\ }\textbf {\bibinfo {volume} {2}},\ \bibinfo {pages}
  {2161} (\bibinfo {year} {1969})}\BibitemShut {NoStop}%
\end{thebibliography}%

\end{document}